%% file: ms_glow.tex
\documentclass[12pt,twocolumn]{aastex701}
\usepackage{graphics}
\usepackage{color}
\usepackage{amsmath}
\usepackage{empheq} 
\usepackage{longtable}

\newcommand{\ltsimeq}{\la}
\newcommand{\gtsimeq}{\ga}
\newcommand{\lsun}{L$_{\odot}$}
\newcommand{\msun}{M$_{\odot}$}
\newcommand{\mstar}{M$_*$}
\newcommand{\mhi}{M$_{HI}$}

\newcommand{\hi}{H{\sc i}}
\newcommand{\hii}{H{\sc ii}}

\newcommand{\mlr}{$\Upsilon_*^{3.6\mu m}$}
\newcommand{\gals}{37}

\shortauthors{McQuinn et al.}
\shorttitle{GLOW: Paper I}

\begin{document}
\title{GLOW I: Comprehensive Measurements of Gas-Rich, Star-Forming, Low-Mass Galaxies in the Nearby Universe\footnote{Based on observations made with the NASA/ESA Hubble Space Telescope, obtained at the Space Telescope Science Institute, which is operated by the Association of Universities for Research in Astronomy, Inc., under NASA contract NAS 5-26555.}}

\author[0000-0001-5538-2614]{Kristen.~B.~W. McQuinn}
\affiliation{Space Telescope Science Institute, 3700 San Martin Drive, Baltimore, MD, 21218, USA}
\affiliation{Rutgers University, Department of Physics and Astronomy, 136 Frelinghuysen Road, Piscataway, NJ 08854, USA} 
\email[show]{kmcquinn@stsci.edu}

\author[0000-0003-4122-7749]{O. Grace Telford}
\affiliation{Department of Physics and Astronomy, University of Utah, 270 S 1400 E, Salt Lake City, UT 84112, USA} 
\email{grace.telford@utah.edu}
\affiliation{Rutgers University, Department of Physics and Astronomy, 136 Frelinghuysen Road, Piscataway, NJ 08854, USA} 

\author[0000-0002-2970-7435]{Roger Cohen}
\affiliation{Rutgers University, Department of Physics and Astronomy, 136 Frelinghuysen Road, Piscataway, NJ 08854, USA} 
\email{}

\author{Liese van Zee}
\affiliation{Department of Astronomy, Indiana University, 727 East Third Street, Bloomington, IN 47405, USA}
\email{}

\author[0000-0001-5368-3632]{Laura C. Hunter}
\affiliation{Department of Physics and Astronomy, Dartmouth College, 6127 Wilder Laboratory, Hanover, NH 03755, USA}
\affiliation{Department of Astronomy, Indiana University, 727 East Third Street, Bloomington, IN 47405, USA}
\email{}

\author[0000-0002-1813-4053]{Madison Smith}
\affiliation{Department of Astronomy, Indiana University, 727 East Third Street, Bloomington, IN 47405, USA}
\email{}

\author[0000-0003-0605-8732]{Evan D. Skillman}
\affiliation{University of Minnesota, Minnesota Institute for Astrophysics, School of Physics and Astronomy, 116 Church Street, S.E., Minneapolis, MN 55455, USA} 
\email{}

\author[0000-0002-7502-0597]{Benjamin Williams}
\affiliation{Department of Astronomy, University of Washington, Box 351580, Seattle, WA 98195, USA}
\email{}

\author[0000-0002-1821-7019]{John M. Cannon}
\affiliation{Department of Physics and Astronomy, Macalester College, Saint Paul, MN 55105, USA}
\email{}

\author[0000-0002-1979-2197]{Joseph Burchett}
\affiliation{Department of Astronomy, New Mexico State University, 1320 Frenger Mall, Las Cruces, NM 88003, USA}
\email{}

\author[0000-0002-4153-053X]{Danielle A. Berg}
\affiliation{Department of Astronomy, The University of Texas at Austin, 2515 Speedway, Stop C1400, Austin, TX 78712, USA}
\affiliation{Cosmic Frontier Center, The University of Texas at Austin, Austin, TX 78712, USA} 
\email{}

\author[0000-0002-1264-2006]{Julianne J.\ Dalcanton}
\affiliation{Center for Computational Astrophysics, Flatiron Institute, 162 Fifth Avenue, New York, NY 10010, USA}
\affiliation{Department of Astronomy, University of Washington, Box 351580, Seattle, WA 98195, USA}
\email{}

\author[0000-0001-8416-4093]{Andrew E.~Dolphin}
\affiliation{Raytheon, 1151 E. Hermans Road, Tucson, AZ 85756, USA}
\affiliation{University of Arizona, Steward Observatory, 933 North Cherry Avenue, Tucson, AZ 85721, USA}
\email{}
\author[0009-0007-6658-0318]{Suchindram Dasgupta}
\affiliation{West Virginia University, Department of Physics and Astronomy, 135 Willey St, Morgantwon, WV 26505, USA} 
\email{}

\author[0009-0000-9367-0983]{Avery Kiihne}
\affiliation{Rutgers University, Department of Physics and Astronomy, 136 Frelinghuysen Road, Piscataway, NJ 08854, USA} 
\email{}

\begin{abstract}
Gas-rich, star-forming, low-mass galaxies in the nearby universe are powerful laboratories for studying baryonic physics in detail including: stellar mass assembly, stellar feedback, chemical enrichment, and the interplay of the interstellar medium with star formation. Investigating these disparate yet interconnected processes requires data obtained by myriad observatories. Here, we present a comprehensive atlas of uniformly processed data on \gals\ low-mass galaxies within 6 Mpc. The atlas includes archival data on (i) the \hi\ from the Very Large Array observatory; (ii) resolved stars from Hubble Space Telescope optical imaging; (iii) Spitzer Space Telescope 3.6$\micron$ imaging; and (iv) optical imaging from ground-based telescopes. We also compile measurements of (i) tip-of-the-red-giant-branch (TRGB) distances to the galaxies; (ii) direct method gas-phase oxygen abundances and nitrogen to oxygen abundance ratios; (iii) constraints on the local environment around each galaxy; and (iv) other measurements from the literature. We supplement the data with new observations where needed to complete the measurements for all galaxies in the sample. From these data, we find good agreement between stellar masses measured from color-magnitude diagrams and those estimated from 3.6$\micron$ imaging by assuming a mass-to-light ratio. We also provide the first mapping of the \hi\ profiles as a function of structural parameters. These data sets and measurements are the foundation for the Galaxies Losing Oxygen via Winds (GLOW) project whose main aim is to characterize the star formation $-$ chemical enrichment cycle of low-mass galaxies by measuring the production, distribution, and retention of oxygen on a galaxy-by-galaxy basis.
\end{abstract} 
\keywords{Chemical enrichment (225), Stellar populations (1622), Dwarf irregular galaxies (417), Interstellar atomic gas (833), Circumgalactic medium (1879), Galaxy environments (2029)}

\section{Introduction}\label{sec:intro}
Star formation activity and the chemical evolution of galaxies are intrinsically linked. Stars not only produce chemical elements through nucleosynthesis, but feedback from stars is the primary agent for the dispersal of metals. Stellar feedback injects newly formed elements back into the interstellar medium (ISM), expels chemical elements into the circumgalactic medium (CGM), and, given sufficient energy and momentum of the feedback, can eject material out of a galaxy’s gravitational potential and into the intergalactic medium (IGM).

The cumulative effects of star formation and chemical enrichment on a galaxy are readily quantified on global scales via the mass$-$gas-phase metallicity (MZ) relation, which is a simple measure of the total mass in stars formed over the lifetime of a galaxy and the present-day census of the chemical element oxygen in the ISM. More detailed study of the interplay of star formation and chemical enrichment is challenging given the different channels metals can follow and the vast spatial and temporal scales involved (i.e., from local star-forming regions in the disks of galaxies to the IGM and from the early epochs of cosmic time to the present-day). While challenging, teasing apart the different components and factors responsible for the global MZ scaling relation can yield deep insights into galaxy growth and the baryon cycle, as well as providing clear and strong constraints for theoretical models of galaxies and the sub-grid physics employed in large cosmological simulations.

The Galaxies Losing Oxygen via Winds (GLOW) project aims to delve deeply into the star formation $-$ chemical enrichment cycle of low-mass galaxies by measuring the production, distribution, and retention of metals, traced by oxygen on a galaxy-by-galaxy basis. We focus on oxygen as a tracer of the metals in the galaxies. Oxygen has a relatively simple nucleosynthetic pathway that is well-understood (e.g., via stars that undergo core-collapse supernovae), is the most abundant heavy element, can be readily measured with optical spectra, and is considered a direct tracer of material expelled in supernovae ejecta. 

For the ``accounting'' of oxygen in GLOW, we focus on galaxies that are nearby (D$\ltsimeq$6 Mpc), given that the most detailed physical constraints on galaxies come from studying spatially resolved systems where the observations are the best. Nearby galaxies offer the added advantage of being well-studied at many wavelengths. Indeed, much of the data needed to quantify the star formation history of a system and constrain the oxygen content in different galaxy components already exist in data archives.

The GLOW project includes a statistically significant sample size of star-forming dwarf galaxies spanning three decades in stellar mass (\mstar\ $\sim10^{6.5} - 10^{9.5}$ \msun), thus enabling us to identify and explore trends on various axes of galaxy properties (e.g., stellar mass, gas content, metallicity, environment, etc.). We focus on low-mass galaxies specifically to sample the stellar mass range where the greatest changes in metal retention are expected: namely the mass regime where the primary driver of metal loss, the energetics of star formation activity, can more readily overcome the gravitational binding energy of the shallow potential wells \citep[e.g.,][]{MacLow1999}. We intentionally include galaxies that straddle the boundary between dwarf and more massive galaxies as this upper mass boundary (\mstar\ $\sim10^9$ \msun) overlaps with previous studies of more massive galaxies \citep[e.g.,][]{Peeples2014, Telford2019}. Together, these studies enable analysis of galaxies across nearly 5 orders of magnitude in stellar mass, and can be extended to even lower-masses \citep[e.g.,][]{McQuinn2015f}. Moreover, the overlap in galaxy properties across studies enables a comparison of the metal retention fractions in galaxies in the same mass regime but determined from very different techniques and using disparate galaxy samples as a check on potential systematic uncertainties or biases.

Here, we present the data and data processing needed for the detailed measurements of the GLOW project including: characterizing the properties of the sample, deriving the full star formation histories and age-metallicity relations based on resolved stellar populations work, and quantifying the environment of the galaxies. We also provide novel measurements of the spatially resolved \hi\ mass profiles of the galaxies and a careful comparison of stellar masses derived from integrated light vs.\ color-magnitude diagram (CMD) derivations. The suite of data products, summarized in Table~\ref{tab:data_chart}, is a rich resource for many future studies of nearby dwarf galaxies. In a companion paper, we present the census of oxygen in the sample and a measure of the mean metal retention fraction as a function of galaxy properties (McQuinn et al. submitted; Paper~II).

The paper is organized as follows. The galaxy sample and overview of the workflow for the GLOW project are described in Section~\ref{sec:sample}. We then jointly describe the data handling and analysis on each of the  data sets. Section~\ref{sec:structural} presents the ground-based optical imaging and {\it Spitzer Space Telescope} IRAC 3.6 $\mu$m imaging with determinations of the geometry and structural parameters. Section~\ref{sec:masses} details the calculation of stellar masses from {\it Spitzer Space Telescope} IRAC 3.6 $\mu$m imaging and compares those values with CMD-based stellar masses. Section~\ref{sec:sfhs} describes the data handling of the HST optical imaging and the derivation of the star formation histories (SFHs) and age-metallicity relations (AMRs). Section~\ref{sec:hi} focuses on the overall gas content and the radial distribution of \hi\ flux from in the galaxies from analysis of VLA data. Finally, as the evolution of galaxies can be impacted by environment, we use existing catalogs of nearby galaxies to quantify the environment of each galaxy using a number of metrics, described in Section~\ref{sec:env}. A summary of the work is presented in Section~\ref{sec:conclusions}. Atlases of the HST data including the HST fields of view, CMDs from the HST imaging, and the SFHs and AMRs reconstructed from the CMDs, are provided in the Appendix. 

\section{Galaxy Sample and Overview of Work Flow}\label{sec:sample}
\subsection{The GLOW Sample}
The GLOW sample includes \gals\ low-mass, star-forming galaxies in the distance range 0.75 $-$ 6 Mpc, with a mean distance of 3.3 Mpc. Since GLOW is primarily constructed using archival observations, nearby galaxies were selected based on the availability of data, including: (i) optical imaging from HST suitable for SFH work, namely imaging in either the F606W, F475W, or F555W filter and in the F814W filter reaching a minimum photometric depth of $\sim$2 mag below the tip of the red giant branch (TRGB). This requirement imposed the outer 6 Mpc distance limit, and ensured that the stellar populations in the galaxies are well-resolved with HST imaging instruments; (ii) direct-method gas-phase oxygen abundances measured from \hii\ regions, which dictated that all galaxies in the sample are actively star-forming, and this excluded gas-poor galaxies within the Local Group; (iii) {\it Spitzer Space Telescope} IRAC 3.6$\mu$m imaging; (iv) \hi\ interferometric data mapping the spatially resolved neutral hydrogen; and (v) ground-based optical imaging. There were a few galaxies without the necessary \hi\ data and one system without a reported gas-phase oxygen abundance in the literature. For these cases, we obtained the necessary data sets including VLA B and C- configuration observations for NGC~4068, NGC~6789, and UGC~08638, previously published in \citet{Richards2018, Hunter2022, Sacchi2024, Hunter2025}, and optical spectra of NGC~4068 obtained on the MMT telescope from which we measured the gas-phase oxygen and, for completeness, nitrogen abundance.

The archival data used for GLOW have been accumulated via numerous science programs executed over decades and include observations spanning from the radio to the ultraviolet. It is an impressive accomplishment by the community that such a wide range of comprehensive data sets exist on nearby galaxies. 

The final GLOW sample spans the following range in mass, luminosity, and gas-phase oxygen abundances:
\begin{itemize}
\item[] Stellar Mass: {$6.5 < $ log (M$_*$/\msun) $< 9.5$}
\item[] \hi\ Mass: {$6.5 < $ log (M$_{HI}$/\msun) $<9.5$}
\item[] Absolute B Mag: $-15 < M_B < -11.5$
\item[] Oxygen Abundance: $7.3 < 12 +$ log(O/H) $<8.3$
\end{itemize}

\noindent For comparison, the LMC has \mstar\ $=2.7 \times 10^9$ \msun, \mhi\ $=4.8 \times 10^8$ \msun, $M_B=-17.9$ mag, and 12$+$log(O/H) $=8.37$ \citep{vanderMarel2002, StaveleySmith2003, deVaucouleurs1972, DominguezGuzman2022}.

Table~\ref{tab:properties} lists the galaxies, ordered by Right Ascension (J2000), with their coordinates, distances, gas-phase oxygen abundances, and the ratios of their nitrogen to oxygen abundances. The distances are all based on the TRGB method measured from HST imaging of the resolved stellar populations. The oxygen abundances were determined using the temperature sensitive [\ion{O}{3}] $\lambda$4363 auroral line with the ``direct-method''. References for the distances and abundance measurements are provided in the table.

\begin{figure}
\begin{center}
\includegraphics[width=0.4\textwidth]{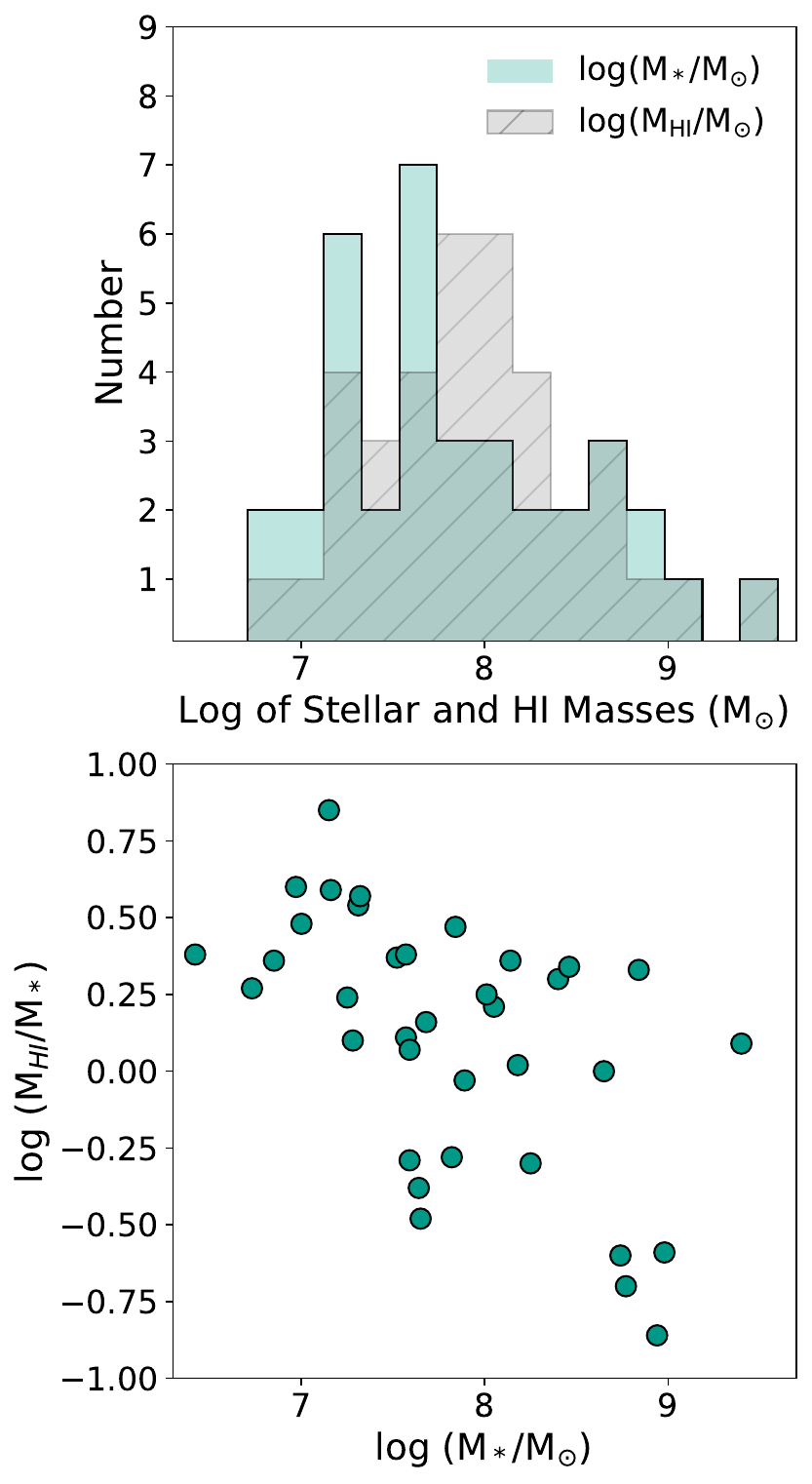}
\end{center}
\caption{Top panel: The GLOW galaxies span 6.5 $<$ log(Mass/\msun) $<$9.5 in both stars and gas. Bottom panel: The galaxies have a range of \mhi/\mstar\ ratios but all are gas-rich and actively star-forming, enabling robust gas-phase oxygen abundances to be measured from their \hii\ regions. }
\label{fig:histo_mass}
\end{figure}

Figure~\ref{fig:histo_mass} presents the distributions of the stellar mass and \hi\ mass in the upper panel and the ratio of \hi\ to stellar mass ($M_{HI} / M_*$) as a function of \mstar\ in the lower panel. The stellar masses are based on the integrated 3.6$\micron$ fluxes of the galaxies and an adopted mass-to-light ratio, described in detail in Section~\ref{sec:masses} and listed in the associated Table~\ref{tab:stellar_masses}. The \hi\ masses were calculated from \hi\ fluxes measured from VLA and ATCA interferometric observations of the 21-cm line. Single dish fluxes were used in cases where the interferometer data and published single dish fluxes were discrepant by more than 25\%. The \hi\ analysis is described in detail in Section~\ref{sec:hi} and quantities are listed in Table~\ref{tab:hi} in that section. The \hi\ fluxes are converted to masses using the TRGB distances provided in Table~\ref{tab:properties}. 

As expected for actively star-forming systems, the galaxies are gas-rich. The majority have gas-to-star ratios typical of dwarfs: $0.5 \ltsimeq M_{HI}/M_* \ltsimeq 4$. The sample also contains two outliers with a much larger \hi\ content relative to the stars (\mhi/\mstar\ $\sim 7-13$) and four systems with somewhat lower \hi\ content (\mhi/\mstar\ $ \sim 0.1-0.5$).

\begin{figure}
\includegraphics[width=0.48\textwidth]{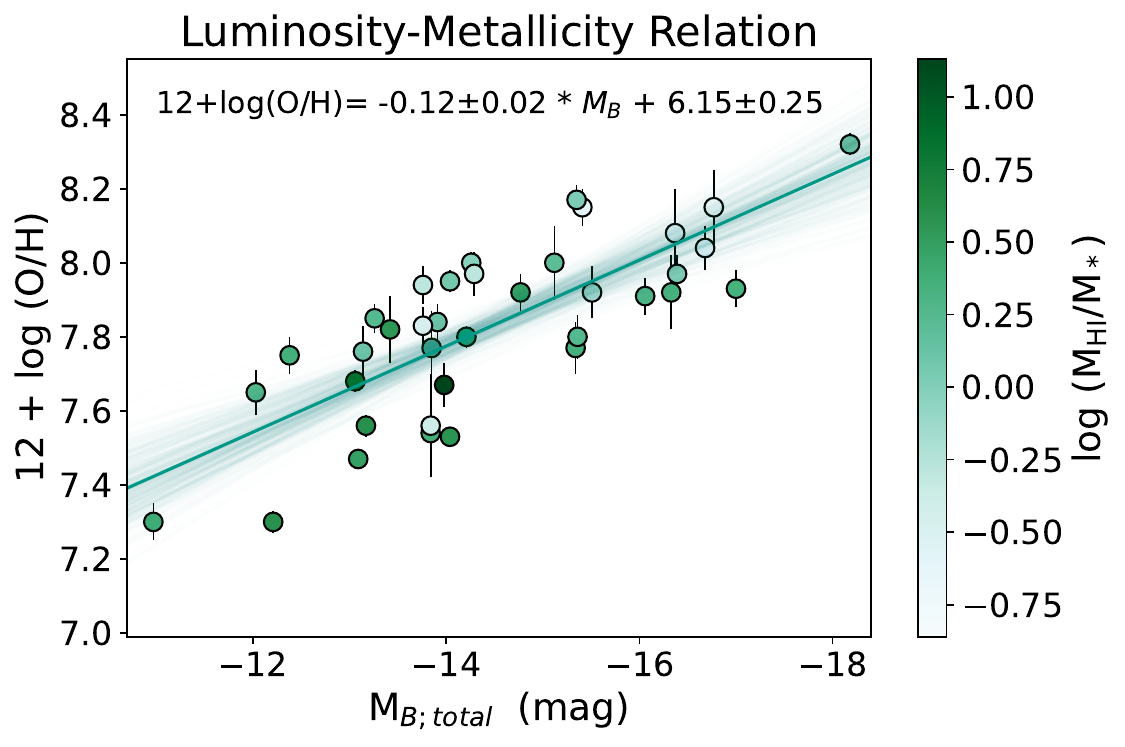}
\includegraphics[width=0.48\textwidth]{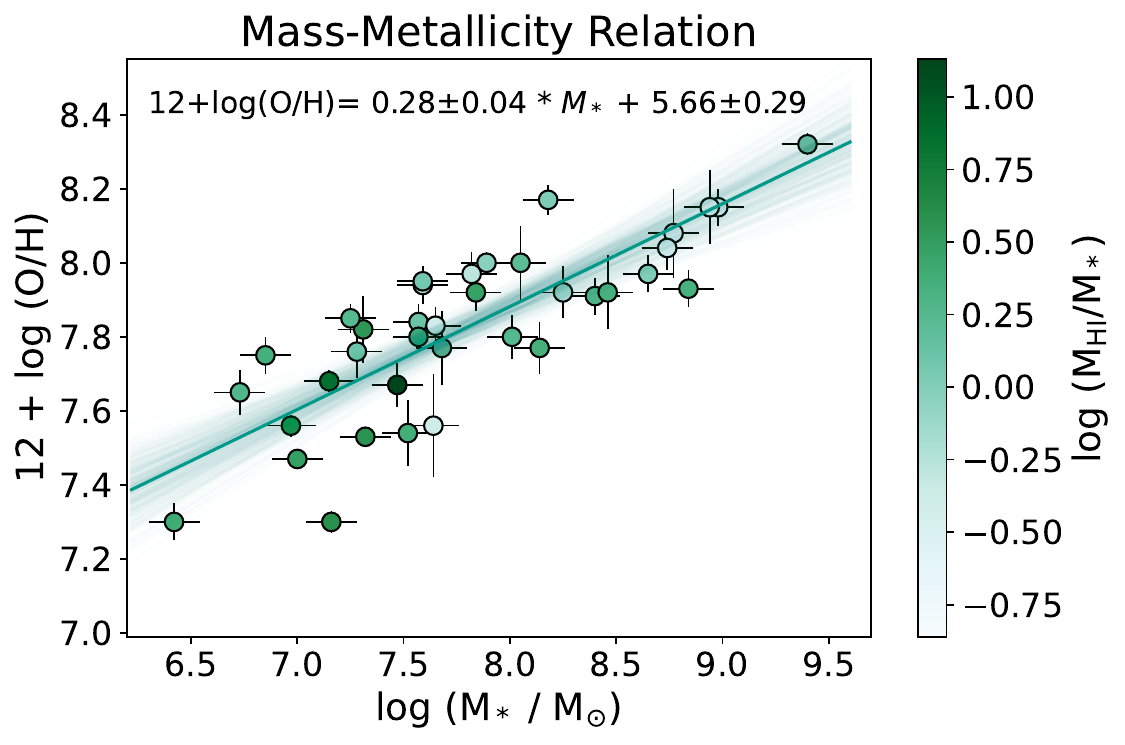}
\includegraphics[width=0.48\textwidth]{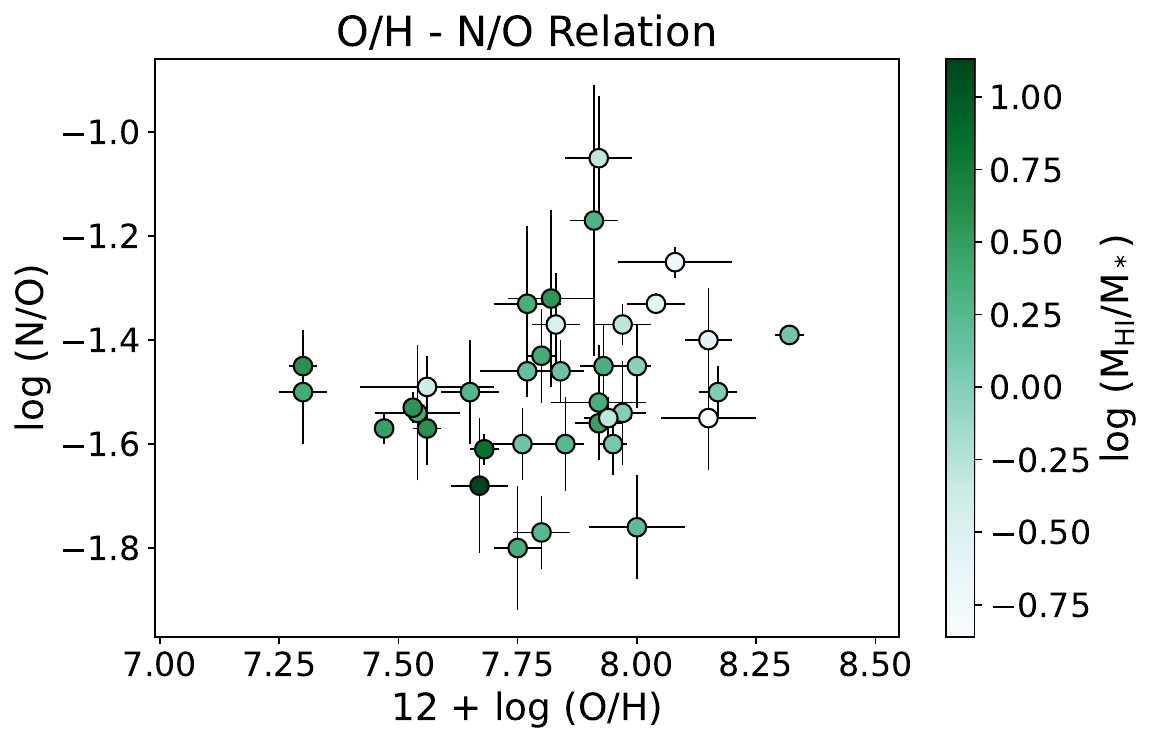}
\caption{LZ relation, MZ relation, and the log of the nitrogen-to-oxygen abundances as a function of oxygen abundance for the GLOW sample. Points are color-coded by the log of their \mhi/\mstar\ ratios. Values of \mstar\ are based on the total 3.6$\micron$ fluxes and adopting a mass-to-light ratio (see \S\ref{sec:masses}). Best fitting lines (thicker line) and realizations within 1$\sigma$ confidence intervals (thinner lines) are shown for the LZ and MZ relations.}

\label{fig:lzr_mzr_no}
\end{figure}

Figure~\ref{fig:lzr_mzr_no} presents the luminosity-metallicity (LZ) relation (top panel) and the MZ relation (middle panel) for the GLOW galaxies with points color-coded by their \mhi/\mstar\ ratios. The best-fitting lines to the LZ and MZ relations are shown as thick solid lines with realizations within the $1\sigma$ confidence intervals as thinner lines. These fits are based on linear regression analysis determined using the python package {\sc linmix} which takes into account uncertainties in both variables. For completeness, the best-fit lines are reproduced in Equations~\ref{eq:LZR} and \ref{eq:MZR}:
\begin{equation}
12+\rm{log(O/H)} = -0.12\pm0.02 \times M_B + 6.15\pm0.25
\label{eq:LZR}
\end{equation}
\begin{equation}
12+\rm{log(O/H)} = 0.28\pm0.04 \times M_* + 5.66\pm0.29
\label{eq:MZR}
\end{equation}

\noindent Our best-fitting lines are in excellent agreement with the high-fidelity LZ and MZ relations of typical field galaxies in the Local Volume reported in Figures~3 and 5, respectively, in \citet{Berg2012}. 

As the LZ and MZ relations of GLOW galaxies agree with the established LZ and MZ relations, their present-day oxygen abundances are likely mainly driven by secular evolution rather than a stochastic external event such as a tidal interaction or the infall of pristine gas. There is growing evidence that galaxies significantly offset from the LZ relation (i.e., galaxies with lower abundances and/or higher luminosities than typical field galaxies) have likely experienced recent infall of less chemically enriched gas, either from an outer disk, another galaxy, or the larger environment. This infall is thought to dilute their gas-phase oxygen abundance and trigger increased star formation activity \citep[see, e.g.,][and references therein]{Ekta2010b, McQuinn2020, Breneman2025}. In a future paper, we will present the MZ relation as a function of time using the SFHs and age-metallicity relations and explore the distribution of galaxies in the MZ relation in more detail (K.~B.~W.~McQuinn, in preparation).

\input{tab1.tex}

Figure~\ref{fig:lzr_mzr_no} also presents the logarithm of the nitrogen-to-oxygen ratio as a function of gas-phase oxygen abundance for the GLOW galaxies (bottom panel), with points color-coded by their \mhi/\mstar\ ratios. The N/O ratios remain approximately constant as a function of oxygen abundance but exhibit significant scatter. No apparent trend is seen between N/O ratios and the \hi-to-star mass ratios.

Nitrogen is thought to have a complex origin, including production as a primary element in the CNO cycle at low metallicities and as a secondary element at high metallicities. The significant scatter at low metallicities suggests either an additional nucleosynthetic pathway for N and/or, given that N and O are produced in stars of different masses and are released on different timescales \citep[e.g.,][]{Vincenzo2016}, a possible time delay in nitrogen enhancement (from e.g., asymptotic giant branch (AGB) stars and Wolf-Rayet stars). Hints of such age sensitivity were reported by \citet{vanZee2006}, who measured lower N/O ratios for blue galaxies and higher N/O ratios for redder galaxies. Correlations between the variation in N/O ratios and the SFHs of the galaxies will be investigated in a future GLOW paper. 

\subsection{Overview of Work Flow}\label{sec:workflow}
As stated above, the primary scientific objectives of the GLOW project include tracking the production, distribution, and retention of metals in galaxies. These goals require empirical constraints from various galaxy components, each necessitating observations at different wavelengths. To provide the reader with an orientation to the complex set of observations, we include high-level descriptions of each data set in our analysis with a schematic in Table~\ref{tab:data_chart}.

Briefly, the global stellar properties of the sample are derived from broad-band photometry. These observations enable the measurement of galaxy geometry and structural parameters (semi-major axis, ellipticity, position angle, scale length, and central surface brightness), integrated magnitudes at various wavelengths, and the total stellar masses of the galaxies. The data sources for these stellar properties include ground-based optical imaging and {\it Spitzer Space Telescope} IRAC 3.6$\mu$m imaging (\S\ref{sec:structural} and \S\ref{sec:masses}). These integrated light measurements and resulting properties will serve as valuable benchmarks for future studies, particularly for anchoring comparisons to more distant galaxies that are too far for resolved studies. 

The quantity of metals contained within the stars and stellar remnants is determined  using CMD fitting techniques which simultaneously constrains the SFH and AMR. The input data for this analysis is HST optical imaging of the resolved stars (\S\ref{sec:sfhs}). The quantity of metals retained in the gas is measured from gas-phase abundances, gas masses, and gas distributions. Data for this analysis include spectroscopic abundances measured from ground-based optical spectra and \hi\ column densities as a function of radius from interferometric data cubes (\S\ref{sec:hi}). The \hi\ column density maps and gas-phase oxygen abundances can also be used to estimate the amount of oxygen retained in the dust of the galaxies.

Constraints on the quantity of metals ejected from the galaxy disk but retained in the halo are based on abundance measurements of the CGM. Data for this purpose include HST COS absorption spectra of the CGM illuminated by background quasars along the line of sight, available for a subset of the sample and published separately in \citet{Zheng2024} and supplemented by other studies including \citet{Johnson2017} and \citet{Tchernyshyov2022}.

Finally, we quantify the density and complexity of the local environment around each system to explore possible environmental effects related to the metal retention fraction using existing galaxy catalogs, specifically the Updated Nearby Galaxy Catalog \citep{Karachentsev2013} (\S\ref{sec:env}).

\input{tab2.tex}

\section{Galaxy Structural Parameters}\label{sec:structural}
The geometry and structural parameters of galaxies are used in our analysis in several ways. First, they define the spatial extent of the galaxies, allowing us to match the same area of an individual system across multiple data sets and establish a consistent physical scale within which we measure integrated fluxes at various wavelengths. Second, these parameters form the basis for binning stellar populations radially, facilitating the reconstruction of SFHs and AMRs within physically motivated annuli, along with a local measure of data completeness and uncertainties (see \S\ref{sec:sfhs}). 

We determined the galaxy shapes through isophotal fitting to IRAC 3.6$\micron$ imaging, which included determining the semi-major, semi-minor, and position angles of the stellar disks. After adopting these parameters, we determined the central surface brightness and scale lengths of the galaxies through surface brightness profile fits in the B-band, R-band, and 3.6$\micron$ data. Finally, we estimated the total flux in each wavelength regime at different spatial extents. This approach allows us to compare the physical characteristics of the galaxies from the optical to the IR and provides a consistency check on our measurements. We first describe the data handling procedures and then elaborate on the isophotal and surface brightness profile fits.

\input{tab3.tex}

\input{tab4.tex}

\subsection{B-band, R-band, 3.6$\micron$ Data Handling}
The ground-based B-band and R-band imaging of the galaxies were previously obtained over numerous observing runs spanning several years from the WIYN 0.9m telescope (PI: Liese van Zee). The ground-based optical imaging was reduced in the standard way and calibrated using spectrophotometric standard stars. 

The IRAC 3.6$\mu$m data were obtained and processed as part of the Local Volume Legacy (LVL) program \citep[for a complete description of the observations and data handling see][]{Dale2009}. There were four galaxies that were not part of LVL (IC~4662, NGC~1569, NGC~6789, UGC~6456) which we processed separately, following the same procedure given in \citet{Dale2009}.

Before determining the structural parameters and fitting surface brightness profiles, foreground stars and background galaxy contaminants were removed from the B, R, and 3.6$\micron$ images. This is particularly important for the 3.6$\micron$ data as IRAC is a sensitive instrument, and given the lower surface brightnesses of the galaxies, contamination from sources that are not part of the target system can significantly impact the integrated fluxes used in our stellar mass calculations. Briefly, we carefully excised foreground stars and background galaxies in the regions of the target galaxies. We identified brighter contaminating sources based on morphologies using the high-resolution HST imaging that overlaps with the IRAC fields of view. This identification of contaminants was not just visual; we also used the PSF-fit metrics produced by the photometry run on the HST images (see Section~\ref{sec:phot} below) to help identify sources that were not star-like in their light distribution. For regions outside of the HST footprint, we relied on ground-based optical imaging; while these data are not as high-resolution as the HST imaging, they still exceed that of the 3.6$\micron$ data. Once identified, we removed these sources with a simple interpolation from surrounding pixels using the {\sc iraf} task {\sc imedit}.

For fainter sources, it becomes more challenging to distinguish between a star in the galaxy and a contaminating source. Because of this ambiguity, we take a statistical approach. Instead of trying to discern whether a faint source is a bona fide member of the galaxy, we measure the average flux per unit area of fainter sources in an off-target region of the images, similar to determining a sky background flux, and subtract the flux scaled to the appropriate area from the global values to account for the contribution of faint contaminants to the total 3.6$\micron$ flux. 

\begin{figure}
\includegraphics[width=0.48\textwidth]{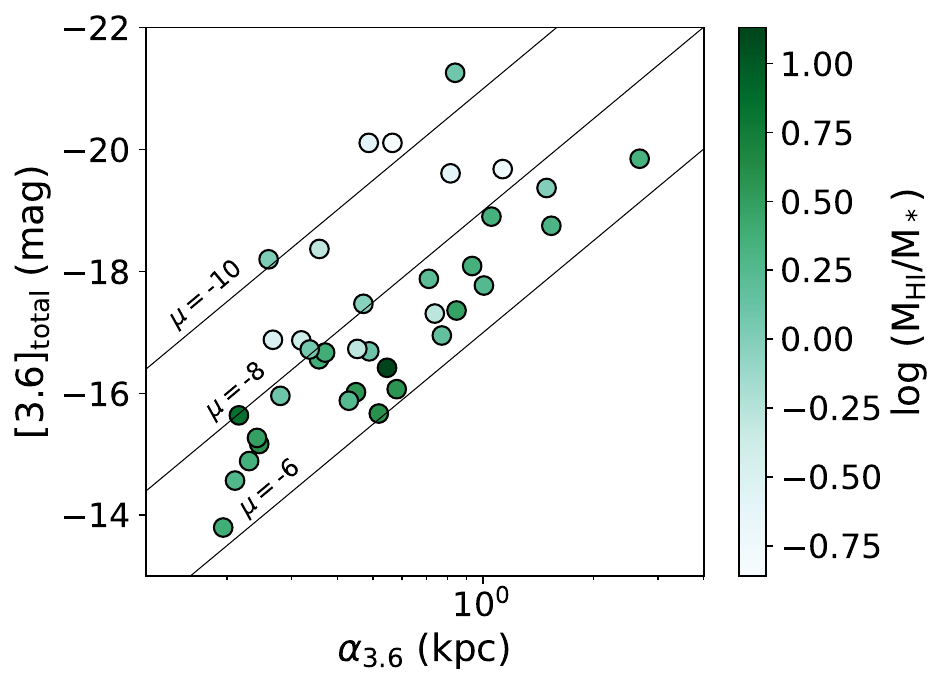}
\includegraphics[width=0.48\textwidth]{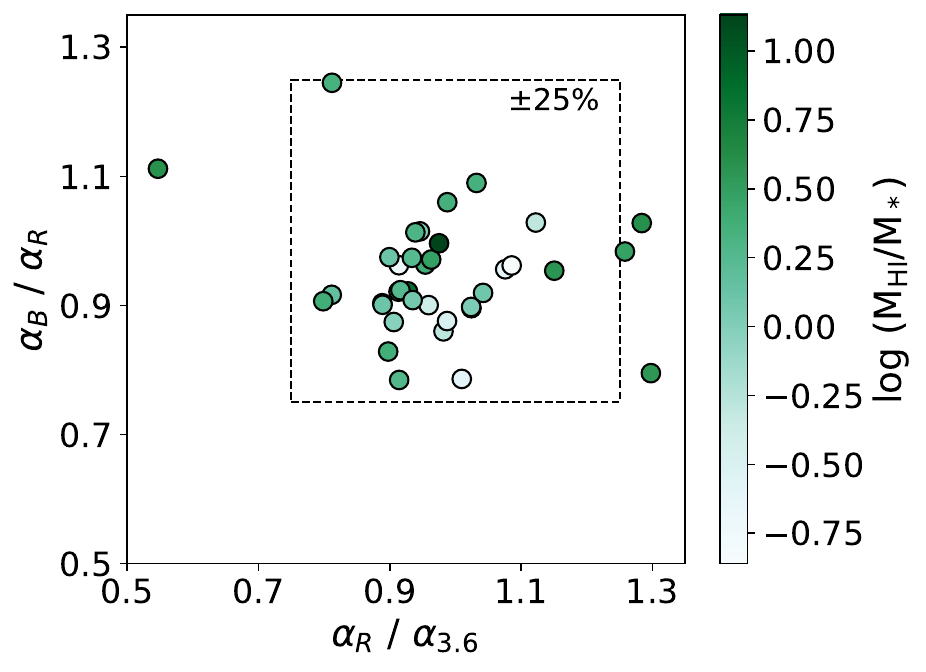}
\caption{Top panel: Total 3.6$\micron$ magnitudes as a function of scale length $\alpha$. Diagonal lines present lines of constant surface brightness $\mu =-10, -8, -6$ mag per \arcsec$^2$, from top to bottom, assuming the mean sample distance of 3.3 Mpc. The general trend is that larger galaxies are also brighter and more massive, as expected. Bottom panel: Comparison of the ratios of B, R, and 3.6$\micron$ scale lengths based on the cleaned images. The dashed box encompasses the points that agree within $\pm$25\%. While there are a few outliers, the overall good agreement between the ratios is an indication that the removal of contaminants from the imaging has not introduced significant biases when measuring the structural parameters of the galaxies, particularly since gradients in the stellar populations are expected. Points in both panels are color-coded by the ratio of \mhi/\mstar.}
\label{fig:scalelengths}
\end{figure}

\subsection{Structural Parameters and Surface Brightness Fits}
Once the images were cleaned, we characterized the shape of a galaxy by fitting isophotes to the outer disk in the 3.6$\micron$ images. Focusing on the outer regions helps avoid the often irregular morphology of higher surface brightness, star-forming regions typical of the inner disks. Note that, despite focusing on the outer disks of the galaxies to define ellipses and measure the structural parameters, the light distribution outside the central regions in star-forming dwarfs may still deviate from an elliptical shape. This is an intrinsic uncertainty in fitting light profiles to dwarf irregular galaxies but evidence suggests that the resulting impact on structural parameters and the related measurements (e.g., total magnitudes) is modest and smaller than other dominant sources of uncertainty (such as contaminant removal) \citep[e.g.,][]{Hunter2006, Pohlen2006}. The best-fitting elliptical parameters, including the position angle in the direction of the redshifted side of the galaxy, are listed in Table~\ref{tab:structural}. 

\begin{figure}
\includegraphics[width=0.48\textwidth]{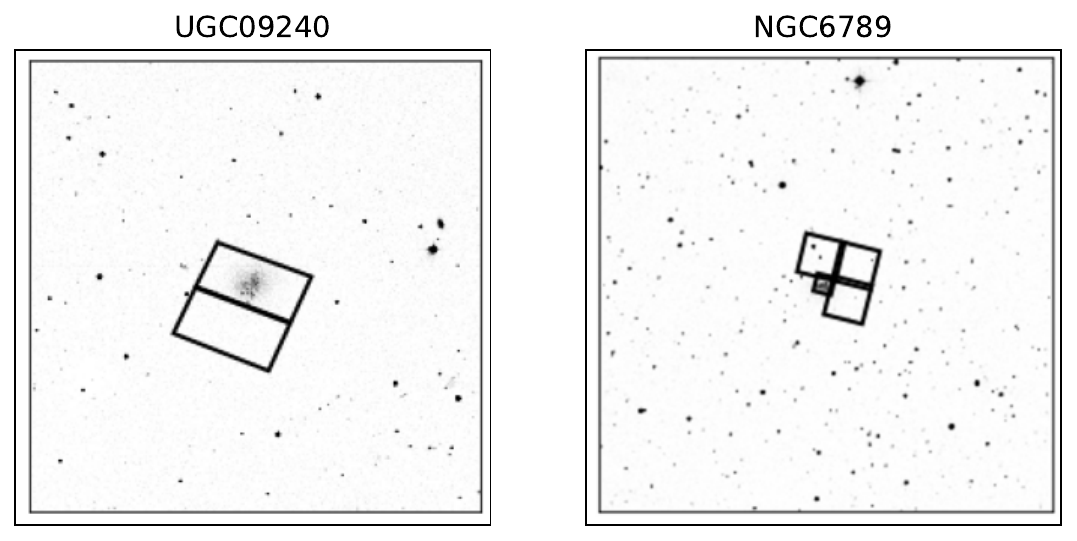}
\caption{HST footprints overlaid on 15\arcmin $\times$ 15\arcmin\ Digital Sky Survey (DSS) images on two galaxies, UGC~09240 and NGC~6789, showing there is a range in areal coverage in the HST data. For the first system, the HST imaging provides excellent coverage of the main stellar disk whereas, for the second system, the coverage is incomplete. To ensure our stellar mass calculations uniformly represent a fuller extent of stellar components across the sample, we opted to use IRAC 3.6$\micron$ imaging and adopt a \mlr\ approach for estimating the stellar masses. A comprehensive atlas of DSS images of the galaxies with HST footprints overlaid for the rest of the sample is provided in Appendix~\ref{sec:atlas_fov}.}
\label{fig:atlas_example}
\end{figure}

After fixing the shape of the outer isophote from the 3.6$\micron$ imaging, ellipses were grown over larger annuli to create extended surface brightness profiles individually in the B-band, R-band, and 3.6$\micron$ data. Surface brightness profiles were then fit with an exponential function and extrapolated based on the curves of growth. Whenever possible, isophotes fit to larger radii were used to help constrain the surface brightness fit. Note that we did not include the central, higher surface brightness regions in the profiles as they are not always well-fit with an exponential disk model. From the surface brightness fits, we extrapolated inward to identify the central surface brightness ($\mu$; corrected by the axial ratio) and scale length ($\alpha$) for each wavelength regime; we also determined the half-light radii. All values are reported in Table~\ref{tab:structural}.

Figure~\ref{fig:scalelengths} (top panel) presents a summary of the 3.6$\micron$ scale lengths as a function of the absolute magnitudes, M$_{3.6\mu m; total}$, color-coded by \mhi/\mstar. The faintest galaxies have the smallest scale lengths and are most gas-rich while the brighter galaxies are more extended as expected. We also overplot lines of constant surface brightness assuming the mean distance to the sample of 3.3 Mpc. Figure~\ref{fig:scalelengths} (bottom panel) compares the ratio of the R-band to 3.6$\micron$ scale lengths with the ratio of the B-band to R-band scale lengths determined from surface brightness profile fits to the cleaned images. The ratios of the scale lengths scatter around a value of unity with deviations most likely reflecting that gradients in stellar population gradients are present. Note that there is an extreme outlier, UGCA~292 (located in the upper far left of the plot), which is a low enough surface brightness galaxy that the imprecise nature of contaminant removal (particularly in the 3.6$\micron$ images) has a larger impact on the recovered properties. In the downstream analysis on stellar masses, UGCA~292, as well as another low-surface brightness galaxy subject to higher uncertainties in contaminant cleaning, UGC~08638, are noted with open plot symbols to denote the higher uncertainties on the stellar extents and masses. 

As a consistency check on our surface brightness fits and derived parameters, we compared our scale lengths and central surface brightness values measured from the R-band and 3.6$\micron$ images to previously published values. We focus on results from \citet{Herrmann2013} who performed extensive analysis on surface brightness profiles as a function of wavelength and whose sample include galaxies that overlap with GLOW. We found overall good agreement between our values and those reported that were derived from the outer parts of the galaxies, particularly in the 3.6$\micron$ results, adding confidence to the robustness of our fits.

\subsection{B-band, R-band, 3.6$\micron$ fluxes}\label{sec:fluxes}
We measured the fluxes in three different apertures for all bandpasses. First, the flux was measured internal to the elliptical isophote corresponding to the extent of the galaxy identifiable by eye in the images. This extent is roughly equivalent to a B-band surface brightness of 25 mag per arcsec$^2$ and a 3.6$\micron$ surface brightness of 22.5 mag per arcsec$^2$ (similar to the apertures used in the LVL program). The IRAC data reach this depth for the majority of the galaxies; it is also a high enough surface brightness that background uncertainties do not significantly impact the flux measurements, and, thus, the flux measurements are robust. In a few cases (WLM, NGC~2366, UGC~05364, NGC~4449, NGC~5253), the 3.6$\micron$ surface brightness did not reach the faint 22.5 mag per arcsec$^2$ within the IRAC footprints and we therefore defined the ellipse for isophote fitting to be as large as possible in the available field of view. The 3.6$\micron$ fluxes were then converted to Vega magnitudes assuming the 3.6$\mu$m zeropoint of 280.9 Jy \citep{Reach2005}. 

The B-band and R-band magnitudes were calculated based on calibrations from spectrophotometric standard stars data. These 3.6$\micron$, B-band, and R-band apparent magnitudes are a direct measure from the data and are simply labeled as magnitudes in Table~\ref{tab:fluxes}. 

Second, the total flux from each galaxy was estimated by extrapolating the surface brightness profiles to infinity. The fluxes measured in this way include emission from the extended stellar disks and represent the total flux from the stellar populations. Magnitudes computed from these fluxes are labeled as {\em total} in Table~\ref{tab:fluxes}.  We conservatively adopt an uncertainty of 0.1 mag on the extrapolated magnitudes, consistent with values reported by previous studies of surface brightness profiles \citep[e.g.,][]{Graham2005, Hunter2006, Pohlen2006, MunozMateos2015}. Our values from these galaxies agree overall with those measured using aperture photometry in the LVL program \citep{Dale2007}, providing confirmation of our extrapolation approach and that the cleaned 3.6$\micron$ images yield an accurate measure of the flux within the galaxies, despite the challenges inherent in contamination removal.

Third, the fluxes were measured within the field of view of the HST observations by matching an aperture to the footprint of the HST images. The 3.6$\mu$m fluxes in these apertures enable a one-to-one comparison of the stellar masses based on the 3.6$\mu$m mass-to-light ratio approach with the CMD-based stellar masses derived from the HST data. For completeness, we also calculate the B-band and R-band fluxes in these same regions. Magnitudes computed from these fluxes are labelled as {\sc hst FoV} in Table~\ref{tab:fluxes}.

For a few galaxies, the HST field of view is notably larger than the stellar disks and, thus, the footprint of the observations also includes significant sky coverage.\footnote{NGC~3741, NGC~4459, UGC~00683, UGC~08024, UGC~09128, UGC~09240, and UGCA~281.} In these cases, we checked that the IRAC fluxes summed in the matched HST apertures were not dominated by the IRAC background or uncertainties in the IRAC background using two approaches. First, we compared the 3.6$\mu$m fluxes from the HST apertures with the total 3.6$\mu$m fluxes extrapolated from the surface-brightness fitting. We found the values were comparable in all cases, which indicates that the summed flux from the larger HST apertures are not biased high due to the background. Second, we compared the stellar masses based on the 3.6$\mu$m and SFHs for these galaxies and found the differences between the values consistent with the rest of the sample.

\begin{figure}
\begin{center}
\includegraphics[width=0.35\textwidth]{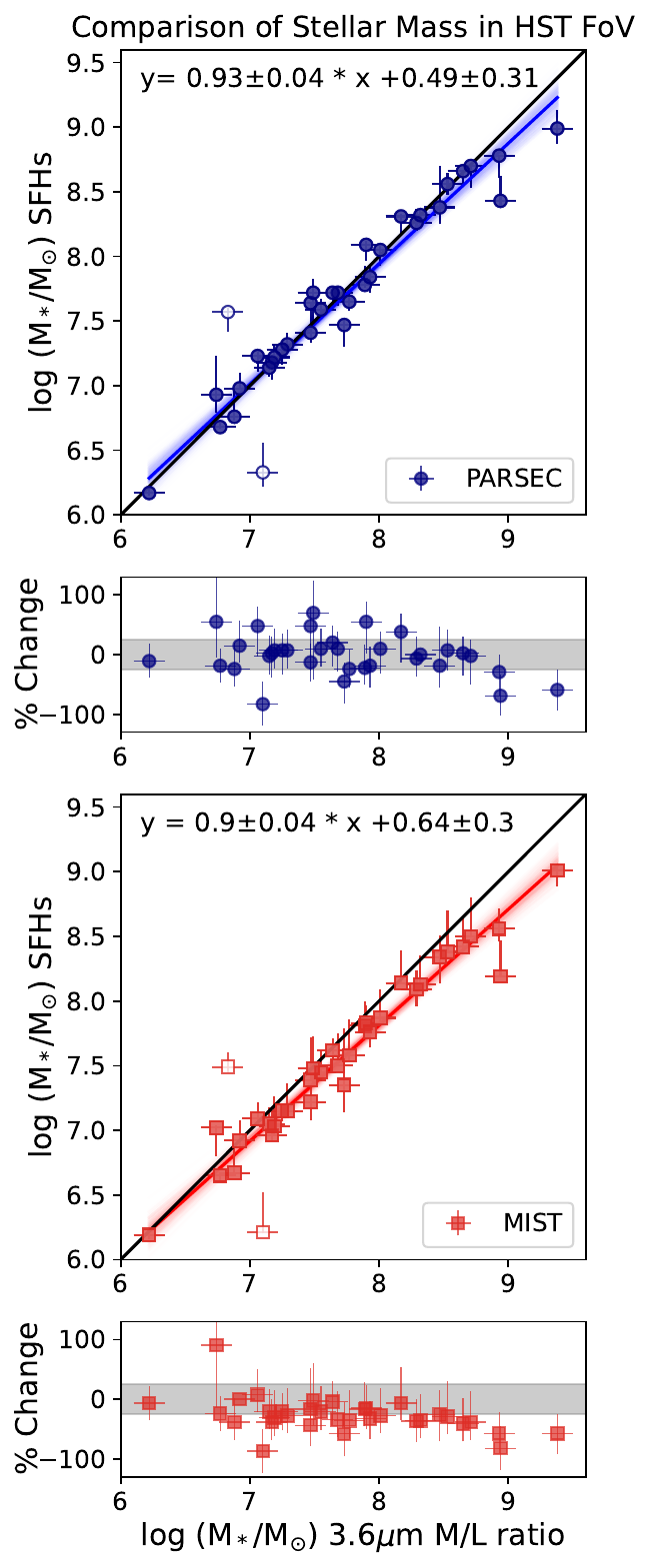}
\vspace{-0.25in}
\end{center}
\caption{Comparison of stellar masses and their relative change based on the 3.6$\micron$ fluxes and assuming a M/L ratio with those derived from the CMD-fitting using the PARSEC models (top two panels) and MIST models (bottom two panels) in matched fields of view. For the mass comparison, we also show a line of unity (black) and the best-fitting line to the points taking into account the uncertainties (line fits are given in the legends). Galaxies where the removal of contamination in the 3.6$\micron$ imaging had higher uncertainties are shown as unfilled points and were not included in the line fits and are off the scale of the relative change plots. The shaded gray bands highlight $\pm$25\% in the relative change in masses using the 3.6$\micron$-based masses as a reference. The masses derived from these very disparate methods are in very good agreement with each other. The MIST models shows a slight bias towards lower values relative to masses calculated from the 3.6$\micron$ fluxes.}
\label{fig:mstar}
\end{figure}

\section{Stellar Masses from from 3.6$\micron$ imaging}\label{sec:masses}
We determined the stellar masses of the galaxies based on the 3.6$\micron$ fluxes and assuming a stellar mass-to-light ratio (\mlr). As the production of oxygen depends on the measurement of the total stellar mass in a galaxy, extra care was taken to ensure that the stellar mass measurements were performed uniformly and encompassed the full stellar disk of the galaxies. Ideally, we would use the star counts from the CMDs and best-fitting SFHs to determine the stellar masses. However, the angular extent of many galaxies is larger than the HST footprint. Demonstrating this, Figure~\ref{fig:atlas_example} presents two representative examples of the areal coverage of the HST data overlaid on 15\arcmin $\times$ 15\arcmin\ DSS images \citep{Bonnarel2000}. While the HST footprint for the first galaxy encompasses the stellar disk of the galaxy, the coverage of the second galaxy is not sufficient to provide a census of the stellar mass. Thus, we sought an alternate approach to determining the stellar masses by using the IRAC imaging that covers a fuller extent of the galaxies. 

The m$_{3.6\mu m;\rm total}$ magnitudes (based on the flux extrapolated to infinity using the surface brightness profiles; see Section~\ref{sec:fluxes}) were converted to stellar masses. First, we convert the magnitudes to absolute solar luminosity units (\lsun) by adopting the solar luminosity at 3.6$\mu$m of 3.24 mag \citep{Oh2008} and the TRGB-based distances listed in Table~\ref{tab:properties}. Second, we assume a constant value for \mlr. Of course, the mass-to-light ratio of galaxies depends on a number of variables, including the age and metallicity of the stellar populations. These dependencies are more acute at bluer wavelengths where the light is dominated by hot, massive stars that make up only a small fraction of the overall mass in a galaxy. In the infrared, these dependencies are somewhat mitigated as the dominant contributors to the longer wavelengths of light, namely evolved, lower mass stars, are also the dominant contributor to the stellar mass. Note, also, that the GLOW galaxies all have similar SFHs (in the sense that they are all still star-forming), so there’s little reason to think that the \mlr\ ratio would vary strongly within the sample. Thus, we chose to adopt a \mlr\ value of 0.47 \msun/\lsun, which was the mean value derived from low-mass galaxies based on the model of \citet{Bell2003} and assuming a `diet' Salpeter IMF. \citep{McGaugh2014}. However, taking note that a constant \mlr\ value is a simplification, we follow the approach from \citet{McQuinn2021} and adopt an uncertainty of 0.12 dex in log space. This systematic uncertainty is based on the range in literature mass-to-infrared light ratios at 3.6$\micron$ of 0.45$-$0.6 (i.e., a range of 0.15, which translates to an uncertainty of 33\% in stellar mass or 0.12 dex); it also corresponds to the upper end of the uncertainties quoted for various mass-to-light ratio calibrations at 3.6$\micron$ \citep[e.g.,][]{Zhu2010, McGaugh2014, Meidt2014, Schombert2019}, and is slightly higher than the 0.1 dex deviations from the mean reported in \citet{McGaugh2014}.

To enable a comparison with the CMD-based stellar masses, which were derived using a Kroupa IMF, we also apply a factor of 0.85 to convert from the scaled Salpeter IMF to a Kroupa IMF \citep[e.g.,][]{Telford2020}. Note that there are still differences between the methods in determining stellar mass (e.g., resolved star fitting includes a treatment of binary survivors which is absent from integrated light methods), but these differences are not significant. Nearly all the galaxies have Galactic extinction measurements that are $\leq$ 0.01 mag in the $L^{\prime}-$band at 3.78 $\mu$m \citep{Schlafly2011} and, thus, no extinction corrections were applied. The exception is NGC~1569 where we subtract the value of $A_{L^{\prime}} = 0.11$ mag from the measured 3.6$\mu$m magnitude prior to calculating stellar masses. We report stellar mass values based on the m$_{3.6\mu m;\rm total}$ in Table~\ref{tab:stellar_masses}. 

As a consistency check on the derived present-day stellar masses, we also calculate the stellar mass from 3.6$\mu$m fluxes in the apertures matched to the HST fields of view and compare those values with the stellar masses from the SFHs from the HST data (see \S~\ref{sec:sfhs} for methodology).\footnote{Two galaxies (WLM, UGC~04483) were omitted from the comparison as the HST data used in GLOW are based on new observations that were not available when the stellar mass comparison with 3.6$\micron$ fluxes in matched apertures was performed.} 

Figure~\ref{fig:mstar} compares $M_{*,3.6\mu m;\rm HST}$ with $M_{*, SFH}$ derived using the PARSEC and MIST stellar libraries. $M_{*, SFH}$ is calculated from the {\it total} mass formed over the lifetime of the galaxy inferred by the SFHs and assuming a fraction of the mass is recycled back to the ISM over the lifetime of the stars \citep[i.e., the gas recycling fraction, “R”, see, e.g., ][]{Kennicutt1994}. We adopt the recycling fraction R of 0.43 from \citet{Vincenzo2016} based on the Kroupa IMF for low metallicity stellar populations. As noted in \citet{Vincenzo2016}, the fraction of gas returned can vary slightly as a function of metallicity of the stellar populations. However, for the range of properties of the GLOW sample, this has less than a 5\% impact on the derived masses, which is well within our mass uncertainties and does not impact our results. Thus, the adoption of a constant value is justified. Also shown are black lines representing unity and colored lines representing the best-fitting lines with dispersions for each comparison. Two outlier galaxies are shown with open plot symbols and were omitted from the fits as described in Section~\ref{sec:structural}. 

\input{tab5.tex}

Figure~\ref{fig:mstar} shows that the stellar masses calculated from m$_{3.6\mu m; \rm HST}$ (i.e., $M_{*,3.6\mu m;\rm HST}$) are in very good agreement with the present-day stellar masses derived from the CMD-based SFHs (i.e., $M_{*, CMD}$), especially when considering the disparate methods, different data sets, and different assumptions used in the calculations. Specifically, stellar masses agree within their uncertainties for 84\% of the sample regardless of the stellar library used. At larger stellar masses, the 3.6$\micron$-based values trend higher than the CMD-based masses which is likely due to the greater presence of young, AGB stars that increase the flux in the near-infrared \citep{Melbourne2012} and change the expected \mlr\ value. The lower panels show the percent difference between the \mlr\ and the CMD-based stellar masses derived from the two stellar models. The shaded grey regions encompass $\pm$25\% differences in the values. The masses derived using the PARSEC models are a closer match to the \mlr\ derivations with a mean percent difference of only $-2$\%, compared to $24$\% for the MIST models, although the PARSEC-based results show greater scatter. The close agreement with the PARSEC models may be attributed to the fact that the \mlr\ is based on population synthesis modeling \citep[e.g.,][]{Bruzual2003} that uses the Padova stellar evolution models, an earlier version of PARSEC. 

$M_{*,3.6\mu m; \rm total}$ is used to determine how much oxygen was produced by nucleosynthesis, whereas the SFHs and AMRs from the CMD-fitting are used to determine the oxygen content in stars and stellar remnants within the HST fields of view. To ensure consistency and to account for oxygen content in stars beyond the HST footprints, we introduce the ratio $M_{*,3.6\mu m;\text{total}} / M_{*, \text{CMD}}$ to scale the SFHs and, thus, the oxygen content locked in the stellar component. The robust agreement observed between $M_{*,3.6\mu m;\text{HST}}$ and $M_{*, \text{CMD}}$ provides evidence that using $M_{*,3.6\mu m;\text{total}}$ to scale the CMD-based masses is appropriate and does not introduce a large systematic uncertainty or bias in our calculation. Adopting this approach also allows for this same method to be applied uniformly to the larger GLOW sample of galaxies ensuring robust inter-sample comparisons. We provide the scaling values to account for the stars outside the HST footprints in Table~\ref{tab:stellar_masses}. These scaling values will be applied in the metal retention calculation (Paper~II).

\input{tab6.tex}

\section{Star Formation Histories and Age-Metallicity Relations}\label{sec:sfhs}
Here, we describe the data processing of the HST images and the steps taken to reconstruct the SFHs and AMRs. We follow the well-established methodology used for simultaneously reconstructing the star formation and chemical enrichment histories of nearby galaxies from CMDs \citep[e.g.,][]{Dolphin2002a, Aparicio2004, Tolstoy2009, Cignoni2015}. The SFHs and AMRs are used to the quantify the amount of oxygen locked in the stars and stellar remnants. In lieu of detailed spectroscopic abundances of a large sample of individual stars in each system, which is challenging with current observational facilities given the distances to the galaxies, the AMR, which quantities the average metallicity of stars formed at different times, can be determined from CMD-fitting techniques and yields the total metal content of the stars when combined with the SFR(t). The AMR derived using this method has been shown to be in good agreement with metallicity trends measured from spectra of stars of different ages in the Local Group galaxies WLM, LMC, and SMC \citep[][respectively]{McQuinn2024a, Cohen2024a, Cohen2024b}. The WLM study \citep{McQuinn2024a} also showed that the metallicity distribution function (MDF) of stars generated based on the best-fitting AMR was an excellent match to the spectroscopically determined MDF based on stars in a similar region of WLM, providing validation of using the AMR as representative of stellar metallicities in a galaxy. 

Note that we performed the CMD-fitting on the stellar catalogs divided into four annuli per galaxy and combined the results for a global solution. We opted for this approach for two reasons. First, the stellar density in galaxies can vary significantly from the inner to the outer regions of the disks. Thus, the completeness function in these different areas also varies which can impact the SFH recovery process. Second, the SFHs of the annuli provide additional information on the history of the galaxies, such as radial trends in the mass assembly, which is investigated in a separate study Cohen et al.\ (submitted). 

\subsection{Photometry of Sources in the HST Images}\label{sec:phot}
The main input to the CMD-fitting technique is high-fidelity photometry of resolved stars in the HST imaging and artificial star tests that measure the completeness of the photometry. As stated above, we required that the HST optical imaging include data in both the F606W\footnote{Or $V-band$ alternative filters such as F555W or F475W} and F814W that reached at least 2 mag below the TRGB in the F814W filter. This photometric depth ensures we can accurately constrain the cumulative SFH of a galaxy with enough precision to constrain the metals retained in the stars over the lifetime of a galaxy \citep[e.g.,][]{Dolphin2002a, Dohm-Palmer2002, McQuinn2010a}. The data were obtained from the Mikulski Archive for Space Telescopes (MAST) at the Space Telescope Science Institute. The specific observations analyzed can be accessed via \dataset[doi:10.17909/1yyh-1q11]{https://doi.org/DOI}.

The imaging for much of the sample was previously processed either as part of the ACS Nearby Galaxy Survey Treasury (ANGST) program \citep{Dalcanton2009} or the STARBurst IRregular Dwarfs Survey (STARBIRDS) program \citep{McQuinn2010a, McQuinn2015b, McQuinn2015c}. However, as we require photometry that was performed uniformly for all data sets and, importantly, a large enough number of artificial stars to significantly populate each annulus for the SFH fits, we re-reduced all data sets.

For consistency with previous works, we adopt the methodology and photometric processing steps from ANGST. Here, we provide a brief description of the data reduction and refer the interested reader to \citet{Dalcanton2009} for an in-depth discussion of the data handling techniques and Cohen et al.\ (submitted) for updates to the approach. Photometry was performed on the standard HST pipeline processed images using the {\sc dolphot} photometry package \citep{Dolphin2000, Dolphin2016} with modules specific to each HST imaging instrument (i.e., ACS or WFPC2). A reference image for cross-referencing point source positions was created by combining the raw images in the bluer filter using {\sc astrodrizzle} software \citep{Gonzaga2012}. Once the sources were identified in the individual images, photometric measurements were made on the individual files (i.e., {\sc flc.fits} for ACS and {\sc c0m.fits} for WFPC2). 

Nine galaxies, identified in Table~\ref{tab:hst_data}, were observed with multiple pointings. In cases where the fields of view have small overlaps or do not overlap at all (e.g., Sextans~A; see Appendix~\ref{sec:atlas_fov}), we performed photometry of each field separately and then eliminated stars in any overlapping area from the photometry of the shallower field. In cases where the pointings overlapped significantly (e.g., NGC~2366; see Appendix~\ref{sec:atlas_fov}), we used {\sc astrodrizzle} to combine the data and performed photometry on the total field of view to capitalize on the increased exposure times on the overlapping areas. Note that in all cases where the imaging from multiple pointings were combined and processed together, exposure times for all pointings were comparable ensuring consistent S/N in the input images. 

\begin{figure*}
\begin{center}
\includegraphics[width=0.9\textwidth]{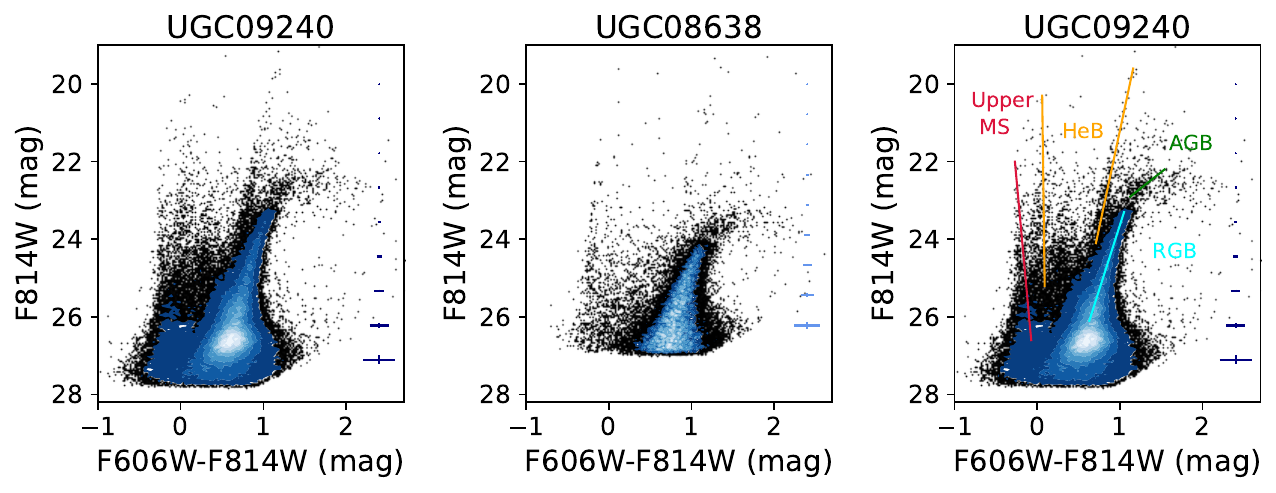}
\end{center}
\caption{Representative CMDs of the sample reaching depths below the red clump (UGC~9240; left) and $\sim$2 mag below the TRGB (UGC08638; middle). The right panel re-plots UGC~09240 with labels identifying populations in the CMD in different stages of stellar evolution (MS: main sequence; HeB: helium burning sequences; AGB: asymptotic giant branch stars; RGB: red giant branch stars). Error bars on the right of each panel indicate typical photometric errors in each magnitude bin. }
\label{fig:cmd_examples}
\end{figure*}

The photometric output from {\sc dolphot} was uniformly filtered for high fidelity point sources ensuring homogeneous star catalogs. Specifically, we rejected sources with low signal-to-noise ratios (S/N$\leq4$ in either filter), object types $>2$ (i.e., elongated, sharp, or extended objects), and error flags signifying that a source was not recovered extremely well (i.e, a source was close to a chip edge or saturated). In addition, we filtered to only allow objects with low sharpness {\sc (sharp$_1 +$ sharp$_2$)}$^2 \leq 0.075$), which eliminates narrow cosmic rays and the most extended sources such as background galaxies that may have been missed in previous processing, and low crowding ({\sc (crowd$_1 +$ crowd$_2$}) $\leq 0.1$), which eliminates sources whose photometry has been strongly impacted by neighboring sources. We refer the interested reader to \citet{Dolphin2000} for a detailed explanation of these quality metrics and photometric output parameters and \citet{Dalcanton2009} for a discussion of the quality filters imposed.

Following standard procedure, artificial star tests were performed on the images to measure the completeness due to both the depth of the images and the stellar density using the same photometric software {\sc dolphot}. To ensure sufficient artificial star counts in each SFH annulus, we injected approximately 5\,M artificial stars for each galaxy. These stars were distributed across the fields according to the point source density measured from the non-filtered photometry and spanning the magnitude and color range that was $\sim1.5$ mag deeper and $\sim1$ mag brighter than the high-fidelity point sources. Results from the artificial star tests were filtered using the same quality cuts as the photometry. The measured 50\% completeness limits of the full stellar catalogs are provided in Table~\ref{tab:hst_data}. The completeness limits for the outer annuli are up to $\sim1$ mag deeper due to lower stellar crowding. 

Figure~\ref{fig:cmd_examples} presents two representative CMDs plotted to the $\sim50$\% completeness limit. The photometry reaches just below the red clump in the left panel from ACS data and $\sim2$ mag below the TRGB in the F814W filter in the middle panel from WFPC2 data. This range of photometric depth brackets the depth of most of the datasets, although there are a few systems with significantly deeper photometry (e.g., WLM, LeoA).  Representative photometric errors per magnitude are shown. CMDs for the complete sample are shown in Appendix~\ref{sec:atlas_cmds}. 

For didactic purposes, in the right panel we also show the CMD for UGC~09240 with stellar evolutionary sequences labeled. In addition to a well-populated upper main-sequence, UGC~09240 hosts well-defined blue and red helium burning sequences, indicative of significant star formation in the last few 100 Myr \citep{McQuinn2011}. There is also a population of AGB stars above the red giant branch (RGB) and extending to redder colors (i.e., F606W$-$F814W $\gtsimeq 1.2$), which provide evidence of star formation at intermediate times. The RGB stars have an F606W$-$F814W color of $\sim1.2$, indicative of metal-poor stellar populations \citep{Bressan2012}, and provide evidence of star formation at older lookback times. The same sequences are present in the CMD of UGC~08638, albeit with different characteristics (e.g., colors, stellar densities, extent of the young sequences, etc.), reflecting the different stellar mass, metallicity, and SFH of the galaxy. Finally, the red clump overdensity of stars on the lower RGB (not labeled) is also evident in the deeper CMD for UGC~09240. 

For the SFH fits, following the approach in Cohen et al.\ (submitted), the photometry and artificial star catalogs were separated into four concentric annuli. The annuli were centered on the RA and Dec coordinates in Table~\ref{tab:properties}, and based on the 3.6$\mu$m structural parameters in Table~\ref{tab:structural}. The widths of the annuli were set such that each contained approximately the same number of observed stars, as the number of observed stars is the primary driver of statistical uncertainties in the SFHs. The SFHs were then separately reconstructed from each sub-catalog and the solutions combined for a global SFH and AMR. See Cohen et al.\ (submitted) for more details.

\begin{figure*}
\includegraphics[width=0.5\textwidth]{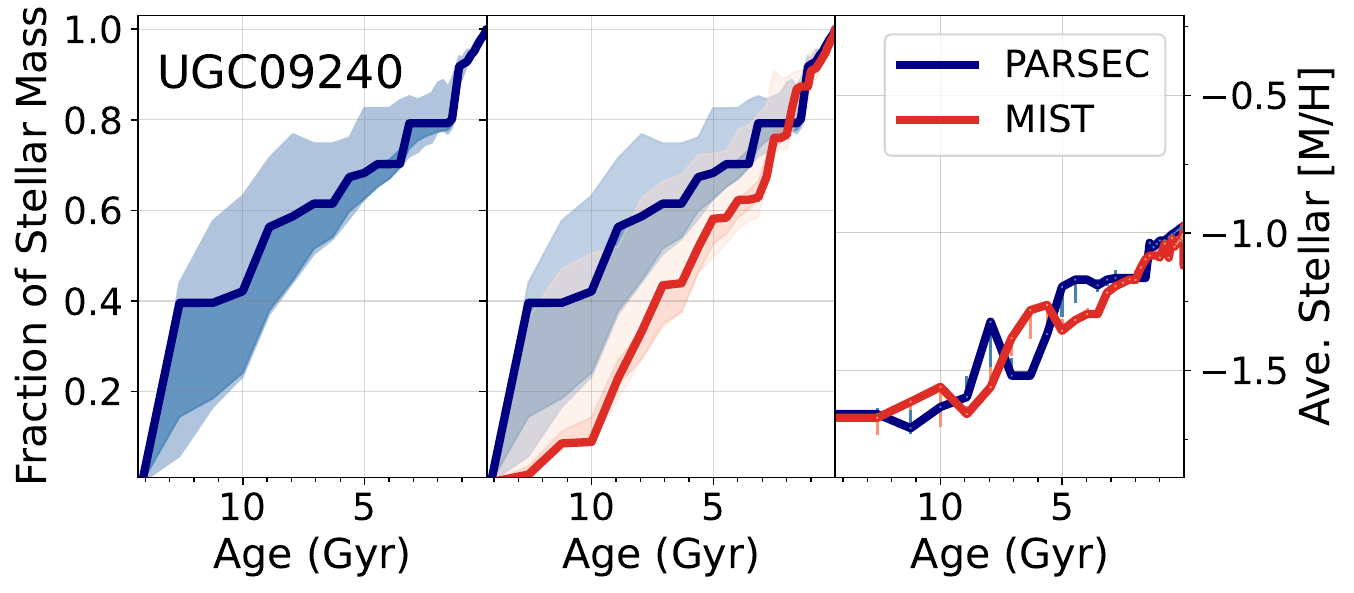}
\includegraphics[width=0.5\textwidth]{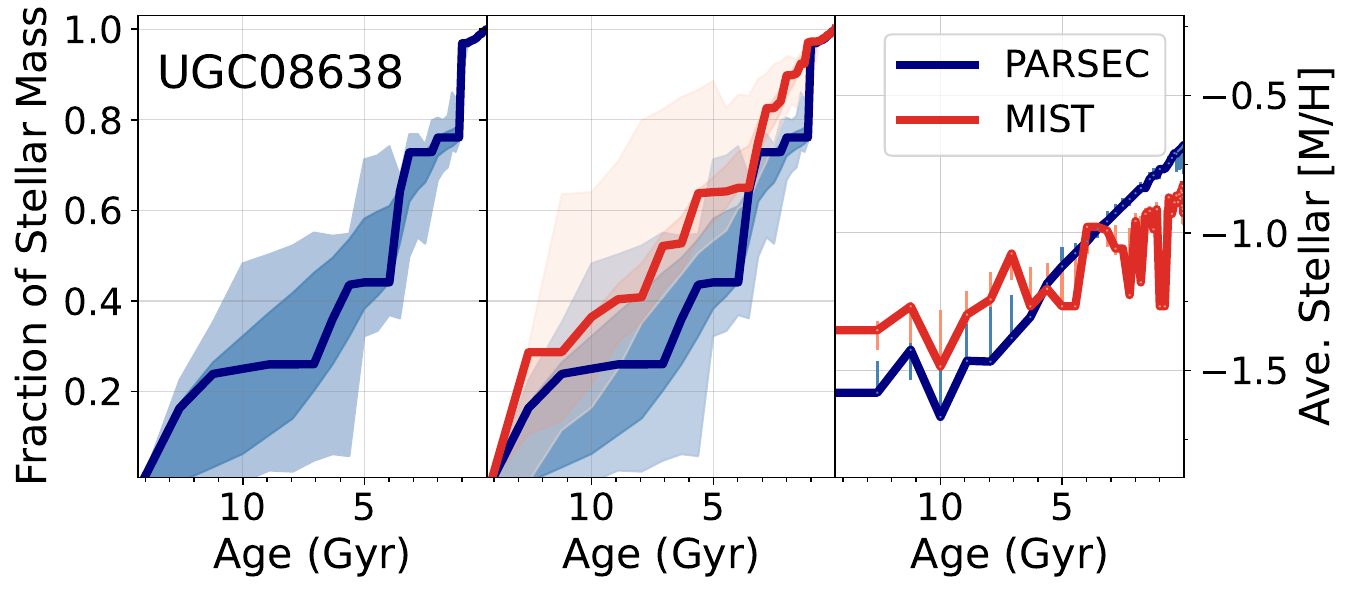}
\caption{Representative SFHs and AMRs of the sample derived from the CMDs shown in Figure~\ref{fig:cmd_examples}. The left panels show the best-fitting solution from the PARSEC stellar library with statistical uncertainties in darker blue shading and the total (statistical and systematic) uncertainties in the lighter blue shading. The total uncertainties are lower for the UGC~09240 which is primarily due to the deeper photometry used in the SFH derivation. The middle panels compare the PARSEC SFH solution with the solution from the MIST stellar libraries (statistical uncertainties only). The final panels presents the AMRs from both libraries.}
\label{fig:sfh_amr_examples}
\end{figure*}

\subsection{Reconstructing the SFHs and AMRs}
The SFH (i.e., SFR(t)) and AMR (i.e., Z(t)) can be reconstructed by synthesizing a CMD with stellar evolution libraries \citep[see e.g.,][for a review of the technique]{Tolstoy2009}. We use the CMD-fitting technique {\sc match} \citep{Dolphin2002a}, which is the same code employed in the metal retention analysis for Leo~P \citep{McQuinn2015f} and for M~31 \citep{Telford2019}, allowing for a direct comparison with these previous results. As described in detail in \citet{McQuinn2015e}, {\sc match} generates synthetic CMDs using stellar evolution libraries, an assumed initial mass function (IMF), a stellar binary fraction with a flat mass distribution, and a set of galaxy-specific parameters (such as the distance to a galaxy, foreground extinction, and metallicity constraints). To account for observational uncertainties, the synthetic CMD is convolved with the completeness function and uncertainties from the artificial star tests. Modelled CMDs are iteratively generated using different combinations of age and metallicity histories and compared to the observed CMD using a Poisson likelihood function. The modelled CMD with the maximum likelihood solution represents the best-fitting SFH and AMR for the observed data. 

We derived the solutions using two different stellar evolution models, namely the Padova-Trieste models \citep[PARSEC;][]{Bressan2012, Chen2014, Tang2014}, and the Mesa Isochrones and Stellar Tracks \citep[MIST;][]{Choi2016}. We chose these libraries as they include isochrones for massive stars which are essential to recover the stars in the upper main sequence and the massive helium burning stars present in many of the galaxies. The resulting range in solutions help quantify the dependency of our final metal retention fraction on the choice of stellar models. 

In applying the CMD-fitting technique, we assumed a Kroupa IMF with mass limits of $0.1-100$ \msun\ \citep{Kroupa2001} and a binary fraction of 0.35. As described in detail in Cohen et al.\ (submitted), we first fit for the SFH in the outer annuli where crowding and internal extinction are lower than in the central regions of the galaxies. For these annuli, we set priors on the distances based on the values determined from the TRGB method (see Table~\ref{tab:properties}) and the foreground extinction determined from the dust maps of Schlegel et al.\ (1998) 
with recalibration from \citet[][listed in Table~\ref{tab:properties}]{Schlafly2011}, and allowed the solutions to float within 0.1 mag of each value (in step sizes of 0.05 mag). The 0.1 mag range is a conservative range of the distance moduli uncertainties and variations in foreground extinction. These values are solved for independently for each of the two stellar libraries, which allows the models to find the best match to the data regardless of small systematic discrepancies between them. We then adopted the best-fitting distance modulus and foreground extinction values to fit for the SFHs in the remaining annuli. 

Internal extinction ($\delta A_V$) was estimated by iteratively fitting the CMDs over a range of values, in intervals of 0.05 mag, until the best-fit was found in each ellipse for each stellar library. As expected for low-metallicity galaxies, the internal extinction was modest, with a median value of 0.25 mag.  The final global values are listed in Table~\ref{tab:hst_data}. We also assumed the default value of an additional 0.5 mag of internal extinction for young ($<40$ Myr) populations. The inclusion of additional extinction for the youngest stars is justified as these sources have often not fully cleared their nascent gas clouds. While assuming higher extinction for young stars is a commonly used approach for CMD-fitting work \citep[e.g.,][]{Dolphin2002a}, we found through testing that this assumption had little impact on our final SFHs. We adopted an age grid of log(t)$ = 6.6-10.15$ using a time resolution log($\delta$t$) = 0.1$ dex for ages less than log(t)$ = 9$ and log($\delta$t$) = 0.05$ dex for older ages. The chemical enrichment history was assumed to be a continuous, non-decreasing function with a minimum [M/H] of $-2.2$. 

\begin{figure}
\includegraphics[width=0.4\textwidth]{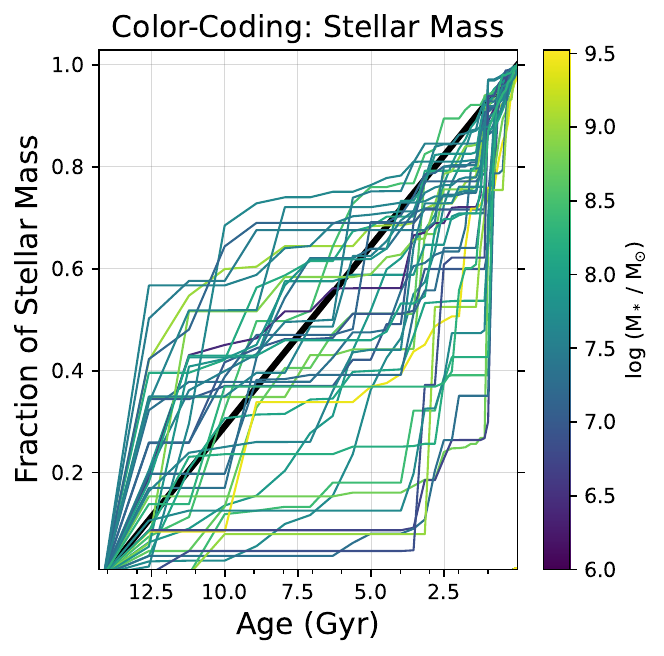}
\includegraphics[width=0.4\textwidth]{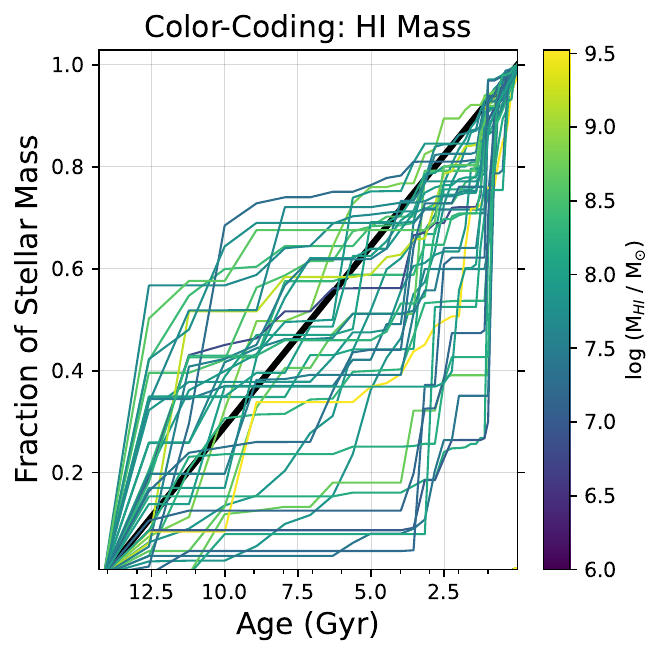}
\includegraphics[width=0.4\textwidth]{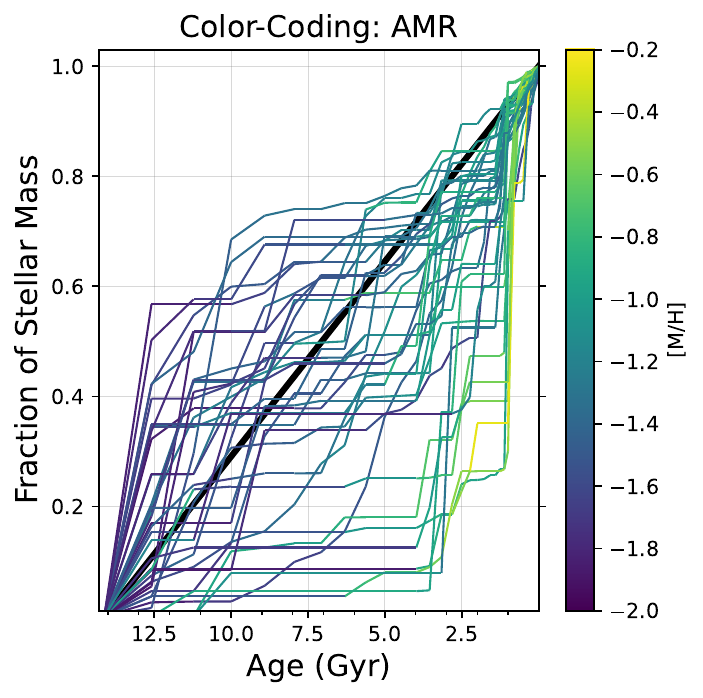}
\caption{The SFHs for the GLOW sample colored coded by stellar mass (top panel), \hi\ mass (middle panel) and each galaxy's AMR (bottom panel). A solid black line of constant star formation is overplotted for comparison. }
\label{fig:sfh_stacked_mass}
\end{figure}

We estimated both random and systematic uncertainties on the best-fit SFR(t) and AMR. Random uncertainties on the best-fitting solutions due to the finite number of stars and signal-to-noise of the data were estimated using a Hamiltonian Monte Carlo approach \citep{Dolphin2013}. Systematic uncertainties from the stellar evolution models are estimated using 50 Monte Carlo simulations \citep{Dolphin2012}. Both the range of solutions between the models and the more formally estimated systematic uncertainties provide a measure of uncertainty from the choice of stellar models \citep[see, e.g.,][for a discussion]{Skillman2017}.

Figure~\ref{fig:sfh_amr_examples} presents examples of the SFH and AMR for two galaxies, UGC~09240 and UGC~08638, derived from the example CMDs  in Figure~\ref{fig:cmd_examples}. For each galaxy, the left panel shows the cumulative SFH solutions using the PARSEC library with the darker shading representing the statistical uncertainties and the lighter shading representing the systematic uncertainties. The middle panel compares the SFH solutions from PARSEC with those from the MIST library, with the shading representing statistical uncertainties only. Finally, the right panel shows the AMRs derived using the two libraries with combined statistical and systematic uncertainties. The uncertainties for UGC~09240 are smaller than for UGC~08638 as expected given both larger number of stars (which can reduce the statistical uncertainties) and the deeper photometry (which can reduce the systematic uncertainties). 

From Figure~\ref{fig:sfh_amr_examples}, the SFH of UGC~09240 shows an early rise in star formation whereas the MIST models prefer a slower start to star formation, but the differences are within the systematic uncertainties between the libraries. The SFH of UGC~08638 derived from the two libaries are also consistent with each other within the uncertainties.

Focusing on the AMRs, we note that while we imposed a constraint of monotonically increasing metallicity for the fits, the final combined AMR solutions do show variations (e.g., the more spikey features in right panels of Figure~\ref{fig:sfh_amr_examples}). The AMRs are derived {\it per annulus} and are based on the different stellar populations and total mass within each region. Thus, when combined, the AMRs can exhibit decreases as a function of time. For the examples shown, UGC~08638 shows a greater increase in stellar metallicity as a function of time compared to UGC~09240. The AMR solutions from the two stellar libraries are in good agreement for both systems. Similar plots for the full sample are presented in Appendix~\ref{sec:atlas_sfhs_amrs_plots}. 

For completeness, Figure~\ref{fig:sfh_stacked_mass} compares the cumulative SFHs of the full GLOW sample color-coded by log(\mstar/\msun) (top panel), and log(\mhi/\msun) (middle panel). The SFHs show a range of patterns but the details, particularly at early epochs, should not be over-interpreted as ancient SFHs can only be derived with precision from deeper photometry that reaches below the oldest main sequence turn-off. The bottom panel in Figure~\ref{fig:sfh_stacked_mass} shows the SFHs now color-coded by the AMRs such that we see the time-evolution in the stellar metallicity for each galaxy. 

\begin{figure}
\includegraphics[width=0.48\textwidth]{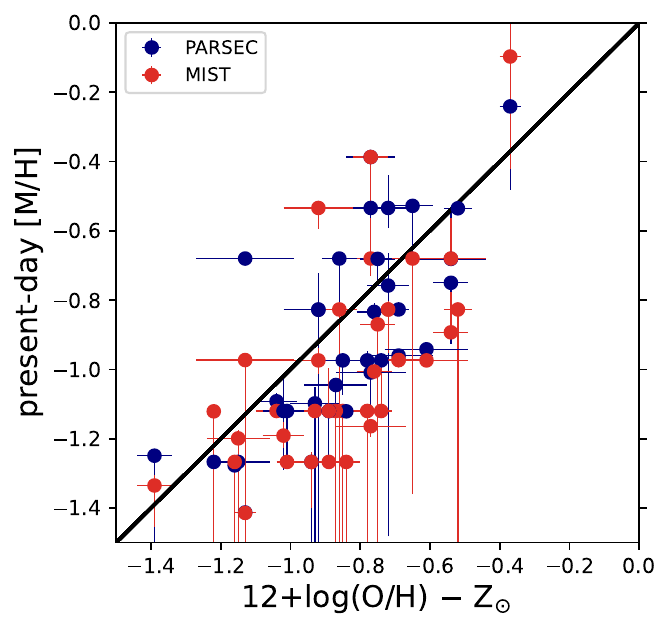}
\caption{Comparison of the present-day [M/H] from the AMRs based on the PARSEC and MIST models with the gas-phase oxygen abundances relative to a solar abundance of 12+log(O/H) $=$ 8.69 \citep{Asplund2021}. While absolute agreement is not expected between [M/H] in the young stars and the gas-phase oxygen abundance, the enrichment of these different components should be correlated, as confirmed in these data.}
\label{fig:metallicities}
\end{figure}

Figure~\ref{fig:metallicities} compares the best-fitting present-day [M/H] values from the CMD from both the PARSEC and MIST models with the spectroscopically measured gas-phase oxygen abundances. While the stellar and gas metallicities trace different elements and slightly different timescales, the expectation is that they should correlate. This is confirmed in the figure overall, with the stellar metallicity estimates slightly offset to lower values than the gas-phase measurements.

\section{Gas Content of the Galaxies}\label{sec:hi}
Here, we focus on quantifying the atomic gas content in the ISM of the galaxies, which is thought to contain the largest reservoir of oxygen retained in low-metallicity galaxies \citep[e.g.,][]{McQuinn2015f}. Given the paucity of detections of molecular gas in low-metallicity galaxies and the uncertainties associated with molecular gas mass estimates, we do not include detailed analysis of the possible molecular content in the galaxies here. Instead we focus solely on the atomic gas component traced by the 21 cm emission line of hydrogen. Similarly, while some fraction of oxygen will deplete onto dust grains, the amount is expected to be sub-dominant to the oxygen mixed in the gas and has been noted to decrease with decreasing metallicity \citep[e.g.,][]{Draine2007, RomanDuval2022, Hamanowicz2024}. For example, the depletion of oxygen, primarily onto silicate, aluminum, and magnesium dust grains, has been measured to range from 0.08 dex at 12+log(O/H) of $\sim$7.3 to 0.12 dex for 12$+$log(O/H) of $\sim$8.8 \citep[e.g.,][]{Peimbert2010}. Thus, given the range of metallicity of the GLOW galaxies, a modest depletion factor of $\sim0.1$ dex for oxygen is expected. We defer a more detailed discussion of the possible oxygen content in dust in the GLOW sample to the companion paper. In the remainder of this section, we present the measurements of the total \hi\ masses in the galaxies and the radial \hi\ mass profiles. 

\subsection{\hi\ Masses}
We determined the \hi\ masses based on the total fluxes measured from 21-cm line observations and adopting the TRGB distances to the galaxies from Table~\ref{tab:properties}. The majority of the \hi\ fluxes were measured from VLA data that we uniformly reprocessed. In seven cases, the \hi\ fluxes from the VLA data were insufficient and fluxes were taken from literature values. For NGC~4449 and UGC~5666, the large spatial extent of the galaxies resulted in poor and uncertain flux recovery for the outskirts of the VLA data. The \hi\ content of NGC~4449 is also complicated by the likely interaction with DDO~125 which has now been mapped to much lower column densities \citep{Mei2023}. For NGC~0784, UGC~06817, UGC~08201, UGC~08508, and UGC~09240, we adopt \hi\ flux measurements based on observations from single-dish telescopes that are available in the literature. Finally, for the two systems not visible by the VLA (IC~4662; IC~5152), we adopt the measurements from the Australia Telescope Compact Array (ATCA) facility. Table~\ref{tab:hi} provides the fluxes and the resulting \mhi, along with the heloiocentric velocities, the full width at half-maximum (FWHM) of the 21-cm line (W$_{50}$), and references.

\subsection{\hi\ Mass Profiles}
The \hi\ content in gas-rich, low-mass galaxies typically extends slightly beyond the main stellar disk. Yet, in some dwarf galaxies, the gas extends significantly farther than the stellar component. A clear example is NGC~3741, where the \hi\ has been noted to extend to $\sim8.3\times$ the Holmberg radius, or approximately 38 optical scale lengths \citep{Begum2005}. This galaxy's \mhi/\mstar\ ratio ($\sim7$) is larger than the majority of dwarf galaxies in the local universe and the estimated virial mass is also high \citep[$\sim10^{11}$ \msun;][]{Gentile2007}, leading to questions as to why the gas in NGC~3741 has been unable to collapse to form stars. While NGC~3741 is one of the more extreme cases, there are other known very gas-rich galaxies in the nearby universe, a few of which are part of the GLOW sample (e.g., UGC~08024, IC~4662, IC~1512). We speculate that the high gas-fractions and spatially-extended nature of the \hi\ mass profiles are a result of high specific angular momentum in the gas that has greater stability against gravitational collapse \citep[e.g.,][]{Obreschkow2016, Lutz2018, ManceraPina2021}.

In addition to the curious nature of these systems, the extent of the \hi\ relative to the stars impacts how much gas we can reasonably assume has been enriched by the by-products of star formation and stellar feedback. For typical low-mass galaxies where the \hi\ disk is only marginally larger than the stellar disk, the expectation is that the gas will be relatively uniformly enriched. Indeed, in the majority of cases where multiple gas-phase oxygen measurements across the disks of low-mass galaxies have been made, the galaxies show little differences in their oxygen abundances as a function of radius \citep[e.g.,][]{Skillman1989, Kobulnicky1996, Lee2006, Croxall2009, Berg2012}. In contrast, for galaxies where the \hi\ extends significantly farther than the stellar component, it is unlikely that the outer gaseous disk has experienced the same significant chemical enrichment by the galaxy's stars as the inner regions. Thus, understanding the spatial extent of the gas relative to the stars is critical to accurately estimating the oxygen content in the ISM for GLOW.

Here, we quantify the \hi\ flux as a function of $3.6\micron$ scale lengths. This mapping of the spatial distribution of \hi\ relative to the stars clearly identifies which galaxies have extended gaseous disks, while also providing a uniform measure of how much \hi\ lies beyond an `enrichment radius'. We measure the fraction of \hi\ flux as a function of 3.6$\micron$ scale lengths in each galaxy using spatially resolved moment-0 maps from VLA and ATCA data and adopting the galaxy geometry determined from the 3.6$\micron$ imaging. 

We made use of archival VLA data from various programs and uniformly reprocessed the data using standard \hi\ data reduction techniques for the majority of galaxies. Briefly, the VLA data were processed using \emph{AIPS}\footnote{The Astronomical Image Processing System (AIPS) was developed by the NRAO.} following the procedures in \citet{Richards2018, Hunter2022}. The individual observing blocks were flagged for RFI, phase and bandpass calibrated, Hanning smoothed, and Doppler corrected. After the individual blocks were processed, they were combined and continuum subtracted. For all galaxies, the data cubes were generated using the task \textit{imagr} with a robust of 5 weight and no uvtaper or uvrange limits. The  0$^{th}$ moment maps were created in GIPSY \citep{GIPSY} following the procedure in \citet{Hunter2022}. The exceptions are for UGC~06817 and UGC~08508 where the natural weighted 0$^{th}$ moment maps were downloaded from the VLA-ANGST Survey\footnote{https://science.nrao.edu/science/surveys/vla-angst/copy\_of\_the-little-things-survey} \citep{Ott2012}. The ATCA data were downloaded as moment 0 maps from the Local Volume \hi\ Survey\footnote{https://www.narrabri.atnf.csiro.au/research/LVHIS/index.html} \citep[LVHIS;][]{Koribalski2018}.

With the \hi\ moment-0 maps in hand, we measure the fraction of \hi\ flux in concentric annuli in intervals of 1.1 scale lengths out to 8.8 scale lengths, which encompasses the \hi\ flux in nearly all the galaxies, but also include \hi\ flux measurements within 10, 12, 15, 20, 25, 30 scale lengths as a few galaxies have extremely extended \hi\ components. The cumulative percentage of flux per radial annuli are listed in Table~\ref{tab:hi_profiles}.

\begin{figure*}
\includegraphics[width=0.95\textwidth]{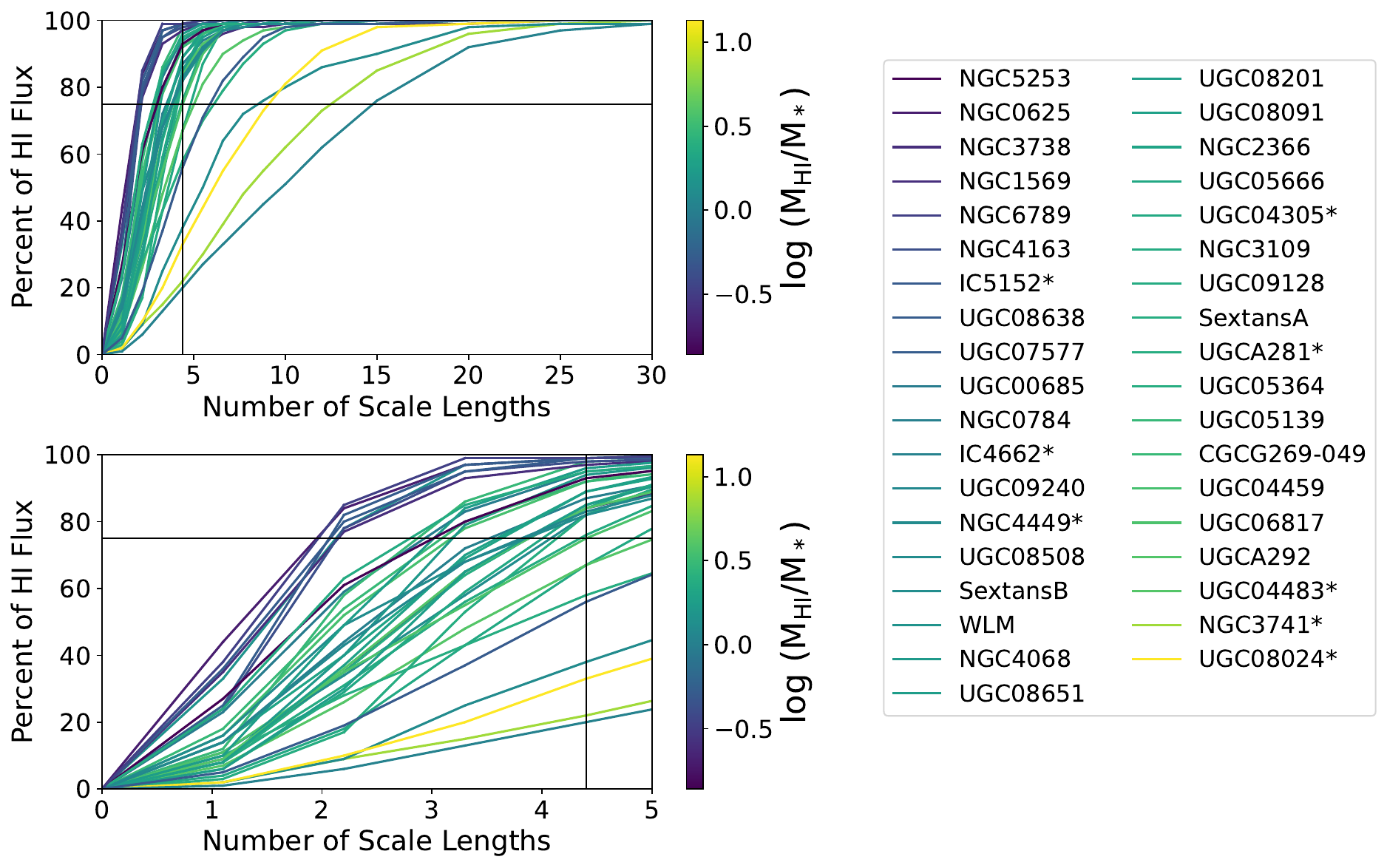}
\caption{The percent of the \hi\ flux as a function of scale lengths color-coded by \mhi/\mstar\ ratios. The top panel shows the full extent of the \hi\ on the x-axis while the bottom panel zooms in on the first 5 scale lengths. The vertical black lines marks 4.4 scale lengths and the horizontal black lines marks the 75\% level of the flux. The side legend lists the galaxies names and the colors associated with their \mhi/\mstar\ values in the color-bar. Most of the galaxies have the majority of their \hi\ flux contained within 4.4 scale lengths based on the stellar light with the exception of the most gas-rich systems. Galaxies with the most extended \hi\ (i.e., $<75$\% of their \hi\ flux contained withing 4.4 scale lengths) are marked with an asterisk. In the extreme cases, the \hi\ flux extends to $10-30$ scale lengths.}
\label{fig:HI_scalelen}
\end{figure*}

\input{tab7.tex}

Figure~\ref{fig:HI_scalelen} shows the radial profiles of the cumulative \hi\ in percentage of total flux, color coded by the logarithm of \mhi/\mstar\ ratios. Using the stellar catalogs from the HST imaging as a guide, we find the stellar components routinely extend to 4.4 scale lengths. Examining the \hi\ distributions, the majority of galaxies have 75\% of the \hi\ flux contained within 4.4 scale lengths, indicating that the ISM in these galaxies have likely been chemically enriched from stellar nucleosynthetic by-products. These galaxies predominantly have \mhi/\mstar\ $\ltsimeq$ 2, typical of gas-rich low-mass systems. Thus, for galaxies where $\geq$75\% of the \hi\ is within 4.4 scale lengths, we assume all the \hi\ mass has been enriched with oxygen. 

Eight galaxies have \hi\ that is more extended (i.e., less than 75\% of the \hi\ flux is within 4.4 scale lengths), with \hi\ detected out to an extreme value of $\sim30$ scale lengths for NGC~3741; these systems also correspond to the galaxies with higher \mhi/\mstar\ values. Given the different extents of the \hi\ and stellar components, it is unlikely that all the \hi\ has been equally enriched with by-products from stellar nucleosynthesis.  This analysis also allows us to determine the amount of \hi\ gas to be included in our metal retention calculation.\footnote{The oxygen mass in the ISM is determined using gas-phase oxygen abundances measured relative to the \hi. To convert these values to an oxygen mass, we require an accurate measure of the \hi\ content in the sample.} We use the amount of \hi\ mass contained within 4.4 scale lengths as representing the gas mass that has been significantly chemically enriched by stellar nucleosynthesis processes and apply a uniform cut to the \hi\ flux; values are provided in Table~\ref{tab:hi}.

\subsection{\hi\ Line-Widths and Estimates of Rotation Speeds}
We estimate gas rotation velocities for the galaxies from the 21 cm line observations. These rotation velocities constrain the total mass (i.e., baryonic and dark matter) within the gaseous disks of the galaxies and provides a basis to explore the oxygen retention fractions as a function of the gravitational potential depths. We take a simple approach to estimating the rotational velocities by using half-width at half-maximum (W50) of the 21 cm line measured from single dish observations from the literature and correcting for the inclination angle:

\begin{equation}
V_{\rm rot} = \frac{W50 \times sin(i)}{2} \label{eq:vrot}
\end{equation}

\noindent where the inclination angle, $i$, is estimated from the 3.6$\mu$m geometry ($q = \frac{b}{a}$ and by assuming a disk thickness, $q_0$, of 0.4, typical of a low-mass galaxy:

\begin{equation}
cos^2i = \frac{q^2 - q_0^2}{1-q_0^2} \label{eq:incl}
\end{equation}

\noindent All values are provided in Table~\ref{tab:hi}. We note that while these rotational velocities are not based on detailed kinematic modelling and should therefore be considered only approximate, they do provide a useful representation of the gravitational potential of the galaxies and, thus, enable a meaningful comparison of the GLOW galaxies abilities to retain metals as a function of total galaxy mass. 

\section{Characterizing Galaxy Environments}\label{sec:env}
Environment is thought to play a role in how many metals are retained in galaxies. Specifically, for galaxies in denser environments, a larger fraction of the expelled metals is expected to be recycled onto galaxies \citep[e.g.,][]{Angles-Alcazar2017} compared to isolated galaxies where metals ejected to the CGM may be partially lost to the IGM. The GLOW galaxies sample a range of environments which allows us to explore the connection between metal retention and the local environment. Indeed, the sample includes galaxies located in the field (i.e., highly isolated), residing in loose associations of low-mass systems \citep[e.g., the NGC3109 association;][]{Tully2006}, and located within a group of galaxies (e.g., the M81 group). In addition, within their local environments, the galaxies can be found at different distances to their nearest neighbors.  

\input{tab8.tex}

Here, we quantitatively characterize the environment around the sample using the Updated Nearby Galaxy Catalog \citep{Karachentsev2013}, which is an updated list of known galaxies within the Local Volume (D$\sim10$ Mpc). For each GLOW galaxy, we calculate four metrics. First, we determine the tidal index, $\Theta$, which relates the tidal force experienced by a galaxy ($i$) from another galaxy ($n$):
\begin{equation}
    \Theta = \textrm{max}[\textrm{log}(M_{n} / D_{in}^{3})] + C  \textrm{, $n$ = 1, 2, ..., N}
\end{equation}\label{eq:tidal_index}
\noindent where $M_{n}$ is the mass of the $n$th galaxy, $D_{in}$ is the distance between the two galaxies, 
and $C$ is a constant set to $-10.96$ so that galaxies with $\Theta <$ 0 represents isolated systems. The tidal index corresponding to the largest tidal force we label $\Theta_1$, and the associated neighboring galaxy we label as the main disturber or MD. If no neighbors were identified within a 1 Mpc radius, no value of $\Theta_1$ is assigned and MD is set to `field'. 

Second, we quantify the overall density of the environment by summing the tidal indices for the five most significant neighbors of each galaxy, which we label $\Theta_5$. Third, we calculate the distance to the nearest neighbor (NN). Finally, we calculate the distance to the nearest large galaxy (NLG), defined as a system with \mstar $>10^9$ \msun. Table~\ref{tab:environment} lists all four metrics including the names of their MDs, NNs, and NLGs; NLGs that are farther than 1 Mpc are marked with an asterisk. When applicable, we also include the name of the association or galaxy group to which each system is identified. 

\begin{figure}
\includegraphics[width=0.48\textwidth]{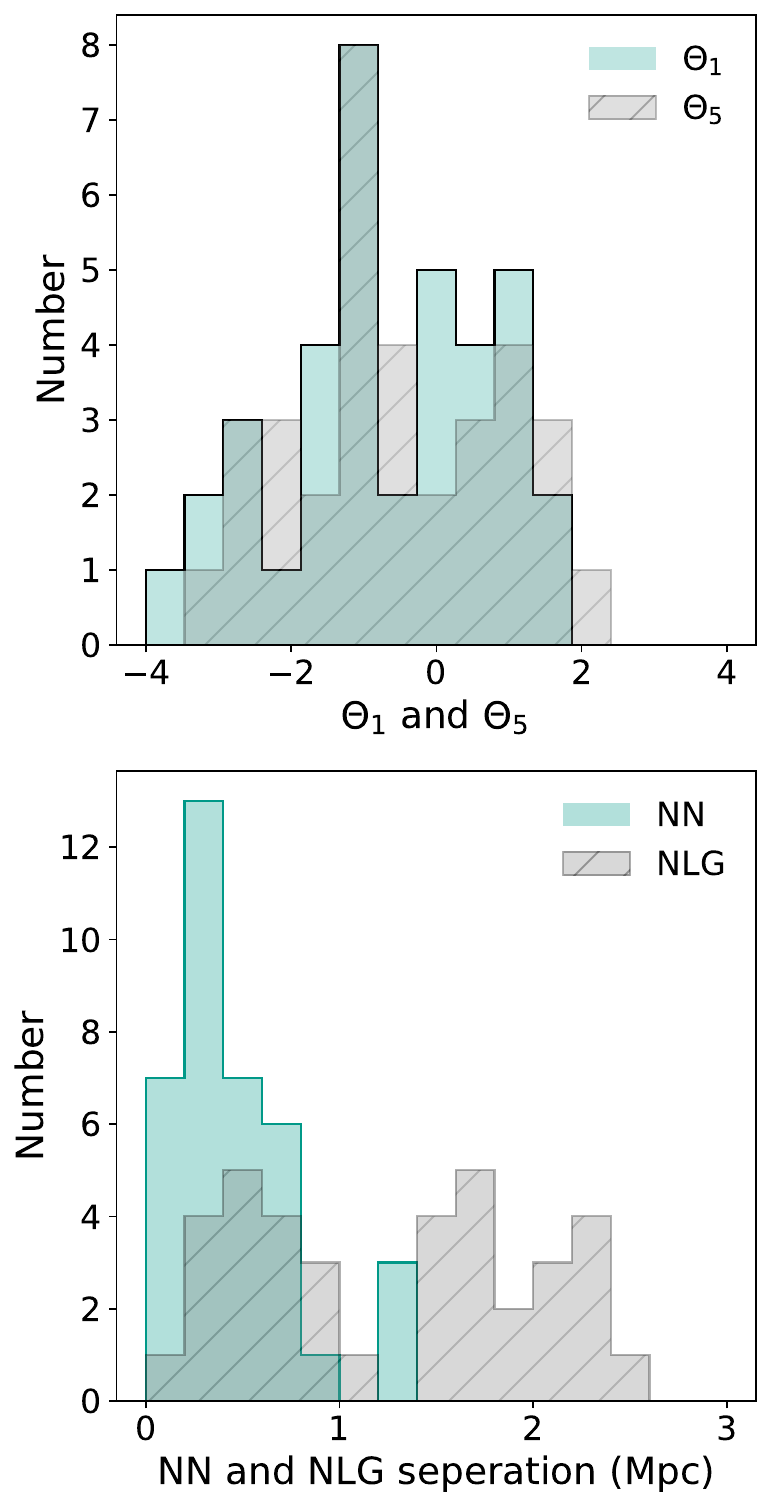}
\caption{Distribution of environment properties.}
\label{fig:environment_histo}
\end{figure}

Figure~\ref{fig:environment_histo} presents histograms of the four metrics for the sample. The top panel shows $\Theta_1$ and $\Theta_5$ where it is clear that the GLOW sample spans a range of environment, but that the majority of galaxies are relatively isolated (i.e., have negative $\Theta$ values). Quantitatively, the mean values of $\Theta_1$ and $\Theta_5$ are $-0.75$ and $-0.85$, respectively. The bottom panel shows the separations of the galaxies from their nearest neighbor and their nearest large galaxy. The mean NN and NLG separations are 0.45 Mpc and 1.25 Mpc, respectively, with standard deviations of 0.33 Mpc and 0.71 Mpc, supporting the same interpretation from the $\Theta$ metrics that the majority of galaxies are isolated but now also showing that a range of environments are represented. We explored each of the four metrics as a function of galaxy metallicity and gas fraction but found no strong correlations. We will investigate whether these metrics correlate with the metal retention fraction in the companion paper. 

\input{tab9}

\section{Summary}\label{sec:conclusions}
The GLOW project is aimed at measuring the production of oxygen from star formation, tracking the distribution of the oxygen in different galaxy components (stellar disk, ISM, CGM), and determining the retention fraction of oxygen in a statistically significant sample of nearby low-mass galaxies. The data sets for GLOW include primarily archival observations from myriad facilities including the VLA, {\em Spitzer Space Telescope}, HST, and ground-based optical telescopes, and incorporate numerous measurements from the literature from optical spectroscopy and single-dish \hi\ studies on 37 galaxies with $6.5 < $ log(\mstar/\msun)\ $ <9.5$ within $\sim6$ Mpc. The galaxies in the sample are all gas-rich, star-forming, and relatively isolated (i.e., none are satellites of a massive galaxy). 

In this first paper, we presented a full suite of measurements on the galaxy properties including:
\begin{itemize}
\item Geometry and structural parameters from B, R, and 3.6$\micron$ imaging data cleaned of foreground and background contaminants.
\item Stellar masses (\mstar) calculated from (i) the 3.6$\micron$ fluxes and adopting a mass-to-light ratio in three different apertures; and (ii) fitting the CMDs of resolved stellar populations with two different stellar libraries (PARSEC; MIST). The 3.6$\micron$-based \mstar\ values were found to be in remarkable agreement with the CMD-based \mstar\ values, although the MIST values show a slight systematic bias towards lower masses. 

\item Star formation histories (SFHs) and age-metallicity relations (AMRs) derived from the HST imaging of the resolved stars using the PARSEC and MIST stellar libraries. The solutions are in overall agreement across the majority of the sample, although notable differences are seen with some galaxies reflecting the uncertainties that remain in precisely modeling stellar evolution. The AMRs typically show a steady increase in metallicity with time although lower mass galaxies show less chemical evolution within the galaxies' stellar populations than the more massive dwarfs. A similar trend was noted in a previous study where the degree of chemical evolution correlates strongly with galaxy \citep{Skillman2014}.

\item \hi\ masses of the galaxies and the spatial distribution of the \hi\ flux as a function of measured scale lengths. The \hi\ mass relative to the \mstar\ for most of the galaxies is typical of nearby star-forming dwarfs (i.e., \mhi/\mstar\ $\ltsimeq2$), although there are several outliers with mass ratios as high as 13.  We also find that 75\% of the spatial extent of the neutral hydrogen (\hi) is contained with 4.4 scale lengths, with the exception of eight galaxies where the \hi\ can be significantly more extended (including a few systems with \hi\ out to 10-30 scale lengths). Most of the galaxies with the extended \hi\ are also the systems with the highest \mhi/\mstar\ values, although exceptions to this trend are also noted. 
\item Using existing catalogs of nearby galaxies, we quantified how isolated the galaxies are using several metrics (distance to nearest neighbor(s) and the distance to a massive galaxy). We find the galaxies reside in a range of environments, but the environments have lower density and the galaxies are relatively isolated.
\end{itemize}

\section{Acknowledgments}
This manuscript is dedicated to the memory of Professor Liese van Zee, a world-class observer and scientist, an excellent colleague, and a trusted friend. This work was supported by the National Science Foundation under grant numbers 1806926 and 1940800 (PI McQuinn). Support for programs HST-GO-16144 (PI McQuinn) and HST-GO-15227 (PI Burchett) was provided by NASA through a grant from the Space Telescope Science Institute, which is operated by the Associations of Universities for Research in Astronomy, Incorporated, under NASA contract NAS5- 26555. O.\ G.\ T.\ acknowledges support from grant HST-AR-16155 from the Space Telescope Science Institute and from a Carnegie-Princeton Fellowship through Princeton University and the Carnegie Observatories. This research has made use of NASA’s Astrophysics Data System Bibliographic Services and the NASA/IPAC Extragalactic Database (NED), which is operated by the Jet Propulsion Laboratory, California Institute of Technology, under contract with the National Aeronautics and Space Administration. The Flatiron Institute is funded by the Simons Foundation.

\facility{Hubble Space Telescope, Very Large Array, Australian Telescope Compact Array, Spitzer Space Telescope}

\appendix
We present three atlases that include the footprints of the HST observations, the CMDs from the HST observations, and the SFHs and AMRs derived from the HST data below. The galaxies are ordered by increasing heliocentric distance. 

\section{Atlas of HST Fields of View}\label{sec:atlas_fov}
\begin{figure*}
\includegraphics[width=0.98\textwidth]{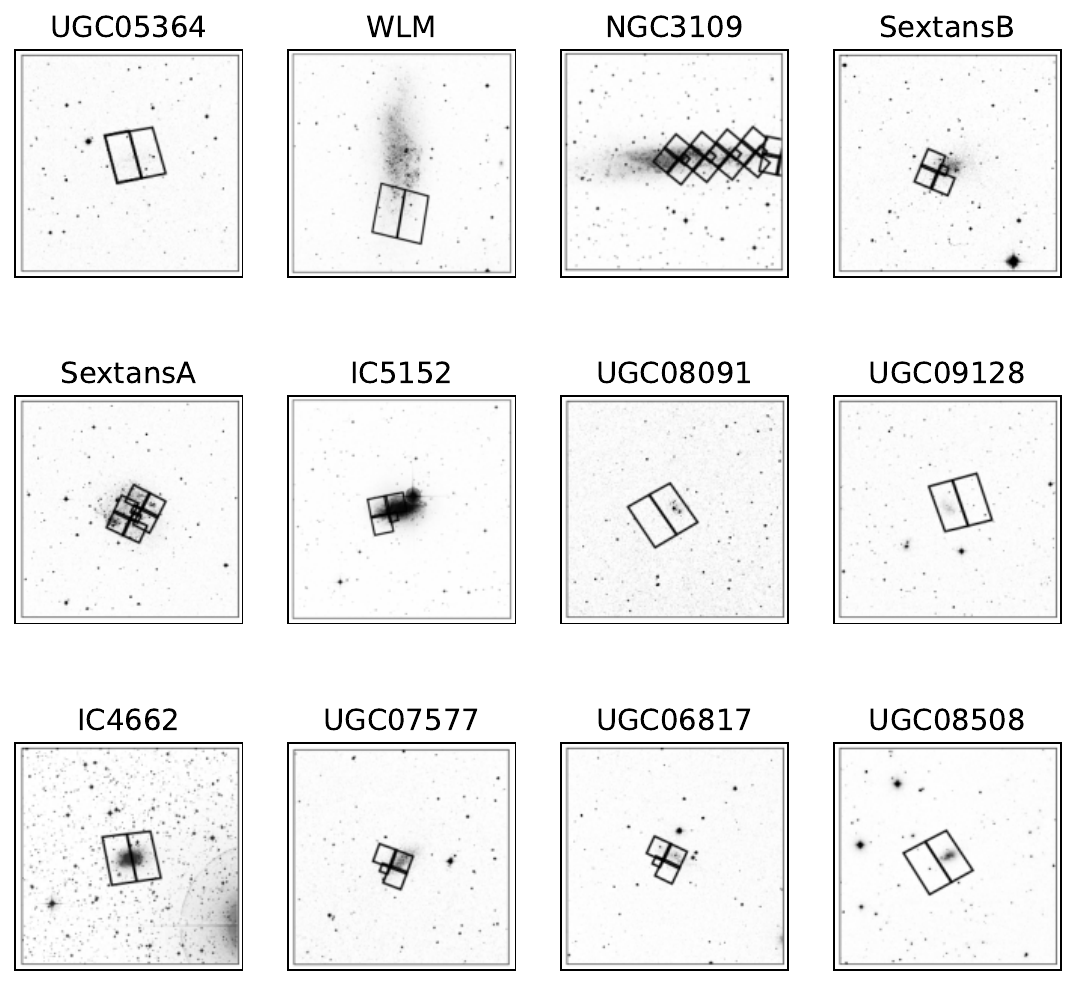}
\caption{HST footprints overlaid on DSS images on a subsample of the GLOW galaxies.}
\label{fig:atlas1}
\end{figure*}

\begin{figure*}
\includegraphics[width=0.98\textwidth]{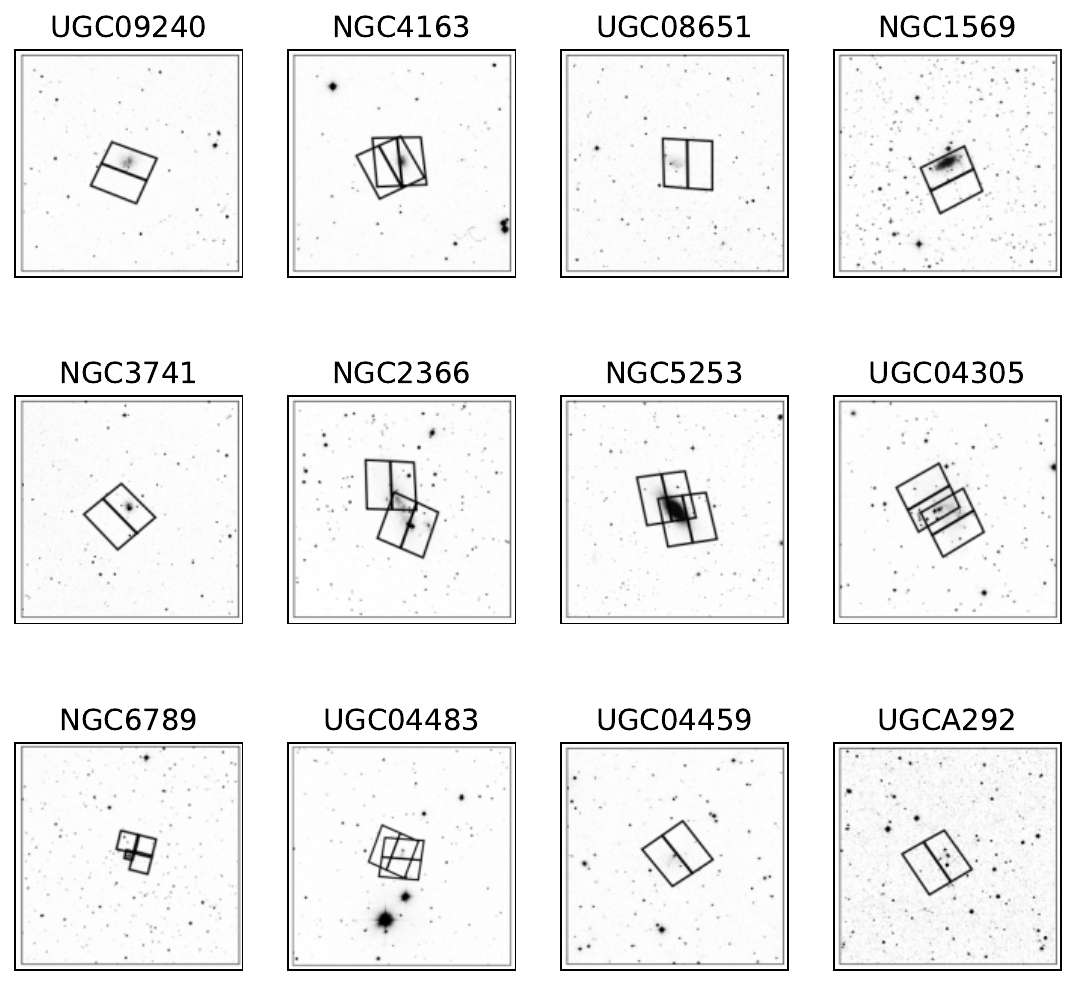}
\caption{HST footprints overlaid on DSS images on a subsample of the GLOW galaxies.}
\label{fig:atlas2}
\end{figure*}

\begin{figure*}
\includegraphics[width=0.98\textwidth]{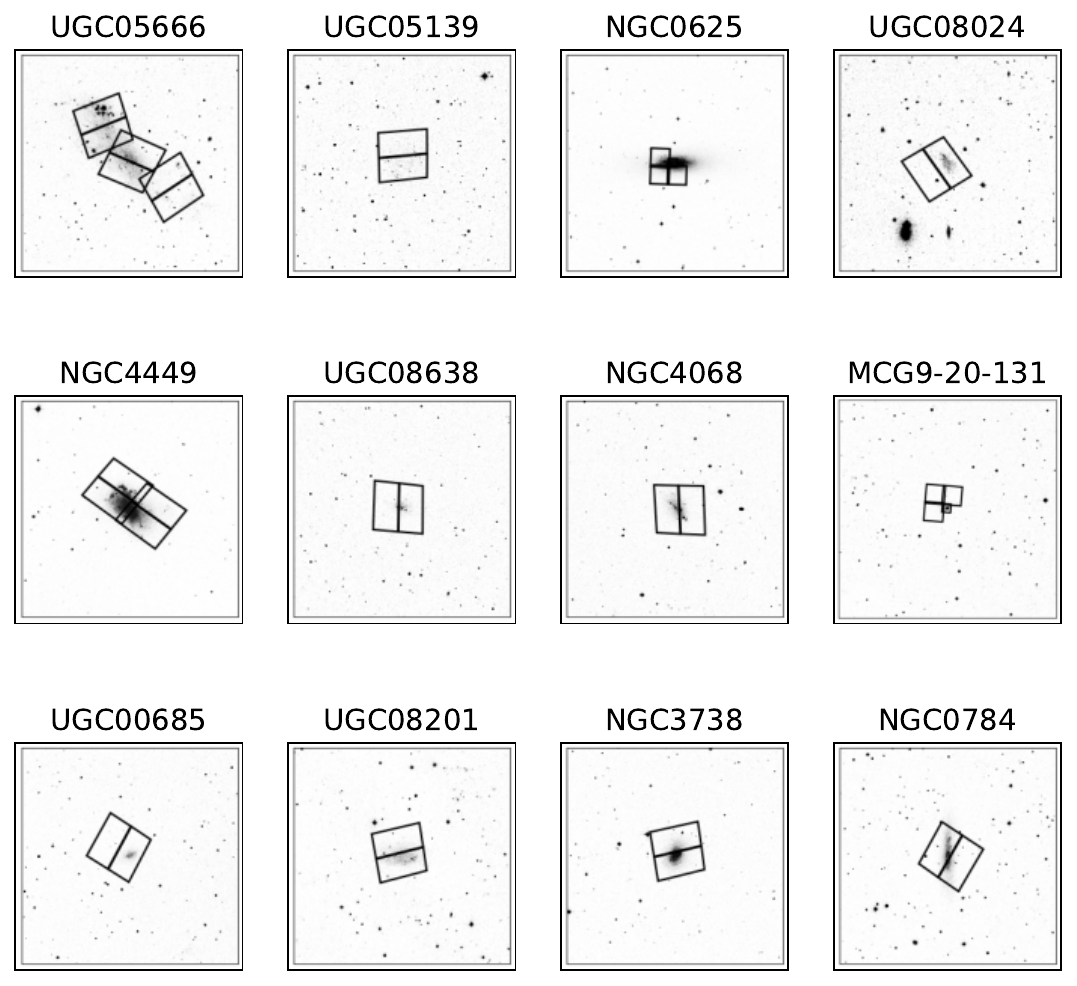}
\caption{HST footprints overlaid on DSS images on a subsample of the GLOW galaxies.}
\label{fig:atlas3}
\end{figure*}

\begin{figure}
\includegraphics[width=0.23\textwidth]{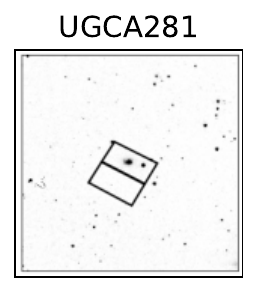}
\caption{HST footprints overlaid on DSS images on a subsample of the GLOW galaxies.}
\label{fig:atlas4}
\end{figure}

Figures~\ref{fig:atlas1}$-$\ref{fig:atlas4} present 15\arcmin $\times$ 15\arcmin\ DSS images of the sample with the HST footprints overlaid. For the majority of the sample, the HST imaging covers most of the stellar disks of the galaxies, providing comprehensive measurement of the global SFHs.

\section{Color-Magnitude Diagrams of GLOW Galaxies}\label{sec:atlas_cmds}
\begin{figure*}
\includegraphics[width=0.98\textwidth]{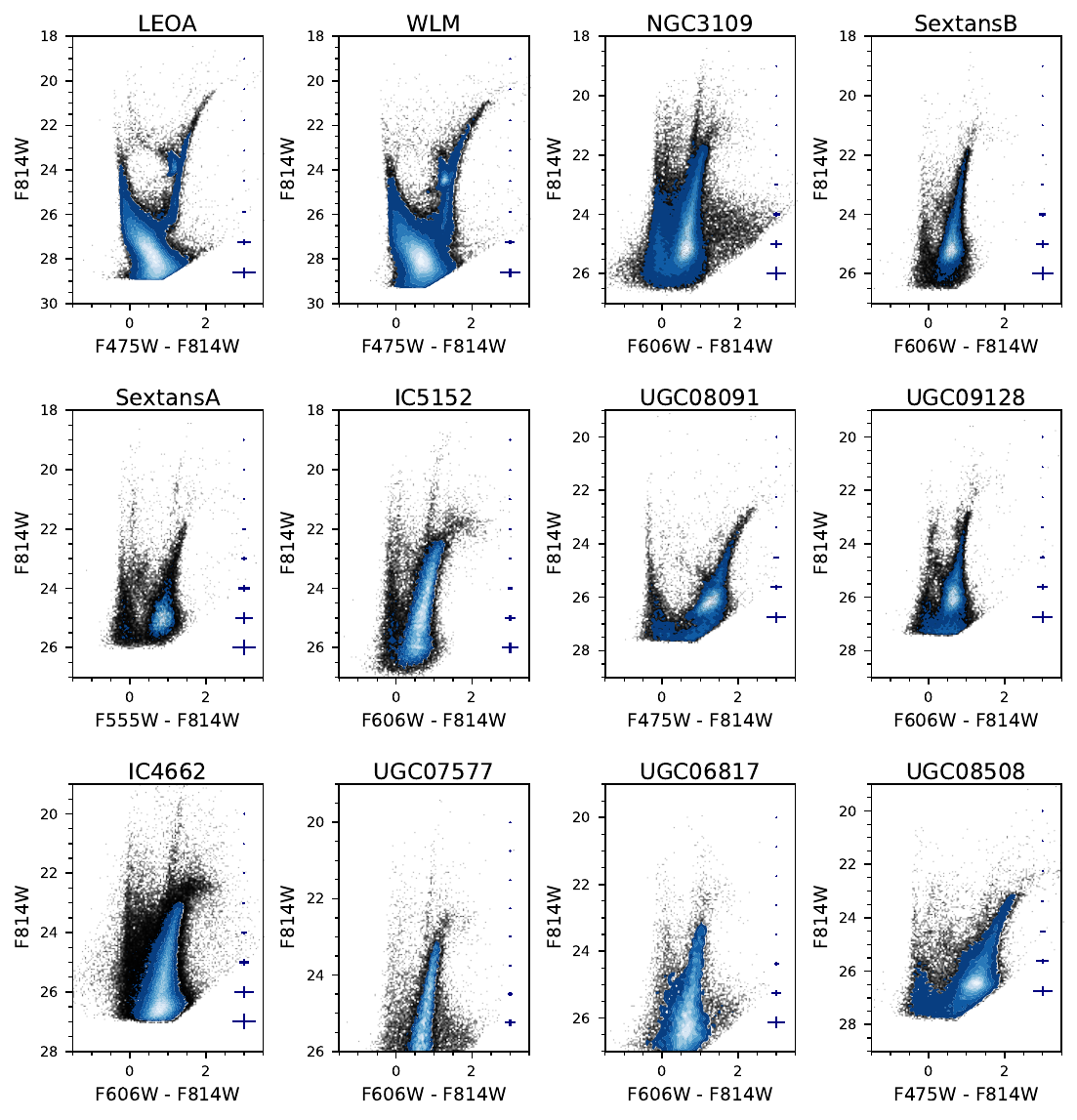}
\caption{CMDs of a subsample of the GLOW galaxies.}
\label{fig:cmd1}
\end{figure*}

\begin{figure*}
\includegraphics[width=0.98\textwidth]{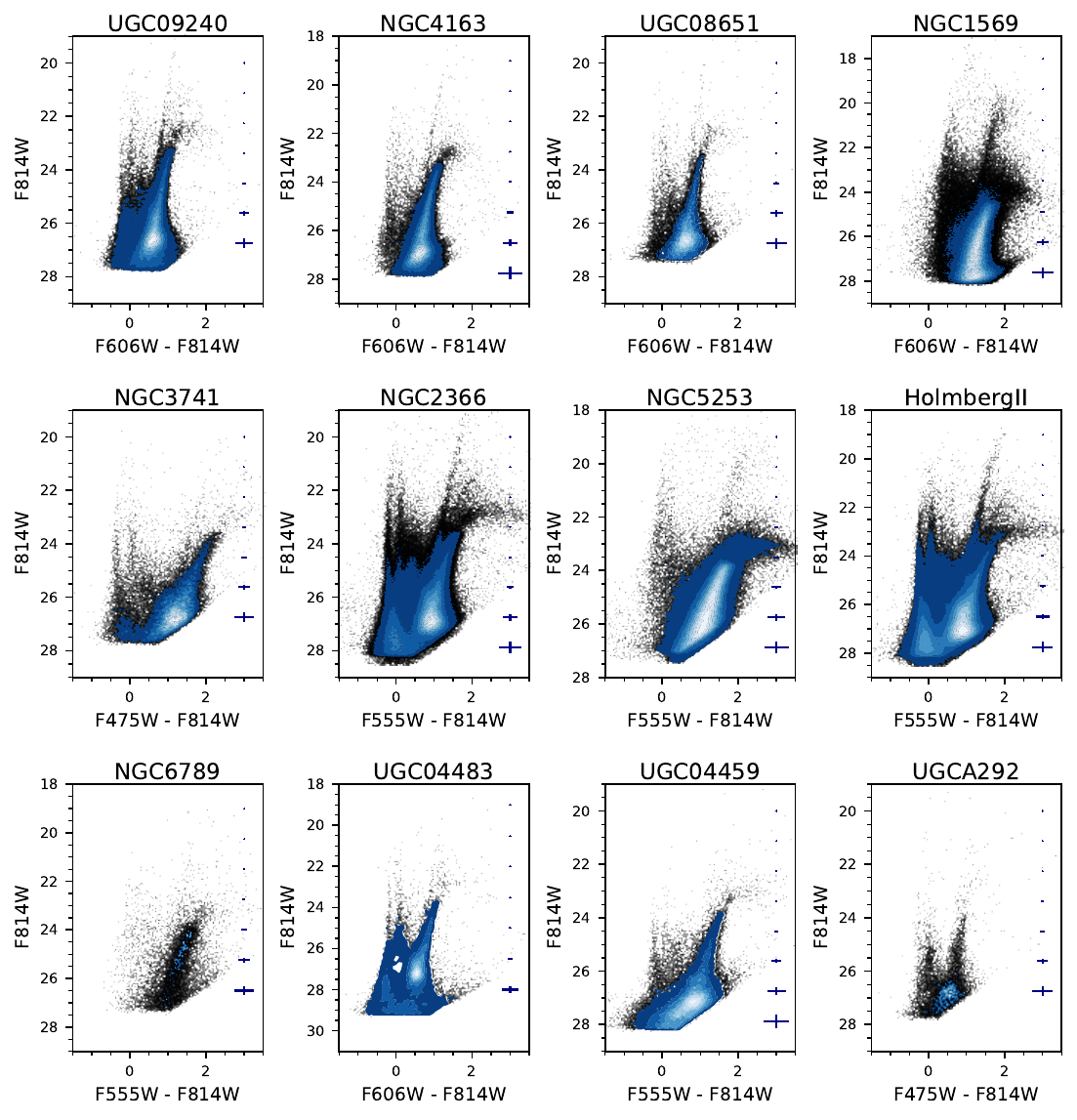}
\caption{CMDs of a subsample of the GLOW galaxies.}
\label{fig:cmd2}
\end{figure*}

\begin{figure*}
\includegraphics[width=0.98\textwidth]{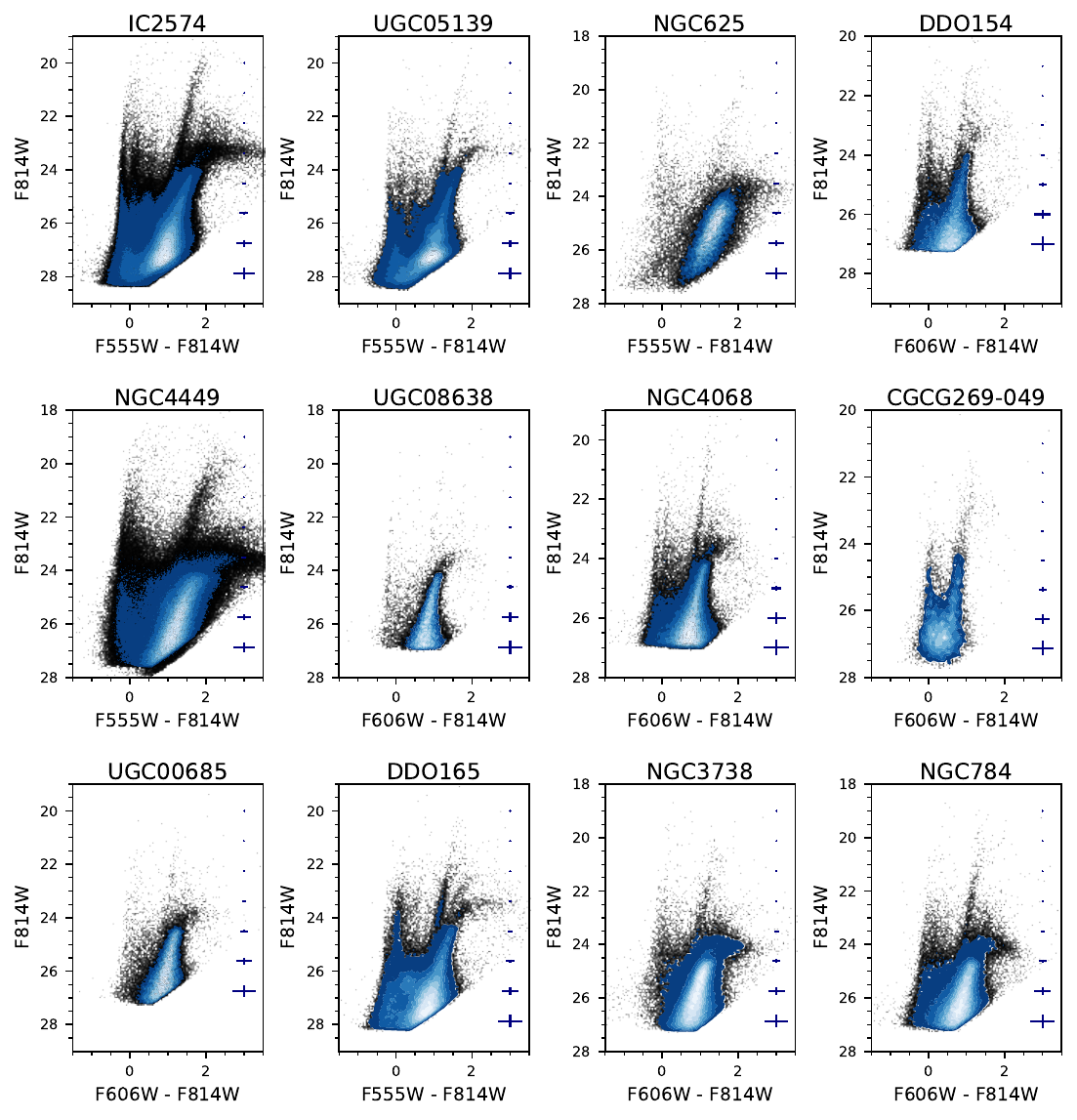}
\caption{CMDs of a subsample of the GLOW galaxies.}
\label{fig:cmd3}
\end{figure*}

\begin{figure}
\includegraphics[width=0.26\textwidth]{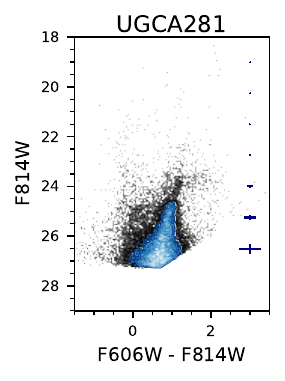}
\caption{CMDs of a subsample of the GLOW galaxies.}
\label{fig:cmd4}
\end{figure}

Figures~\ref{fig:cmd1}$-$\ref{fig:cmd4} present an atlas of the HST optical CMDs for the sample. The F814W magnitudes are shown on the y-axis for all galaxies; the bluer filter used in the color is either the F475W, F555W, or the F606W, depending on available observations in the HST archive. Error bars on the right of each panel indicate typical photometric errors in each magnitude bin. 

\section{Star Formation Histories and Age-Metallicity Relations of GLOW Galaxies}\label{sec:atlas_sfhs_amrs_plots}
\begin{figure*}
\includegraphics[width=0.98\textwidth]{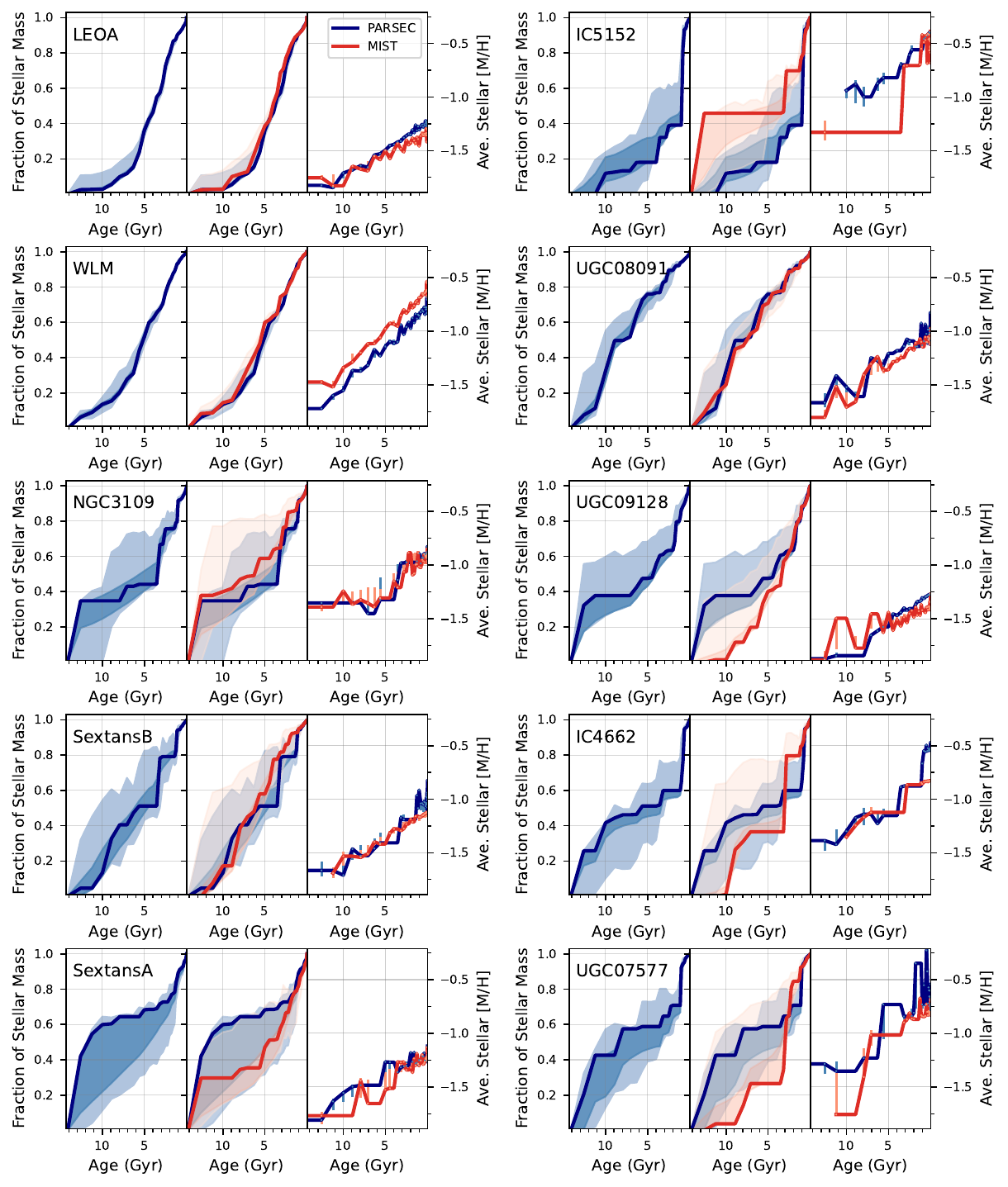}
\caption{SFHs and AMRs for a subset of the GLOW sample.}
\label{fig:sfh1}
\end{figure*}

\begin{figure*}
\includegraphics[width=0.98\textwidth]{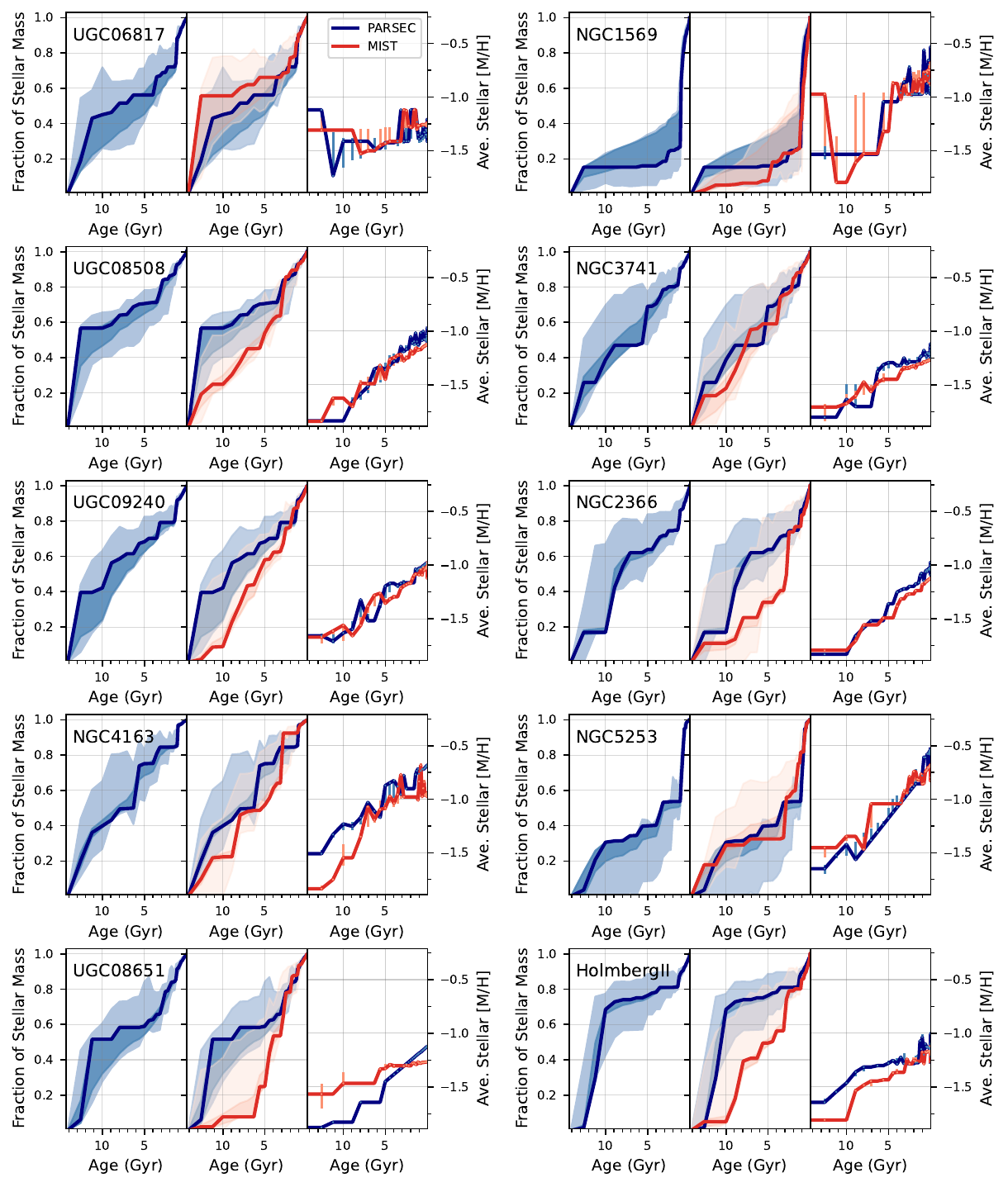}
\caption{FHs and AMRs for a subset of the GLOW sample.}
\label{fig:sfh2}
\end{figure*}

\begin{figure*}
\includegraphics[width=0.98\textwidth]{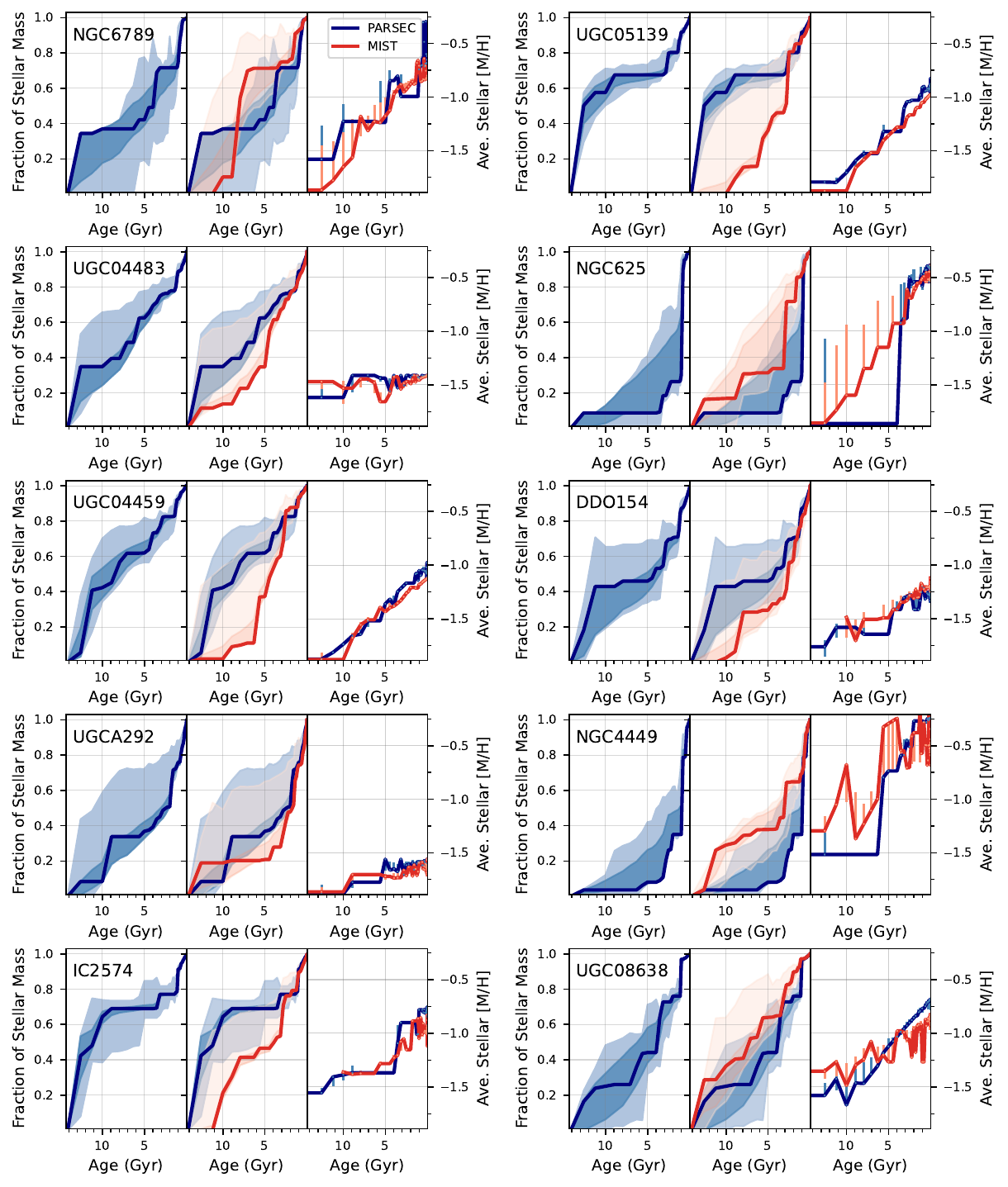}
\caption{FHs and AMRs for a subset of the GLOW sample.}
\label{fig:sfh3}
\end{figure*}

\begin{figure*}
\includegraphics[width=0.98\textwidth]{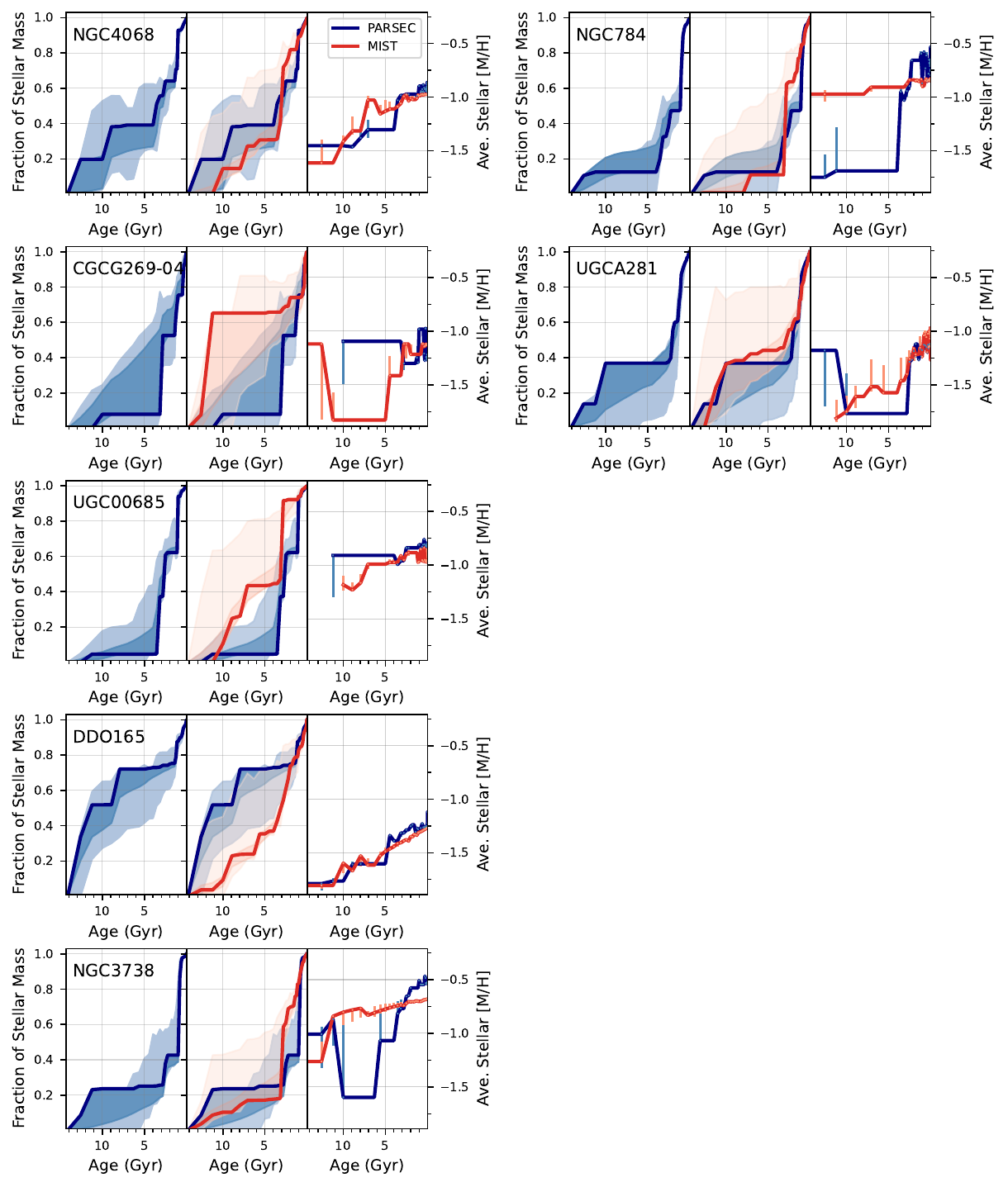}
\caption{FHs and AMRs for a subset of the GLOW sample.}
\label{fig:sfh4}
\end{figure*}

Figures~\ref{fig:sfh1}$-$\ref{fig:sfh4} present the SFHs and AMRs for the sample derived from the CMDs shown in Figures~\ref{fig:cmd1}$-$\ref{fig:cmd4}. For each galaxy, the left panels show the best-fitting solution from the PARSEC stellar library with statistical uncertainties in darker blue shading and the total (statistical and systematic) uncertainties in the lighter blue shading. The middle panels compare the PARSEC SFH solution with the solution from the MIST stellar libraries (statistical uncertainties only). The final panels presents the AMRs from both libraries.

\renewcommand\bibname{{References}}
\bibliography{ms_glow.bib}
\end{document}

%% file: tab1.tex
\begin{table*}
\caption{Data Sets, Measurements, and Their Applications in GLOW}
\label{tab:data_chart}
\begin{center}
\begin{tabular}{l | l}
\hline \hline
		& \\
Data    	& Applications \\
\hline \hline
21-cm line 		& $\bullet$ Spatially resolved \hi\ distribution used to estimate the percent of \hi\  flux within\\
interferometric data cubes (\S~\ref{sec:hi})	& different scale lengths determined primarily from 3.6$\micron$ surface brightness fits \\
\hline
{\it Spitzer Space Telescope}  & $\bullet$ Geometry based on isophote fitting \\ 
3.6 $\micron$ imaging	(\S~\ref{sec:structural}, \S~\ref{sec:masses})					& $\bullet$ Scale lengths measured from 3.6 $\micron$ surface brightness profiles \\
								& $\bullet$ Stellar masses based on 3.6 $\micron$ fluxes and adopting a mass-to-light ratio \\
                                & $\bullet$ Scaling to account for stellar mass outside of HST footprint \\
\hline
HST optical imaging of resolved stars (\S~\ref{sec:sfhs})			& $\bullet$ Star formation histories (SFHs) and age-metallicity relations (AMRs) \\
								& $\bullet$ Metal content locked in stars from SFHs and AMRs\\
								& $\bullet$ Evolution of the MZ relation from SFHs and AMRs \\
								& $\bullet$ Energy injected to drive outflows from the SFHs \\

\hline
Ground-Based B-band and R-band		& $\bullet$ Scale lengths and central surface brightness from surface brightness profiles\\
 imaging (\S~\ref{sec:structural})	& $\bullet$ Integrated light measurements  \\
\hline \hline
								& \\
Measurements from the Literature				& Applications \\		
\hline\hline	
21-cm line fluxes from single dish data			& $\bullet$ Total \hi\ fluxes of the galaxies used to estimate the \hi\ mass  \\
                    			& $\bullet$ 21 cm line widths used to estimate the rotational velocities of the gas \\
\hline
TRGB distances						& $\bullet$ Places myriad galaxy properties on an absolute scale\\		
								& $\bullet$ Determines separation between galaxies and overall density of environment\\	
\hline								
Gas-phase oxygen and nitrogen abundances		& $\bullet$ Determines oxygen content in the ISM and N/O variations across the sample\\
\hline								
Elemental abundance constraints  in CGM 		& $\bullet$ Constraints on the metal content in the CGM \\
\hline
Updated Nearby Galaxy Catalog				& $\bullet$ Quantifies the density and complexity of local environment\\
\hline \hline
\end{tabular}
\end{center}
\tablecomments{The data used in the GLOW project are listed in the top portion of the table. Additional measurements from the literature are listed in the bottom portion; references are provided in the individual data tables below. The high-level workflow is described in Section~\ref{sec:workflow}.}
\end{table*}

%% file: tab2.tex
\begin{deluxetable*}{llCCCCCCc}
\setlength{\tabcolsep}{2pt}
\label{tab:properties}
\tablecaption{GLOW Galaxy Sample}
\tablehead{
\colhead{Galaxy} & \colhead{Alt.} & \colhead{RA} & \colhead{Dec} & \colhead{Distance} & \colhead{$A_V$} & \colhead{12+log(O/H)} & \colhead{log(N/O)} & Ref. \\
\colhead{}	& \colhead{Name}	&  {\rm (J2000)}	& {\rm (J2000)}	& {\rm (Mpc)}	& {\rm (mag)}	&	&	&
}
\colnumbers
\startdata
WLM 		& & 00:01:58.53 & -15:27:27.20 & 0.98 _{- 0.04 } ^{+ 0.02 } & 0.10 & 7.77 \pm 0.10 & -1.46 \pm 0.05 & 1 \\
UGC~00685 	& & 01:07:22.43 & +16:41:04.60 & 4.81 _{- 0.04 } ^{+ 0.04 } & 0.16 & 8.00 \pm 0.03 & -1.45 \pm 0.08 & 1 \\
NGC~0625 	& & 01:35:04.58 & -41:26:13.50 & 4.02 _{- 0.07 } ^{+ 0.06 } & 0.05 & 8.08 \pm 0.12 & -1.25 \pm 0.03 & 1 \\
NGC~0784 	& & 02:01:16.95 & +28:50:06.59 & 5.37 _{- 0.02 } ^{+ 0.05 } & 0.16 & 7.97 \pm 0.05 & -1.54 \pm 0.10 & 2 \\
NGC~1569 	& & 04:30:49.68 & +64:50:54.35 & 3.19 _{- 0.09 } ^{+ 0.09 } & 1.90 & 8.15 \pm 0.05 & -1.40 \pm 0.10 & 3 \\
NGC~2366 	& & 07:28:53.10 & +69:12:37.39 & 3.28 _{- 0.03 } ^{+ 0.05 } & 0.10 & 7.91 \pm 0.05 & -1.17 \pm 0.26 & 1 \\
UGC~04305 	& Ho~II & 08:19:08.19 & +70:43:18.21 & 3.47 _{- 0.03 } ^{+ 0.03 } & 0.09 & 7.92 \pm 0.10 & -1.52 \pm 0.11 & 1 \\
UGC~04459 	& & 08:34:06.85 & +66:10:38.74 & 3.68 _{- 0.03 } ^{+ 0.05 } & 0.10 & 7.82 \pm 0.09 & -1.32 \pm 0.17 & 1 \\
UGC~04483 	& & 08:37:03.27 & +69:46:32.27 & 3.58 _{- 0.15 } ^{+ 0.15 } &0.09 & 7.56 \pm 0.03 & -1.57 \pm 0.07 & 1 \\
UGC~05139 	& Ho~I & 09:40:30.44 & +71:11:05.90 & 4.02 _{- 0.06 } ^{+ 0.04 } & 0.14 & 7.92 \pm 0.05 & -1.56 \pm 0.05 & 2 \\
UGC~05364 	& Leo~A	& 09:59:24.98 & +30:44:49.10 & 0.74 _{- 0.05 } ^{+ 0.16 } & 0.06 & 7.30 \pm 0.05 & -1.50 \pm 0.10 & 1 \\
SextansB 	& UGC~05373; DDO~70 & 09:59:59.75 & +05:19:57.61 & 1.43 _{- 0.02 } ^{+ 0.02 } & 0.09 & 7.84 \pm 0.05 & -1.46 \pm 0.06 & 1 \\
NGC~3109 	& & 10:03:09.66 & -26:09:28.80 & 1.34 _{- 0.05 } ^{+ 0.06 } & 0.18 & 7.77 \pm 0.07 & -1.33 \pm 0.15 & 1 \\
SextansA 	& DDO~75 & 10:11:00.89 & -04:41:37.84 & 1.45 _{- 0.05 } ^{+ 0.05 } & 0.12 & 7.54 \pm 0.09 & -1.54 \pm 0.13 & 1 \\
UGC~05666 	& IC2574 & 10:28:29.88 & +68:25:32.88 & 3.93 _{- 0.04 } ^{+ 0.05 } & 0.10 & 7.93 \pm 0.05 & -1.45 \pm 0.08 & 1 \\
NGC~3738 	& & 11:35:48.52 & +54:31:27.40 & 5.30 _{- 0.05 } ^{+ 0.05 } & 0.03 & 8.04 \pm 0.06 & -1.33 \pm 0.02 & 1 \\
NGC~3741 	& & 11:36:06.08 & +45:17:11.25 & 3.22 _{- 0.18 } ^{+ 0.16 } & 0.07 & 7.68 \pm 0.03 & -1.61 \pm 0.03 & 2 \\
UGC~06817 	& & 11:50:53.63 & +38:52:50.48 & 2.65 _{- 0.10 } ^{+ 0.10 } & 0.07 & 7.53 \pm 0.02 & -1.53 \pm 0.03 & 2 \\
NGC~4068 	& & 12:04:02.40 & +52:35:27.28 & 4.39 _{- 0.04 } ^{+ 0.04 } & 0.06 & 8.00 \pm 0.10 & -1.76 \pm 0.10 & this work \\
NGC~4163 	& & 12:12:09.01 & +36:10:07.99 & 2.99 _{- 0.03 } ^{+ 0.04 } & 0.06 & 7.56 \pm 0.14 & -1.49 \pm 0.06 & 2 \\
CGCG269-049 & & 12:15:47.05 & +52:23:17.53 & 4.61 _{- 0.19 } ^{+ 0.17 } & 0.07 & 7.47 \pm 0.02 & -1.57 \pm 0.03 & 2 \\
UGCA~281 	& & 12:26:16.79 & +48:29:38.15 & 5.70 _{- 0.13 } ^{+ 0.11 } & 0.04 & 7.80 \pm 0.03 & -1.43 \pm 0.09 & 1 \\
UGC~07577 	& DDO125 & 12:27:42.15 & +43:29:32.66 & 2.61 _{- 0.06 } ^{+ 0.05 } & 0.06 & 7.97 \pm 0.06 & -1.37 \pm 0.04 & 2 \\
NGC~4449 	& & 12:28:11.20 & +44:05:39.05 & 4.27 _{- 0.02 } ^{+ 0.02 } & 0.05 & 8.32 \pm 0.03 & -1.39 \pm 0.02 & 2 \\
UGCA~292 	& & 12:38:40.24 & +32:45:46.13 & 3.85 _{- 0.09 } ^{+ 0.55 } & 0.04 & 7.30 \pm 0.03 & -1.45 \pm 0.07 & 1 \\
UGC~08024 	& DDO154 & 12:54:05.47 & +27:08:54.74 & 4.04 _{- 0.06 } ^{+ 0.07 } & 0.03 & 7.67 \pm 0.06 & -1.68 \pm 0.13 & 1 \\
UGC~08091 	& GR8 & 12:58:39.88 & +14:13:06.07 & 2.19 _{- 0.12 } ^{+ 0.09 } & 0.07 & 7.65 \pm 0.06 & -1.50 \pm 0.10 & 1 \\
UGC~08201 	& DDO165 & 13:06:25.46 & +67:42:25.46 & 4.83 _{- 0.04 } ^{+ 0.04 } & 0.07 & 7.80 \pm 0.06 & -1.77 \pm 0.07 & 2 \\
UGC~08508 	& & 13:30:44.28 & +54:54:38.12 & 2.67 _{- 0.10 } ^{+ 0.10 } & 0.04 & 7.76 \pm 0.07 & -1.60 \pm 0.07 & 2 \\
UGC~08638 	& & 13:39:19.29 & +24:46:33.76 & 4.29 _{- 0.04 } ^{+ 0.06 } & 0.04 & 7.94 \pm 0.05 & -1.55 \pm 0.04 & 2 \\
UGC~08651 	& & 13:39:53.85 & +40:44:25.13 & 3.10 _{- 0.06 } ^{+ 0.06 } & 0.02 & 7.85 \pm 0.04 & -1.60 \pm 0.09 & 1 \\
NGC~5253 	& & 13:39:55.92 & -31:38:30.40 & 3.44 _{- 0.02 } ^{+ 0.03 } & 0.15 & 8.15 \pm 0.10 & -1.55 \pm 0.10 & 1 \\
UGC~09128 	& DDO187 & 14:15:56.86 & +23:03:22.83 & 2.30 _{- 0.04 } ^{+ 0.03 } & 0.06 & 7.75 \pm 0.05 & -1.80 \pm 0.12 & 1 \\
UGC~09240 	& DDO190 & 14:24:43.70 & +44:31:36.35 & 2.83 _{- 0.04 } ^{+ 0.05 } & 0.03 & 7.95 \pm 0.03 & -1.60 \pm 0.06 & 1 \\
IC~4662 	& & 17:47:08.49 & -64:38:31.40 & 2.55 _{- 0.02 } ^{+ 0.02 } & 0.19  & 8.17 \pm 0.04 & -1.50 \pm 0.05 & 4,5 \\
NGC~6789 	& & 19:16:41.98 & +63:58:15.38 & 3.55 _{- 0.07 } ^{+ 0.05 } & 0.19 & 7.83 \pm 0.05 & -1.37 \pm 0.10 & 6 \\
IC~5152 	& & 22:02:41.82 & -51:17:47.30 & 1.96 _{- 0.05 } ^{+ 0.03 } & 0.07 & 7.92 \pm 0.07 & -1.05 \pm 0.12 & 1 \\
\enddata
\tablecomments{Galaxy sample, distances, Galactic extinction, gas-phase oxygen abundances, and N/O ratio ordered by RA. Distances are from the Extragalactic Distance Database \citep{Tully2013}. $A_V$ values represent the average Galactic extinction at the coordinates of the galaxy from \cite{Schlafly2011}. Tabulated values for 12 + log(O/H) and log(N/O) are from a combination of individual \hii\ regions and averages over multiple \hii\ regions in each galaxy from the compilations for the Local Volume Legacy Survey reported in the following references}: 1 $-$ \citet{Marble2010} and 2 $-$ \citet{Berg2012} with additional references for galaxies not in LVL: 3 $-$ \citet{Kobulnicky1997}; 4 $-$ \citet{Hidalgo-Gamez2001}; 5 $-$ \citet{Stasinska1986}; 6 $-$ \citet{Garcia-Benito2012}.
\end{deluxetable*}

%% file: tab3.tex
\begin{deluxetable*}{lDDD | DDr | DDr | DDDrD}
\tabletypesize{\footnotesize}
\setlength{\tabcolsep}{3pt}
\label{tab:structural}
\tablecaption{Structural Properties}
\tablehead{
\colhead{Galaxy} & \multicolumn2c{a} & \multicolumn2c{b} &\multicolumn2c{PA} & \multicolumn2c{$\mu_B$} & \multicolumn2c{$\alpha_B$}  & \colhead{r$_{h,B}$} &  \multicolumn2c{$\mu_R$} & \multicolumn2c{$\alpha_R$} & \colhead{r$_{h,R}$} & \multicolumn2c{$\mu_{3.6}$} & \multicolumn4c{$\alpha_{3.6}$} & \multicolumn2c{r$_{h,3.6}$} \\
\colhead{} & \multicolumn2c{(\arcsec)} & \multicolumn2c{(\arcsec)} & \multicolumn2c{($^{\circ}$)} & \multicolumn2c{(mag/\arcsec$^2$)} & \multicolumn2c{(\arcsec)} & \colhead{(\arcsec)} & \multicolumn2c{(mag/\arcsec$^2$)} & \multicolumn2c{(\arcsec)} & \colhead{(\arcsec)} & \multicolumn2c{(mag/\arcsec$^2$)} & \multicolumn2c{(\arcsec)} & \multicolumn2c{(kpc)} &   \colhead{(\arcsec)} & \multicolumn2c{(kpc)}
}
\decimalcolnumbers
\startdata
WLM & 268.5 & 117.9 & 183.58 & 23.50 & 120.7 & 206 & 22.87 & 131.7 & 224 & 21.05 & 162.3 & 0.77 & 278 & 1.32 \\
UGC~00685 & 67.6 & 43.0 & 116.86 & 22.06 & 16.0 & 26 & 21.36 & 18.3 & 30 & 19.35 & 20.2 & 0.47 & 33 & 0.77 \\
NGC~0625 & 221.8 & 89.8 & 272.08 & 22.38 & 51.0 & 60 & 21.33 & 53.0 & 69 & 19.30 & 58.0 & 1.13 & 67 & 1.31 \\
NGC~0784 & 202.6 & 56.4 & 0.94 & 22.89 & 54.9 & 94 & 22.05 & 54.1 & 93 & 19.99 & 57.2 & 1.49 & 95 & 2.47 \\
NGC~1569 & 175.8 & 86.6 & 117.57 & 21.30 & 25.0 & 33 & 20.35 & 31.8 & 34 & 17.29 & 31.5 & 0.49 & 30 & 0.46 \\
NGC~2366 & 191.2 & 65.0 & 31.01 & 23.42 & 91.8 & 138 & 22.76 & 90.6 & 142 & 20.80 & 96.5 & 1.53 & 154 & 2.45 \\
UGC~04305 & 209.9 & 160.7 & 209.62 & 22.57 & 70.4 & 127 & 21.79 & 64.6 & 116 & 19.73 & 62.6 & 1.05 & 110 & 1.85 \\
UGC~04459 & 54.5 & 48.3 & 133.00 & 23.50 & 26.0 & 44 & 22.95 & 32.7 & 58 & 20.78 & 25.2 & 0.45 & 44 & 0.79 \\
UGC~04483 & 42.5 & 25.4 & 351.53 & 22.96 & 18.6 & 28 & 22.44 & 18.1 & 30 & 20.22 & 14.1 & 0.24 & 25 & 0.43 \\
UGC~05139 & 95.9 & 90.6 & 66.84 & 23.81 & 53.7 & 95 & 23.16 & 54.6 & 98 & 20.72 & 43.4 & 0.85 & 82 & 1.60 \\
UGC~05364 & 100.0 & 48.4 & 102.45 & 23.70 & 50.0 & 97 & 23.33 & 51.9 & 100 & 21.04 & 54.4 & 0.20 & 106 & 0.38 \\
SextansB & 170.8 & 114.2 & 97.10 & 22.74 & 61.8 & 105 & 22.04 & 63.4 & 110 & 20.28 & 70.5 & 0.49 & 121 & 0.84 \\
NGC~3109 & 476.3 & 131.7 & 94.32 & 23.12 & 150.3 & 258 & 22.30 & 141.8 & 244 & 20.21 & 143.6 & 0.93 & 242 & 1.57 \\
SextansA & 169.3 & 145.3 & 46.26 & 21.09 & 37.7 & 103 & 21.09 & 45.5 & 109 & 19.35 & 50.7 & 0.36 & 113 & 0.79 \\
UGC~05666 & 411.8 & 153.2 & 45.48 & 23.76 & 141.8 & 262 & 22.72 & 113.9 & 222 & 20.81 & 140.3 & 2.67 & 241 & 4.59 \\
NGC~3738 & 124.1 & 81.8 & 340.76 & 21.82 & 32.6 & 32 & 20.92 & 34.1 & 41 & 18.64 & 31.7 & 0.81 & 39 & 1.00 \\
NGC~3741 & 42.2 & 28.9 & 12.30 & 21.72 & 11.8 & 19 & 21.22 & 12.8 & 22 & 19.56 & 13.8 & 0.22 & 23 & 0.36 \\
UGC~06817 & 91.1 & 54.9 & 238.36 & 23.52 & 49.6 & 86 & 23.06 & 52.0 & 90 & 21.30 & 45.2 & 0.58 & 77 & 0.99 \\
NGC~4068 & 94.4 & 52.5 & 31.36 & 22.27 & 28.1 & 53 & 21.66 & 30.5 & 55 & 19.86 & 33.4 & 0.71 & 57 & 1.21 \\
NGC~4163 & 79.0 & 51.0 & 192.42 & 21.86 & 19.0 & 29 & 21.11 & 21.1 & 34 & 19.13 & 22.0 & 0.32 & 36 & 0.52 \\
CGCG269-049 & 28.5 & 15.9 & 312.15 & 22.15 & 10.1 & 17 & 21.67 & 10.4 & 18 & 19.99 & 10.8 & 0.24 & 17 & 0.38 \\
UGCA~281 & 37.6 & 28.2 & 264.87 & 21.79 & 9.7 & 12 & 21.31 & 10.7 & 14 & 19.84 & 13.4 & 0.37 & 19 & 0.53 \\
UGC~07577 & 169.9 & 84.9 & 127.46 & 23.08 & 46.8 & 82 & 22.40 & 51.8 & 91 & 20.55 & 58.3 & 0.74 & 100 & 1.27 \\
NGC~4449 & 175.0 & 101.3 & 41.02 & 19.83 & 38.8 & 60 & 19.25 & 42.2 & 66 & 16.87 & 40.5 & 0.84 & 64 & 1.32 \\
UGCA~292 & 47.7 & 40.0 & 63.91 & 23.70 & 16.9 & 32 & 23.15 & 15.2 & 31 & 21.43 & 27.8 & 0.52 & 50 & 0.93 \\
UGC~08024 & 73.8 & 39.6 & 215.92 & 23.12 & 27.1 & 48 & 22.51 & 27.2 & 49 & 20.74 & 27.9 & 0.55 & 49 & 0.96 \\
UGC~08091 & 55.0 & 36.6 & 40.66 & 22.16 & 14.2 & 28 & 22.00 & 18.1 & 34 & 20.45 & 19.8 & 0.21 & 36 & 0.38 \\
UGC~08201 & 122.9 & 61.6 & 269.32 & 22.64 & 36.3 & 73 & 22.27 & 39.3 & 76 & 20.64 & 42.9 & 1.00 & 83 & 1.94 \\
UGC~08508 & 63.2 & 39.7 & 118.08 & 22.13 & 17.3 & 30 & 21.58 & 19.2 & 34 & 19.77 & 21.6 & 0.28 & 36 & 0.47 \\
UGC~08638 & 62.2 & 34.3 & 69.17 & 22.65 & 18.4 & 31 & 22.09 & 21.4 & 37 & 20.07 & 21.8 & 0.45 & 38 & 0.79 \\
UGC~08651 & 65.6 & 35.0 & 241.12 & 23.18 & 26.0 & 48 & 22.54 & 26.7 & 49 & 20.77 & 28.6 & 0.43 & 52 & 0.78 \\
NGC~5253 & 164.9 & 80.5 & 42.83 & 20.82 & 35.4 & 35 & 19.91 & 36.8 & 44 & 17.54 & 33.9 & 0.57 & 29 & 0.48 \\
UGC~09128 & 57.7 & 34.4 & 224.70 & 22.63 & 18.9 & 35 & 22.30 & 21.1 & 38 & 20.43 & 20.6 & 0.23 & 36 & 0.40 \\
UGC~09240 & 73.1 & 58.3 & 127.74 & 21.72 & 20.8 & 36 & 21.18 & 22.9 & 40 & 19.43 & 24.5 & 0.34 & 43 & 0.59 \\
IC~4662 & 110.5 & 72.0 & 280.92 & 19.93 & 19.3 & 30 & 19.35 & 21.5 & 34 & 17.26 & 21.0 & 0.26 & 34 & 0.42 \\
NGC~6789 & 57.6 & 48.4 & 269.50 & 21.51 & 13.4 & 21 & 20.74 & 15.3 & 25 & 18.70 & 15.5 & 0.27 & 25 & 0.43 \\
IC~5152 & 180.8 & 105.8 & 279.29 & 21.08 & 43.4 & 73 & 20.23 & 42.2 & 72 & 17.89 & 37.6 & 0.36 & 64 & 0.61 \\
\enddata
\tablecomments{Columns 2, 3, and 4 are the semi-major axes, semi-minor axes, and position angles determined from isophote fits. Columns 5$-$13 are the central surface brightness in units of mag per square arcsec ($\mu$) corrected by the axial ratio of the system, scale lengths ($\alpha$) in units of arcsec, and half-light radii ($r_{h}$) in units of arcsec measured from surface brightness profile fits from the cleaned B-band, R-band, and 3.6$\mu$m images.}
\end{deluxetable*}

%% file: tab4.tex
\begin{deluxetable*}{l D@{$\pm$}DD@{$\pm$}DD@{$\pm$}D | DDD | D@{$\pm$}DD@{$\pm$}DD@{$\pm$}D}
\tabletypesize{\footnotesize}
\setlength{\tabcolsep}{3pt}
\label{tab:fluxes}
\tablecaption{Integrated Apparent Magnitudes in Different Band-passes and Apertures}
\tablehead{
\colhead{Galaxy} & \multicolumn{12}c{Visual Extent} &  \multicolumn6c{Extrapolated Total} & \multicolumn{12}c{HST FoV}\\
\hline
\colhead{}  &  \multicolumn4c{m$_B$}  & \multicolumn4c{m$_R$} & \multicolumn4c{m$_{3.6\mu m}$} & \multicolumn2c{m$_B$} & \multicolumn2c{m$_R$} & \multicolumn2c{m$_{3.6\mu m}$} & \multicolumn4c{m$_B$} & \multicolumn4c{m$_R$} & \multicolumn4c{m$_{3.6\mu m}$} \\
\colhead{} & \multicolumn4c{(mag)} &  \multicolumn4c{(mag)} &  \multicolumn4c{(mag)} &  \multicolumn2c{(mag)} &  \multicolumn2c{(mag)} & \multicolumn2c{(mag)} &  \multicolumn4c{(mag)} &  \multicolumn4c{(mag)}  &  \multicolumn4c{(mag)} 
}  
\decimalcolnumbers
\startdata
WLM & 11.60 & 0.04 & 10.84 & 0.03 & 8.78 & 0.02 & 11.11 & 10.28 & 8.01 & 13.05 & 0.04 & 12.28 & 0.03 & 10.32 & 0.03 \\
UGC~00685 & 14.28 & 0.03 & 13.32 & 0.04 & 11.14 & 0.04 & 14.15 & 13.15 & 10.94 & 14.19 & 0.08 & 13.22 & 0.13 & 10.92 & 0.04 \\
NGC~0625 & 11.74 & 0.02 & 10.71 & 0.02 & 8.44 & 0.02 & 11.65 & 10.61 & 8.34 & 12.40 & 0.02 & 11.41 & 0.02 & 9.10 & 0.02 \\
NGC~0784 & 12.42 & 0.04 & 11.60 & 0.02 & 9.45 & 0.02 & 12.26 & 11.45 & 9.28 & 12.55 & 0.04 & 11.73 & 0.02 & 9.56 & 0.02 \\
NGC~1569 & 12.10 & 0.04 & 10.58 & 0.03 & 7.43 & 0.01 & 12.11 & 10.55 & 7.41 & 12.20 & 0.04 & 10.70 & 0.03 & 7.52 & 0.01 \\
NGC~2366 & 12.00 & 0.02 & 11.42 & 0.02 & 9.37 & 0.02 & 11.52 & 10.93 & 8.83 & 11.73 & 0.02 & 11.16 & 0.02 & 9.10 & 0.02 \\
UGC~04305 & 11.64 & 0.04 & 11.02 & 0.05 & 9.01 & 0.02 & 11.37 & 10.80 & 8.80 & 11.80 & 0.04 & 11.17 & 0.04 & 9.14 & 0.02 \\
UGC~04459 & 14.96 & 0.04 & 14.19 & 0.04 & 12.34 & 0.07 & 14.41 & 13.39 & 11.81 & 14.52 & 0.12 & 13.63 & 0.11 & 11.84 & 0.06 \\
UGC~04483 & 15.04 & 0.03 & 14.58 & 0.04 & 12.93 & 0.09 & 14.6 & 14.13 & 12.60 & 14.34 & 0.06 & 13.91 & 0.10 & 12.92 & 0.09 \\
UGC~05139 & 13.98 & 0.04 & 13.30 & 0.04 & 11.22 & 0.04 & 13.25 & 12.54 & 10.66 & 13.80 & 0.05 & 13.12 & 0.06 & 11.07 & 0.04 \\
UGC~05364 & 14.08 & 0.03 & 13.66 & 0.04 & 11.39 & 0.05 & 13.38 & 12.92 & 10.55 & 13.74 & 0.04 & 13.31 & 0.07 & 11.04 & 0.08 \\
SextansB & 12.21 & 0.04 & 11.45 & 0.05 & 9.52 & 0.02 & 11.87 & 11.08 & 9.09 & 13.43 & 0.04 & 12.66 & 0.05 & 10.73 & 0.03 \\
NGC~3109 & 10.53 & 0.03 & 9.81 & 0.03 & 7.75 & 0.01 & 10.29 & 9.60 & 7.54 & 11.20 & 0.03 & 10.53 & 0.02 & 8.46 & 0.02 \\
SextansA & 12.13 & 0.03 & 11.54 & 0.03 & 9.52 & 0.04 & 11.96 & 11.29 & 9.24 & 12.59 & 0.02 & 12.05 & 0.02 & 10.13 & 0.03 \\
UGC~05666 & 11.22 & 0.04 & 10.56 & 0.04 & 8.38 & 0.02 & 10.97 & 10.44 & 8.12 & 11.43 & 0.04 & 10.75 & 0.34 & 8.58 & 0.02 \\
NGC~3738 & 12.05 & 0.03 & 11.22 & 0.03 & 9.11 & 0.02 & 11.94 & 11.08 & 9.01 & 12.02 & 0.03 & 11.19 & 0.03 & 9.09 & 0.02 \\
NGC~3741 & 14.66 & 0.03 & 14.01 & 0.03 & 12.19 & 0.07 & 14.48 & 13.78 & 11.90 & 14.52 & 0.09 & 13.87 & 0.14 & 11.90 & 0.06 \\
UGC~06817 & 13.76 & 0.04 & 13.22 & 0.03 & 11.62 & 0.05 & 13.08 & 12.49 & 11.05 & 13.87 & 0.04 & 13.35 & 0.04 & 11.71 & 0.05 \\
NGC~4068 & 13.30 & 0.02 & 12.57 & 0.02 & 10.63 & 0.03 & 13.09 & 12.30 & 10.33 & 13.18 & 0.03 & 12.41 & 0.04 & 10.43 & 0.03 \\
NGC~4163 & 13.65 & 0.05 & 12.72 & 0.04 & 10.67 & 0.03 & 13.54 & 12.56 & 10.51 & 13.50 & 0.06 & 12.58 & 0.06 & 10.51 & 0.03 \\
CGCG269-049 & 15.52 & 0.04 & 14.98 & 0.04 & 13.35 & 0.09 & 15.23 & 14.64 & 13.05 & 15.31 & 0.07 & 14.83 & 0.07 & 13.72 & 0.10 \\
UGCA~281 & 14.68 & 0.04 & 14.16 & 0.04 & 12.37 & 0.07 & 14.57 & 14.01 & 12.11 & 14.35 & 0.10 & 13.88 & 0.13 & 11.70 & 0.05 \\
UGC~07577 & 12.99 & 0.04 & 12.12 & 0.04 & 10.06 & 0.03 & 12.79 & 11.87 & 9.77 & 13.48 & 0.03 & 12.64 & 0.03 & 10.65 & 0.03 \\
NGC~4449 & 10.04 & 0.04 & 9.30 & 0.03 & 6.97 & 0.01 & 9.97 & 9.20 & 6.89 & 10.01 & 0.04 & 9.26 & 0.03 & 6.94 & 0.01 \\
UGCA~292 & 16.06 & 0.06 & 15.68 & 0.09 & 13.06 & 0.10 & 15.72 & 15.44 & 12.26 & 15.75 & 0.23 & 15.6 & 0.56 & 12.41 & 0.07 \\
UGC~08024 & 14.41 & 0.03 & 13.79 & 0.04 & 11.96 & 0.06 & 14.05 & 13.41 & 11.61 & 14.18 & 0.06 & 13.53 & 0.11 & 11.60 & 0.05 \\
UGC~08091 & 14.88 & 0.03 & 14.18 & 0.02 & 12.49 & 0.08 & 14.67 & 13.85 & 12.13 & 14.63 & 0.10 & 13.79 & 0.09 & 12.03 & 0.06 \\
UGC~08201 & 13.30 & 0.02 & 12.76 & 0.03 & 10.97 & 0.04 & 13.06 & 12.48 & 10.65 & 13.23 & 0.03 & 12.67 & 0.04 & 10.91 & 0.04 \\
UGC~08508 & 14.16 & 0.04 & 13.42 & 0.03 & 11.46 & 0.05 & 13.99 & 13.2 & 11.17 & 14.12 & 0.08 & 13.33 & 0.11 & 11.26 & 0.04 \\
UGC~08638 & 14.63 & 0.04 & 13.80 & 0.04 & 11.74 & 0.05 & 14.40 & 13.46 & 11.43 & 14.41 & 0.07 & 13.51 & 0.13 & 13.34 & 0.04 \\
UGC~08651 & 14.61 & 0.03 & 13.95 & 0.03 & 12.08 & 0.06 & 14.20 & 13.52 & 11.58 & 14.39 & 0.06 & 13.76 & 0.09 & 11.72 & 0.05 \\
NGC~5253 & 10.96 & 0.03 & 10.09 & 0.02 & 7.61 & 0.01 & 10.91 & 10.02 & 7.57 & 10.92 & 0.03 & 10.05 & 0.02 & 7.59 & 0.01 \\
UGC~09128 & 14.71 & 0.04 & 14.12 & 0.03 & 12.24 & 0.07 & 14.43 & 13.77 & 11.92 & 14.49 & 0.11 & 13.86 & 0.14 & 11.84 & 0.06 \\
UGC~09240 & 13.39 & 0.03 & 12.68 & 0.03 & 10.80 & 0.04 & 13.22 & 12.45 & 10.54 & 13.30 & 0.05 & 12.56 & 0.05 & 10.63 & 0.03 \\
IC~4662 & 11.71 & 0.02 & 10.90 & 0.02 & 8.87 & 0.04 & 11.68 & 10.85 & 8.83 & 11.70 & 0.02 & 10.89 & 0.02 & 8.85 & 0.07 \\
NGC~6789 & 14.09 & 0.04 & 13.08 & 0.02 & 11.02 & 0.02 & 13.99 & 12.92 & 10.87 & 14.30 & 0.04 & 13.26 & 0.03 & 11.27 & 0.02 \\
IC~5152 & 11.05 & 0.02 & 10.26 & 0.02 & 8.15 & 0.02 & 10.95 & 10.17 & 8.09 & 11.84 & 0.02 & 11.04 & 0.02 & 8.88 & 0.01 \\
\enddata
\tablecomments{Apparent magnitudes measured from the cleaned B-band, R-band, and 3.6$\micron$ images in three different apertures: Visual Extent corresponds to approximately the 25 mag per square arcsec surface brightness level for the B and R band images and 22 mag per square arcsec for the 3.6$\micron$ images; Extrapolated Total includes an estimate of the total flux based on surface brightness profile fits extrapolated to infinity; HST FoV includes the flux within an aperture matched to the HST images. In cases where there are multiple HST pointings, the magnitudes are based on the flux from all footprints. We conservatively adopt an uncertainty of 0.1 mag on the extrapolated magntiudes; see text for details.}
\end{deluxetable*}

%% file: tab5.tex
\begin{deluxetable*}{lcccccc}
\label{tab:stellar_masses}
\tablecaption{Stellar Masses}
\tablehead{
\colhead{Galaxy} &  \colhead{[3.6] Total} &  \colhead{[3.6] HST FoV}  &  \colhead{PARSEC HST FoV} &  \colhead{MIST HST FoV} &  \colhead{PARSEC} &  \colhead{MIST} \\
\colhead{} &  \colhead{log(M$_*$/\msun)}  &  \colhead{log(M$_*$/\msun)}  &  \colhead{log(M$_*$/\msun)}  &  \colhead{log(M$_*$/\msun)} &  \colhead{scaling} &  \colhead{scaling} \\
}
\decimalcolnumbers
\startdata
WLM & 7.68$\pm{0.12}$ & \nodata & 6.60$^{+0.01}_{-0.03}$ & 6.60$^{+0.02}_{-0.02}$ & 12.02$^{+3.42}_{-3.33}$ & 12.02$^{+3.37}_{-3.37}$ \\
UGC~00685 & 7.89$\pm{0.12}$ & 7.89$\pm{0.12}$ & 7.78$^{+0.15}_{-0.02}$ & 7.81$^{+0.19}_{-0.02}$ & 1.29$^{+0.36}_{-0.57}$ & 1.20$^{+0.34}_{-0.62}$ \\
NGC~0625 & 8.77$\pm{0.12}$ & 8.47$\pm{0.12}$ & 8.38$^{+0.32}_{-0.13}$ & 8.34$^{+0.17}_{-0.20}$ & 2.45$^{+1.00}_{-1.93}$ & 2.69$^{+1.44}_{-1.29}$ \\
NGC~0784 & 8.65$\pm{0.12}$ & 8.53$\pm{0.12}$ & 8.56$^{+0.09}_{-0.08}$ & 8.38$^{+0.32}_{-0.00}$ & 1.23$^{+0.41}_{-0.42}$ & 1.86$^{+0.51}_{-1.46}$ \\
NGC~1569 & 8.98$\pm{0.12}$ & 8.94$\pm{0.12}$ & 8.43$^{+0.19}_{-0.05}$ & 8.19$^{+0.28}_{-0.00}$ & 3.55$^{+1.06}_{-1.84}$ & 6.17$^{+1.71}_{-4.33}$ \\
NGC~2366 & 8.40$\pm{0.12}$ & 8.29$\pm{0.12}$ & 8.26$^{+0.11}_{-0.06}$ & 8.09$^{+0.15}_{-0.13}$ & 1.38$^{+0.43}_{-0.52}$ & 2.04$^{+0.83}_{-0.90}$ \\
UGC~04305 & 8.46$\pm{0.12}$ & 8.32$\pm{0.12}$ & 8.32$^{+0.05}_{-0.08}$ & 8.13$^{+0.22}_{-0.05}$ & 1.38$^{+0.46}_{-0.41}$ & 2.14$^{+0.64}_{-1.24}$ \\
UGC~04459 & 7.31$\pm{0.12}$ & 7.29$\pm{0.12}$ & 7.32$^{+0.09}_{-0.06}$ & 7.15$^{+0.21}_{-0.00}$ & 0.98$^{+0.30}_{-0.34}$ & 1.45$^{+0.40}_{-0.81}$ \\
UGC~04483 & 6.97$\pm{0.12}$ & \nodata & 6.77$^{+0.12}_{-0.07}$ & 6.67$^{+0.18}_{-0.03}$ & 1.58$^{+0.51}_{-0.62}$ & 2.00$^{+0.57}_{-1.00}$ \\
UGC~05139 & 7.84$\pm{0.12}$ & 7.68$\pm{0.12}$ & 7.72$^{+0.04}_{-0.10}$ & 7.50$^{+0.25}_{-0.00}$ & 1.32$^{+0.47}_{-0.38}$ & 2.19$^{+0.61}_{-1.40}$ \\
UGC~05364 & 6.42$\pm{0.12}$ & 6.22$\pm{0.12}$ & 6.17$^{+0.04}_{-0.02}$ & 6.19$^{+0.03}_{-0.03}$ & 1.78$^{+0.50}_{-0.52}$ & 1.70$^{+0.48}_{-0.48}$ \\
SextansB & 7.57$\pm{0.12}$ & 6.92$\pm{0.12}$ & 6.98$^{+0.12}_{-0.01}$ & 6.92$^{+0.16}_{-0.06}$ & 3.89$^{+1.08}_{-1.52}$ & 4.47$^{+1.38}_{-2.06}$ \\
NGC~3109 & 8.14$\pm{0.12}$ & 7.77$\pm{0.12}$ & 7.65$^{+0.09}_{-0.07}$ & 7.58$^{+0.28}_{-0.04}$ & 3.09$^{+0.99}_{-1.07}$ & 3.63$^{+1.06}_{-2.55}$ \\
SextansA & 7.52$\pm{0.12}$ & 7.17$\pm{0.12}$ & 7.18$^{+0.00}_{-0.13}$ & 6.96$^{+0.18}_{-0.04}$ & 2.19$^{+0.89}_{-0.61}$ & 3.63$^{+1.06}_{-1.81}$ \\
UGC~05666 & 8.84$\pm{0.12}$ & 8.65$\pm{0.12}$ & 8.66$^{+0.01}_{-0.11}$ & 8.42$^{+0.22}_{-0.01}$ & 1.51$^{+0.57}_{-0.42}$ & 2.63$^{+0.73}_{-1.52}$ \\
NGC~3738 & 8.74$\pm{0.12}$ & 8.71$\pm{0.12}$ & 8.70$^{+0.01}_{-0.17}$ & 8.50$^{+0.30}_{-0.00}$ & 1.10$^{+0.53}_{-0.31}$ & 1.74$^{+0.48}_{-1.29}$ \\
NGC~3741 & 7.15$\pm{0.12}$ & 7.15$\pm{0.12}$ & 7.14$^{+0.09}_{-0.07}$ & 7.05$^{+0.13}_{-0.05}$ & 1.02$^{+0.33}_{-0.35}$ & 1.26$^{+0.38}_{-0.51}$ \\
UGC~06817 & 7.32$\pm{0.12}$ & 7.06$\pm{0.12}$ & 7.23$^{+0.03}_{-0.12}$ & 7.09$^{+0.13}_{-0.04}$ & 1.23$^{+0.48}_{-0.35}$ & 1.70$^{+0.50}_{-0.69}$ \\
NGC~4068 & 8.05$\pm{0.12}$ & 8.01$\pm{0.12}$ & 8.05$^{+0.03}_{-0.12}$ & 7.87$^{+0.22}_{-0.03}$ & 1.00$^{+0.39}_{-0.28}$ & 1.51$^{+0.43}_{-0.87}$ \\
NGC~4163 & 7.64$\pm{0.12}$ & 7.64$\pm{0.12}$ & 7.72$^{+0.01}_{-0.10}$ & 7.62$^{+0.09}_{-0.07}$ & 0.83$^{+0.30}_{-0.23}$ & 1.05$^{+0.34}_{-0.36}$ \\
CGCG269-049 & 7.00$\pm{0.12}$ & 6.74$\pm{0.12}$ & 6.93$^{+0.30}_{-0.14}$ & 7.02$^{+0.02}_{-0.22}$ & 1.17$^{+0.50}_{-0.87}$ & 0.95$^{+0.55}_{-0.27}$ \\
UGCA~281 & 7.57$\pm{0.12}$ & 7.73$\pm{0.12}$ & 7.47$^{+0.05}_{-0.17}$ & 7.35$^{+0.44}_{-0.21}$ & 1.26$^{+0.60}_{-0.38}$ & 1.66$^{+0.92}_{-1.74}$ \\
UGC~07577 & 7.82$\pm{0.12}$ & 7.47$\pm{0.12}$ & 7.64$^{+0.00}_{-0.20}$ & 7.39$^{+0.33}_{-0.00}$ & 1.51$^{+0.81}_{-0.42}$ & 2.69$^{+0.74}_{-2.18}$ \\
NGC~4449 & 9.40$\pm{0.12}$ & 9.38$\pm{0.12}$ & 8.99$^{+0.14}_{-0.12}$ & 9.01$^{+0.03}_{-0.12}$ & 2.57$^{+1.00}_{-1.09}$ & 2.45$^{+0.96}_{-0.70}$ \\
UGCA~292 & 7.16$\pm{0.12}$ & 7.10$\pm{0.12}$ & 6.33$^{+0.23}_{-0.11}$ & 6.21$^{+0.31}_{-0.01}$ & 6.76$^{+2.53}_{-4.04}$ & 8.91$^{+2.47}_{-6.82}$ \\
UGC~08024 & 7.47$\pm{0.12}$ & 7.47$\pm{0.12}$ & 7.41$^{+0.08}_{-0.08}$ & 7.22$^{+0.12}_{-0.14}$ & 1.15$^{+0.38}_{-0.38}$ & 1.78$^{+0.76}_{-0.70}$ \\
UGC~08091 & 6.73$\pm{0.12}$ & 6.77$\pm{0.12}$ & 6.68$^{+0.03}_{-0.01}$ & 6.65$^{+0.02}_{-0.04}$ & 1.12$^{+0.31}_{-0.32}$ & 1.20$^{+0.35}_{-0.34}$ \\
UGC~08201 & 8.01$\pm{0.12}$ & 7.90$\pm{0.12}$ & 8.09$^{+0.03}_{-0.12}$ & 7.83$^{+0.14}_{-0.11}$ & 0.83$^{+0.32}_{-0.24}$ & 1.51$^{+0.57}_{-0.64}$ \\
UGC~08508 & 7.28$\pm{0.12}$ & 7.25$\pm{0.12}$ & 7.28$^{+0.00}_{-0.12}$ & 7.15$^{+0.06}_{-0.08}$ & 1.0$^{+0.39}_{-0.28}$ & 1.35$^{+0.45}_{-0.42}$ \\
UGC~08638 & 7.59$\pm{0.12}$ & 6.83$\pm{0.12}$ & 7.57$^{+0.03}_{-0.15}$ & 7.49$^{+0.11}_{-0.04}$ & 1.05$^{+0.46}_{-0.30}$ & 1.26$^{+0.37}_{-0.47}$ \\
UGC~08651 & 7.25$\pm{0.12}$ & 7.19$\pm{0.12}$ & 7.22$^{+0.10}_{-0.05}$ & 7.03$^{+0.23}_{-0.07}$ & 1.07$^{+0.32}_{-0.38}$ & 1.66$^{+0.53}_{-0.99}$ \\
NGC~5253 & 8.94$\pm{0.12}$ & 8.93$\pm{0.12}$ & 8.78$^{+0.00}_{-0.17}$ & 8.56$^{+0.16}_{-0.10}$ & 1.45$^{+0.69}_{-0.40}$ & 2.40$^{+0.86}_{-1.11}$ \\
UGC~09128 & 6.85$\pm{0.12}$ & 6.88$\pm{0.12}$ & 6.76$^{+0.20}_{-0.03}$ & 6.67$^{+0.19}_{-0.00}$ & 1.23$^{+0.35}_{-0.66}$ & 1.51$^{+0.42}_{-0.78}$ \\
UGC~09240 & 7.59$\pm{0.12}$ & 7.55$\pm{0.12}$ & 7.59$^{+0.08}_{-0.05}$ & 7.45$^{+0.17}_{-0.07}$ & 1.00$^{+0.30}_{-0.33}$ & 1.38$^{+0.44}_{-0.66}$ \\
IC~4662 & 8.18$\pm{0.12}$ & 8.17$\pm{0.12}$ & 8.31$^{+0.02}_{-0.13}$ & 8.14$^{+0.25}_{-0.00}$ & 0.74$^{+0.30}_{-0.21}$ & 1.10$^{+0.30}_{-0.70}$ \\
NGC~6789 & 7.65$\pm{0.12}$ & 7.49$\pm{0.12}$ & 7.72$^{+0.11}_{-0.27}$ & 7.48$^{+0.25}_{-0.14}$ & 0.85$^{+0.58}_{-0.32}$ & 1.48$^{+0.63}_{-0.95}$ \\
IC~5152 & 8.25$\pm{0.12}$ & 7.93$\pm{0.12}$ & 7.84$^{+0.08}_{-0.12}$ & 7.76$^{+0.04}_{-0.12}$ & 2.57$^{+1.00}_{-0.85}$ & 3.09$^{+1.21}_{-0.90}$ \\
\enddata
\tablecomments{Stellar masses determined using different methods. Columns 2$-$3: Log of stellar masses and uncertainties based on the 3.6$\micron$ fluxes extrapolated from the surface brightness fits (total) and within the HST footprints (HS FoV). Two galaxies (WLM, UGC~04483) do not have 3.6$\micron$-based stellar masses in the HST field of view as the HST data used in GLOW are based on new observations that were not available when the stellar mass comparison with 3.6$\micron$ fluxes was performed. Columns 4$-$5: Log of the stellar masses and uncertainties derived from fitting the CMDs based on the PARSEC and MIST stellar libraries and assuming a recycling fraction; see text for details. Columns 6$-$7: Scaling factors of the \mstar\ from the total 3.6$\micron$ fluxes compared with \mstar\ measured from the HST data. These scaling factors are used in the calculation of the metal content retained in the stars. }
\end{deluxetable*}

%% file: tab6.tex
\begin{deluxetable*}{lccccccccc}
\label{tab:hst_data}
\tablecaption{Summary of HST Observations}
\tablehead{
\colhead{Galaxy}		& \colhead{Inst.} 	& \colhead{PID} & \colhead{F606W }	& \colhead{F814W}	& \colhead{Mult.\ Ptgs.} & \colhead{F606W 50\%} & \colhead{F814W 50\%} & \colhead{dA$_V$ PARSEC} & \colhead{dA$_V$ MIST} 
}
\decimalcolnumbers
\startdata
WLM & ACS & 13768 & 27360$^a$ & 34050 & \nodata &  29.46 & 28.52 & 0.20 & 0.25  \\
UGC~00685 & ACS & 10210 & 934 & 1226 & \nodata & 27.54 & 26.64 & 0.40 & 0.50 \\
NGC~0625 & WFPC2 & 8708 & 5200$^a$ & 10400 & \nodata & 26.56 & 25.34 & 0.80 & 0.95 \\
NGC~0784 & ACS & 10210 & 933 & 1226 & \nodata & 27.47 & 26.52 & 0.35 & 0.25 \\
NGC~1569 & ACS & 10885 & 19568 & 19568 & \nodata & 27.24 & 26.18 & 0.15 & 0.95 \\
NGC~2366 & ACS & 10605 & 4780$^a$ & 4780 & Y & 28.11 & 27.18 & 0.40 & 0.35 \\
UGC~04305 & ACS & 10605 & 4660$^a$ & 4660 & Y & 28.06 & 27.12 & 0.40 & 0.25 \\
UGC~04459 & ACS & 10605 & 4768$^a$ & 4768 & \nodata & 28.25 & 27.44 & 0.20 & 0.20 \\
UGC~04483 & WFC3 & 15194 & 36120 & 51600 & \nodata & 29.09 & 27.99 & 0.00 & 0.00 \\
UGC~05139 & ACS & 10605 & 5829$^a$ & 5936 & \nodata & 28.47 & 27.65 & 0.25 & 0.25 \\
UGC~05364 & ACS & 10590 & 19200$^a$ & 19520 & \nodata &  29.43 & 28.47 & 0.00 & 0.00 \\
SextansB & WFPC2 & 10915 & 2700 & 3900 & \nodata & 26.09 & 24.99 & 0.25 & 0.25 \\
NGC~3109 & WFPC2 & 11307 & 2400 & 2400 & Y & 25.84 & 24.72 & 0.25 & 0.30 \\
SextansA & WFPC2 & 7496 & 19200$^a$ & 38400 & Y & 26.31 & 25.12 & 0.25 & 0.15 \\
UGC~05666 & ACS & 10605 & 4784$^a$ & 4784 & \nodata & 28.14 & 27.25 & 0.50 & 0.45 \\
NGC~3738 & ACS & 12546 & 900 & 900 & \nodata & 27.51 & 26.51 & 0.55 & 0.15 \\
NGC~3741 & ACS & 10915 & 2262$^a$ & 2331 & \nodata & 28.06 & 27.08 & 0.20 & 0.10 \\
UGC~06817 & WFPC2 & 11986 & 4800 & 7200 & \nodata & 26.76 & 25.63 & 0.50 & 0.20 \\
NGC~4068 & ACS & 9771 & 1200 & 900 & \nodata & 27.64 & 26.61 & 0.30 & 0.10 \\
NGC~4163 & ACS & 10915 & 2292 & 2250 & Y & 28.05 & 27.01 & 0.00 & 0.00 \\
CGCG269-049 & WFPC2 & 11986 & 10700 & 19300 & \nodata & 26.69 & 25.44 & 0.85 & 0.75 \\
UGCA~281 & ACS & 10905 & 938 & 1148 & \nodata & 27.67 & 26.75 & 0.05 & 0.00 \\
UGC~07577 & WFPC2 & 11986 & 4800 & 7200 & \nodata & 26.34 & 25.21 & 0.45 & 0.35 \\
NGC~4449 & ACS & 10585 & 2460$^a$ & 2060 & Y & 27.05 & 25.94 & 0.95 & 0.95 \\
UGCA~292 & ACS & 10915 & 2250$^a$ & 2274 & \nodata & 28.30 & 27.36 & 0.00 & 0.05 \\
UGC~08024 & ACS & 10905 & 924 & 1128 & \nodata & 27.69 & 26.82 & 0.10 & 0.00 \\
UGC~08091 & ACS & 10915 & 5984$^a$ & 5271 & \nodata & 28.24 & 27.28 & 0.15 & 0.10 \\
UGC~08201 & ACS & 10605 & 4768$^a$ & 4768 & \nodata & 28.21 & 27.36 & 0.20 & 0.10 \\
UGC~08508 & ACS & 10915 & 2349$^a$ & 2280 & \nodata & 28.08 & 27.10 & 0.20 & 0.00 \\
UGC~08638 & ACS & 9771 & 1200 & 900 & \nodata & 27.71 & 26.64 & 0.25 & 0.00 \\
UGC~08651 & ACS & 10210 & 1016 & 1209 & \nodata & 27.76 & 26.85 & 0.25 & 0.05 \\
NGC~5253 & ACS & 10765 & 1200$^a$ & 1200 & Y & 26.99 & 25.99 & 0.45 & 0.30 \\
UGC~09128 & ACS & 10210 & 985 & 1174 & \nodata & 27.82 & 26.93 & 0.20 & 0.30 \\
UGC~09240 & ACS & 10915 & 2301 & 2265 & \nodata & 28.06 & 27.06 & 0.20 & 0.00 \\
IC~4662 & ACS & 9771 & 1200 & 900 & \nodata & 26.67 & 25.65 & 0.50 & 0.40 \\
NGC~6789 & WFPC2 & 8122 & 8200$^a$ & 8200 & \nodata & 26.80 & 25.62 & 0.15 & 0.25 \\
IC~5152 & WFPC2 & 11986 & 4800 & 9600 & Y & 25.35 & 24.24 & 0.50 & 0.40 \\
\enddata
\tablecomments{$^a$ Indicates that the data obtained used the F555W or F475W filter instead of the F606W filter. Values for fields with multiple pointings are representative of the data for that galaxy.}
\end{deluxetable*}

%% file: tab7.tex
\begin{deluxetable*}{llDDllDlDDD}
\tabletypesize{\footnotesize}
\setlength{\tabcolsep}{2pt}
\label{tab:hi}
\tablecaption{HI Observations and HI Properties}
\tablehead{
\colhead{Galaxy} 	& \colhead{Telescope} & \multicolumn2c{V$_{\rm helio}$} & \multicolumn2c{W50} 		& \colhead{$i$} & \colhead{V$_{\rm rot}$}		& \multicolumn2c{HI Flux} 		& \colhead{Reference} & \multicolumn2c{log(M$_{HI}$/\msun)} & \multicolumn2c{log(M$_{HI}$/\msun)} & \multicolumn2c{M$_{HI}$/\mstar} \\
\colhead{}		& \colhead{}		   & \multicolumn2c{(km s$^{-1}$)}     & \multicolumn2c{(km s$^{-1}$)}  & \colhead{($^{\circ}$)} & \colhead{(km s$^{-1}$)} & \multicolumn2c{(Jy km s$^{-1}$)} & \colhead{} & \multicolumn2c{total} & \multicolumn2c{4.4 $\alpha_{3.6\mu\rm{m}}$}	    & \multicolumn2c{total}
}
\decimalcolnumbers
\startdata
WLM & VLA & -123.50 & 64.5 & 78 & 32 & 308.39 & This work & 7.84 & 7.84 & 1.45 \\
UGC~00685 & VLA & 154.84 & 73.5 & 57 & 43 & 13.41 & This work & 7.86 & 7.84 & 0.93 \\
NGC~0625 & Parkes 64m & 395.0 & 82.0 & 86 & 37 & 30.67 & \citet{Courtois2009} & 8.07 & 8.06 & 0.20 \\
NGC~0784 & Arecibo 305m & 193.0 & 88.0 & 0 & \nodata & 66.36 & \citet{Hoffman2019} & 8.65 & 8.59 & 1.00 \\
NGC~1569 & VLA & -90.29 & 63.8 & 71 & 33 & 102.19 & This work & 8.39 & 8.31 & 0.26 \\
NGC~2366 & VLA & 101.0 & 99.0 & 0 & \nodata & 198.54 & This work & 8.70 & 8.65 & 2.00 \\
UGC~04305 & VLA & 156.48 & 53.98 & 44 & 38 & 220.34 & \citet{Hunter2023} & 8.80 & 8.62 & 2.19 \\
UGC~04459 & VLA & 19.62 & 30.5 & 30 & \nodata & 22.27 & This work & 7.85 & 7.82 & 3.47 \\
UGC~04483 & VLA & 155.44 & 33.9 & 61 & 19 & 12.33 & This work & 7.57 & 7.40 & 3.98 \\
UGC~05139 & VLA & 140.11 & 31.0 & 20 & \nodata & 53.25 & This work & 8.31 & 8.30 & 2.95 \\
UGC~05364 & VLA & 23.52 & 18.2 & 72 & 9 & 48.90 & This work & 6.80 & 6.73 & 2.40 \\
SextansB & VLA & 298.58 & 41.6 & 54 & 25 & 98.30 & This work & 7.68 & 7.59 & 1.29 \\
NGC~3109 & VLA & 405.08 & 110.8 & 0 & \nodata & 752.79 & This work & 8.50 & 8.38 & 2.29 \\
SextansA & VLA & 324.09 & 46.8 & 34 & 41 & 155.55 & This work & 7.89 & 7.80 & 2.34 \\
UGC~05666 & GBT 300 ft & 57.77 & 108.9 & 0 & \nodata & 406.13 & \citet{Peng2023} & 9.17 & 9.09 & 2.14 \\
NGC~3738 & VLA & 221.27 & 75.2 & 55 & 45 & 20.92 & This work & 8.14 & 8.13 & 0.25 \\
NGC~3741 & VLA & 227.62 & 93.4 &  52 & 58 & 41.37 & This work & 8.00 & 7.35 & 7.08 \\
UGC~06817 & GBT 42m & 243.0 & 36.0 & 60 & 16 & 47.14 & \citet{Springob2005} & 7.89 & 7.77 & 3.72 \\
NGC~4068 & VLA & 210.00 & 56.0 & 65 & 30 & 40.40 & This work & 8.26 & 8.24 & 1.62 \\
NGC~4163 & VLA & 163.80 & 26.7 & 56 & 16 & 8.73 & This work & 7.26 & 7.26 & 0.42 \\
CGCG269-049 & VLA & 159.00 & 29.1 & 64 & 16 & 5.99 & This work & 7.48 & 7.44 & 3.02 \\
UGCA~281 & VLA & 284.09 & 47.5 & 46 & 32 & 11.57 & This work & 7.95 & 7.71 & 2.40 \\
UGC~07577 & VLA & 195.65 & 28.4 & 50 & 15 & 21.84 & This work & 7.54 & 7.54 & 0.52 \\
NGC~4449 & Effelsberg 100m & 214.0 & 136.0 & 62 & 63 & 712.0 & \citet{Bajaja1994} & 9.49 & 9.07 & 1.23 \\
UGCA~292 & VLA & 308.38 & 28.2 &  36 & 23 &15.95 & This work & 7.75 & 7.67 & 3.89 \\
UGC~08024 & VLA & 376.00 & 91.0 & 167 & 49 & 02.69 & This work & 8.60 & 8.11 & 13.49 \\
UGC~08091 & VLA & 213.80 & 26.6 & 54 & 16 & 8.76 & This work & 7.00 & 6.94 & 1.86 \\
UGC~08201 & VLA & 30.10 & 46.7 & 74 & 24 & 33.50 & This work & 8.26 & 8.19 & 1.78 \\
UGC~08508 & Effelsberg 100m & 62.0 & 52.0 & 58 & 28 & 14.2 & \citet{Huchtmeier1989} & 7.38 & 7.29 & 1.26 \\
UGC~08638 & VLA & 274.50 & 31.7 & 65 & 17 & 4.64 & This work & 7.30 & 7.30 & 0.51 \\
UGC~08651 & VLA & 200.78 & 42.4 & 67 & 22 & 13.70 & This work & 7.49 & 7.47 & 1.74 \\
NGC~5253 & VLA & 406.89 & 64.5 & 72 & 33 & 42.73 & This work & 8.08 & 8.04 & 0.14 \\
UGC~09128 & VLA & 153.50 & 35.6 & 61 & 20 & 13.02 & This work & 7.21 & 7.19 & 2.29 \\
UGC~09240 & GBT 100 m & 150.0 & 44.0 & 41 & 34 & 23.96 & \citet{Hogg2007} & 7.66 & 7.57 & 1.17 \\
IC~4662 & ATCA & 302.0 & 86.0 & 55 & 51 & 103.5 & LVHIS & 8.2 & 7.50 & 1.05 \\
NGC~6789 & VLA & -148.80 & 62.5 &  36 & 52 & 4.94 & This work & 7.17 & 7.16 & 0.33 \\
IC~5152 & ATCA & 122.0 & 84.0 & 62 & 47 & 98.6 & LVHIS & 7.95 & 7.70 & 0.50 \\
\enddata
\tablecomments{Summary of the \hi\ data and associated global measurements used in our analysis and relevant references. We also calculate the inclination angles ($i$) based on the 3.6$\mu$m geometry (see Equation~\ref{eq:incl}, and inclination angle corrected rotational velocities (V$_{\rm rot}$) based on the W50 values (see Equation~\ref{eq:vrot}). We also provide the \hi\ mass contained within 4.4 scale lengths ($\alpha_{3.6\mu \rm m}$) in the second to last column. The detailed radial profiles of the \hi\ fluxes are provided in Table~\ref{tab:hi_profiles}. }
\end{deluxetable*}

%% file: tab8.tex
\begin{deluxetable*}{lccccccccccccccc}
\tabletypesize{\footnotesize}
\setlength{\tabcolsep}{5pt}
\label{tab:hi_profiles}
\tablecaption{HI Radial Profiles}
\tablehead{
\colhead{Galaxy} & \colhead{M$_{\rm HI}$/M$_*$} & \multicolumn{14}{c}{Percent of \hi\ Flux within Annuli of 1.1$-$30 3.6$\micron$ Scale Lengths}\\
\colhead{} & \colhead{} & \colhead{1.1$\alpha$ } & \colhead{ 2.2$\alpha$ } & \colhead{ 3.3$\alpha$ } & \colhead{ 4.4$\alpha$ } & \colhead{ 5.5$\alpha$ } & \colhead{ 6.6$\alpha$ } & \colhead{ 7.7$\alpha$ } & \colhead{ 8.8$\alpha$ } & \colhead{ 10$\alpha$ } & \colhead{ 12$\alpha$ } & \colhead{ 15$\alpha$ } & \colhead{ 20$\alpha$ } & \colhead{ 25$\alpha$ } & \colhead{ 30$\alpha$ } \\
}
\decimalcolnumbers
\startdata
WLM & 1.45 & 33 & 78 & 97 & 99 & 100 & 100 & 100 & 100 & 100 & 100 & 100 & 100 & 100 & 100 \\
UGC~00685 & 0.93 & 23 & 59 & 83 & 95 & 98 & 99 & 99 & 99 & 99 & 100 & 100 & 100 & 100 & 100 \\
NGC~0625 & 0.20 & 44 & 84 & 97 & 99 & 99 & 99 & 100 & 100 & 100 & 100 & 100 & 100 & 100 & 100 \\
NGC~0784 & 1.00 & 14 & 44 & 72 & 87 & 94 & 97 & 98 & 99 & 99 & 99 & 99 & 100 & 100 & 100 \\
NGC~1569 & 0.26 & 9 & 35 & 65 & 84 & 92 & 96 & 98 & 98 & 99 & 99 & 99 & 99 & 100 & 100 \\
NGC~2366 & 2.00 & 14 & 40 & 69 & 89 & 96 & 99 & 99 & 99 & 99 & 100 & 100 & 100 & 100 & 100 \\
UGC~04305 & 2.19 & 4 & 18 & 43 & 67 & 87 & 98 & 99 & 99 & 100 & 100 & 100 & 100 & 100 & 100 \\
UGC~04459 & 3.47 & 16 & 52 & 78 & 92 & 98 & 99 & 99 & 100 & 100 & 100 & 100 & 100 & 100 & 100 \\
UGC~04483 & 3.98 & 7 & 26 & 48 & 67 & 81 & 90 & 94 & 97 & 98 & 99 & 99 & 100 & 100 & 100 \\
UGC~05139 & 2.95 & 12 & 58 & 86 & 98 & 99 & 100 & 100 & 100 & 100 & 100 & 100 & 100 & 100 & 100 \\
UGC~05364 & 2.40 & 8 & 35 & 64 & 85 & 96 & 99 & 99 & 99 & 100 & 100 & 100 & 100 & 100 & 100 \\
SextansB & 1.29 & 11 & 34 & 58 & 82 & 96 & 99 & 100 & 100 & 100 & 100 & 100 & 100 & 100 & 100 \\
NGC~3109 & 2.29 & 8 & 31 & 56 & 76 & 92 & 98 & 99 & 99 & 99 & 99 & 100 & 100 & 100 & 100 \\
SextansA & 2.34 & 3 & 17 & 53 & 82 & 96 & 99 & 99 & 99 & 99 & 100 & 100 & 100 & 100 & 100 \\
UGC~05666 & 2.14 & 6 & 29 & 59 & 84 & 97 & 99 & 99 & 100 & 100 & 100 & 100 & 100 & 100 & 100 \\
NGC~3738 & 0.25 & 35 & 77 & 93 & 97 & 99 & 99 & 99 & 99 & 99 & 100 & 100 & 100 & 100 & 100 \\
NGC~3741 & 7.08 & 2 & 9 & 15 & 22 & 30 & 39 & 48 & 55 & 62 & 73 & 85 & 96 & 99 & 100 \\
UGC~06817 & 3.72 & 11 & 35 & 55 & 75 & 90 & 97 & 99 & 99 & 100 & 100 & 100 & 100 & 100 & 100 \\
NGC~4068 & 1.62 & 8 & 40 & 79 & 94 & 98 & 99 & 99 & 99 & 99 & 100 & 100 & 100 & 100 & 100 \\
NGC~4163 & 0.42 & 24 & 78 & 95 & 98 & 99 & 99 & 99 & 100 & 100 & 100 & 100 & 100 & 100 & 100 \\
CGCG269-049 & 3.02 & 18 & 54 & 80 & 92 & 96 & 98 & 99 & 99 & 99 & 100 & 100 & 100 & 100 & 100 \\
UGCA~281 & 2.40 & 10 & 28 & 43 & 58 & 70 & 79 & 87 & 93 & 97 & 99 & 100 & 100 & 100 & 100 \\
UGC~07577 & 0.52 & 25 & 82 & 97 & 99 & 100 & 100 & 100 & 100 & 100 & 100 & 100 & 100 & 100 & 100 \\
NGC~4449 & 1.23 & 2 & 9 & 25 & 38 & 50 & 64 & 72 & 76 & 80 & 86 & 90 & 98 & 99 & 100 \\
UGCA~292 & 3.89 & 9 & 36 & 65 & 84 & 93 & 97 & 99 & 99 & 99 & 99 & 100 & 100 & 100 & 100 \\
UGC~08024 & 13.49 & 2 & 10 & 20 & 33 & 44 & 55 & 64 & 73 & 81 & 91 & 98 & 99 & 100 & 100 \\
UGC~08091 & 1.86 & 8 & 37 & 70 & 89 & 97 & 99 & 99 & 99 & 100 & 100 & 100 & 100 & 100 & 100 \\
UGC~08201 & 1.78 & 6 & 30 & 65 & 85 & 95 & 98 & 99 & 99 & 99 & 100 & 100 & 100 & 100 & 100 \\
UGC~08508 & 1.26 & 16 & 49 & 68 & 82 & 91 & 97 & 99 & 99 & 99 & 100 & 100 & 100 & 100 & 100 \\
UGC~08638 & 0.51 & 36 & 80 & 95 & 99 & 99 & 100 & 100 & 100 & 100 & 100 & 100 & 100 & 100 & 100 \\
UGC~08651 & 1.74 & 10 & 49 & 84 & 96 & 99 & 99 & 100 & 100 & 100 & 100 & 100 & 100 & 100 & 100 \\
NGC~5253 & 0.14 & 27 & 61 & 80 & 93 & 97 & 99 & 99 & 99 & 99 & 100 & 100 & 100 & 100 & 100 \\
UGC~09128 & 2.29 & 24 & 63 & 85 & 95 & 98 & 99 & 99 & 99 & 99 & 100 & 100 & 100 & 100 & 100 \\
UGC~09240 & 1.17 & 14 & 43 & 68 & 83 & 92 & 97 & 98 & 99 & 99 & 99 & 100 & 100 & 100 & 100 \\
IC~4662 & 1.05 & 1 & 6 & 13 & 20 & 27 & 33 & 39 & 45 & 51 & 62 & 76 & 92 & 97 & 100 \\
NGC~6789 & 0.33 & 38 & 85 & 99 & 99 & 100 & 100 & 100 & 100 & 100 & 100 & 100 & 100 & 100 & 100 \\
IC~5152 & 0.50 & 5 & 19 & 37 & 56 & 71 & 82 & 89 & 95 & 98 & 99 & 99 & 100 & 100 & 100 \\
\enddata
\tablecomments{Properties of the \hi\ of the galaxies as a function of radius. Column 2 $-$ Ratio of the total \hi\ to stellar mass in the galaxies. Columns 3 $-$ 16 $-$ Percent of the \hi\ flux measured within annuli defined by 3.6$\micron$ structural parameters as a function of scale lengths in the galaxies. Scale lengths are in increments of 1.1$\alpha$ until 8.8$\alpha$, after which the interval spacing increases.}
\end{deluxetable*}

%% file: tab9.tex
\begin{table*}
\begin{center}
\caption{Characterizing the Environment around the GLOW Galaxies}
\label{tab:environment}
\begin{tabular}{ll clclclc}
\hline \hline
Name &     Group membership &  $\Theta_1$ 	& MD  &  $\Theta_5$ &  NN &  NN sep. 	&  NLG &  NLG sep. \\ 
	&			&			&	&		&	& (Mpc)	&	 & (Mpc)	\\
\hline \hline
WLM &          Local Group &   -1.19 &  M33 &   -1.04 &              Cetus &           0.22 &   M33 &            0.89 \\
UGC~00685 &                Field &     -- &       Field &    -- &          AGC748778 &           1.36 &     NGC~0404* &            2.35 \\
NGC~0625 &             Sculptor &   -0.87 &  ESO245-005 &   -0.87 &         ESO245-005 &           0.30 &     NGC~0253* &            1.43 \\
NGC~0784 &       NGC~0672 Group &   -1.46 &    UGC~01281 &   -1.32 &               KK17 &           0.30 &     NGC~0628* &            2.43 \\
NGC~1569 &          IC342 Group &    1.17 &      IC0342 &    1.18 &               CamB &           0.20 &       IC0342 &            0.31 \\
NGC~2366 &            M81 Group &    0.94 &     NGC~2403 &    1.09 &             DDO044 &           0.14 &      NGC~2403 &            0.21 \\
UGC~04305 &            M81 Group &    0.77 &  M81 &    1.02 &             KDG052 &           0.17 &   M81 &            0.55 \\
UGC~04459 &            M81 Group &    0.81 &  M81 &    0.99 &             KDG052 &           0.34 &   M81 &            0.53 \\
UGC~04483 &            M81 Group &    1.14 &      HoII &    1.48 &             HoII &           0.11 &       HolmII &            0.11 \\
UGC~05139 &            M81 Group &    1.48 &  M81 &    1.62 &  [CKT2009]d0926+70 &           0.10 &   M81 &            0.32 \\
UGC~05364 &          Local Group &   -2.21 &     NGC~3109 &   -1.93 &               LeoT &           0.44 &  M33* &            1.51 \\
SextansB &  NGC~3109 Assoc. &   -1.65 &        SextansA &   -1.51 &               SextansA &           0.26 &     NGC~2403* &            2.14 \\
NGC~3109 &  NGC~3109 Assoc. &   -0.27 &      Antlia &   -0.27 &             Antlia &           0.04 &         LMC* &            2.00 \\
SextansA &  NGC~3109 Assoc. &   -1.33 &        SextansB &   -1.12 &               SextansB &           0.25 &     NGC~5102* &            2.22 \\
UGC~05666 &            M81 Group &    1.75 &  M81 &    1.90 &             DDO078 &           0.09 &   M81 &            0.26 \\
NGC~3738 &           Can Ven I  &   -3.11 &     UGC~A281 &   -2.74 &       SBS 1224+533 &           0.71 &     NGC~4605* &            1.03 \\
NGC~3741 &            Can Ven I &   -1.58 &     NGC~4214 &   -1.42 &           UGC~06541 &           0.69 &     NGC~2976* &            1.52 \\
UGC~06817 &           Can Ven I  &   -0.87 &     NGC~4214 &   -0.69 &            NGC~4190 &           0.32 &     NGC~2976* &            1.72 \\
NGC~4068 &            Can Ven I &   -0.90 &     NGC~4449 &   -0.67 &           UGC~07298 &           0.23 &      NGC~4449 &            0.75 \\
NGC~4163 &            Can Ven I &    1.27 &      KDG090 &    1.62 &             KDG090 &           0.03 &     NGC~2976* &            1.94 \\
CGCG269-049 &            Can Ven I &   -2.92 &    UGC~08245 &   -2.83 &           UGC~04879 &           0.81 &     NGC~2403* &            1.78 \\
UGCA~281 &        NGC~5128 Group &   -0.20 &     NGC~4490 &   -0.16 &            NGC~4707 &           0.45 &      NGC~4490 &            0.69 \\
UGC~07577 &            Can Ven I &   -1.05 &     NGC~4214 &   -0.92 &             DDO099 &           0.38 &  M82* &            1.67 \\
 NGC~4449 &            Can Ven I &    0.40 &     NGC~4736 &    0.51 &      LV J1228+4358 &           0.20 &      NGC~4736 &            0.54 \\
UGCA~292 &            Can Ven I &   -1.18 &     NGC~4449 &   -0.86 &             DDO154 &           0.59 &      NGC~4449 &            0.93 \\
UGC~08024 &           Can Ven I  &    0.42 &     NGC~4826 &    0.43 &            NGC~4826 &           0.51 &      NGC~4826 &            0.51 \\
UGC~08091 &               Field  &   -3.51 &      UGC~09128 &   -3.39 &             DDO187 &           0.74 &     NGC~5253* &            2.10 \\
UGC~08201 &            M81 Group &   -0.60 &     NGC~4236 &   -0.45 &            NGC~4236 &           0.56 &      NGC~4236 &            0.56 \\
UGC~08508 &           M101 Group &   -2.41 &      UGC~09420 &   -2.03 &             DDO190 &           0.62 &  M82* &            1.51 \\
UGC~08638 &            Can Ven I &   -0.13 &     NGC~4826 &   -0.13 &              KK177 &           0.67 &      NGC~4826 &            0.77 \\
NGC~5253 &        NGC~5128 Group &    0.31 &     NGC~5128 &    0.52 &        [KK2000] 58 &           0.30 &      NGC~5128 &            0.75 \\
UGC~09128 &               Field  &   -2.97 &      UGC~09420 &   -2.77 &              KK230 &           0.48 &     NGC~4449* &            2.23 \\
UGC~09240 &            Can Ven I &   -2.47 &      DDO181 &   -2.10 &             DDO181 &           0.49 &     NGC~4449* &            1.69 \\
IC~4662 &                Field &     -- &       Field &    -- &             IC3104 &           1.21 &  ESO270-017* &            1.92 \\
NGC~6789 &                Field &     -- &       Field &    -- &           UGC~11411 &           1.24 &     NGC~4236* &            2.36 \\
IC~5152 &          Local Group &   -1.33 &     NGC~0055 &   -1.32 &              KK258 &           0.74 &      NGC~0055 &            0.89 \\
\hline \hline
\end{tabular}
\end{center}
\tablecomments{Environment descriptors for galaxies in the GLOW sample. Columns: (1) Galaxy name, (2) galaxy group membership based on literature search, (3) tidal index for the largest tidal force, (4) the main disturber (MD), (5) sum of the tidal indices for the five most important neighbors, (6) the nearest neighbor (NN), (7) the separation between the galaxy and NN, (8) the nearest large (\mstar $>10^9$ \msun) galaxy (NLG); any NLG separated from the GLOW galaxy by more than 1 Mpc are marked with an asterisk, and (9) distance from the galaxy to the NLG.} 
\end{table*}